\documentclass[10pt,aps,prb,twocolumn,floatfix,floats,showpacs,superscriptaddress,raggedbottom]{revtex4-1}
\usepackage{graphicx,latexsym}
\usepackage{dcolumn}
\usepackage{amssymb,amsmath,bm}
\usepackage{subfigure}

\newcommand{\brames}[1]{( #1|}
\newcommand{\ketmes}[1]{|#1)}

\usepackage{hyperref}
\hypersetup{
    pdfnewwindow=true,       
    colorlinks=true,         
    linkcolor=blue,          
    citecolor=blue,          
    filecolor=magenta,       
    urlcolor=black           
}

\def\eq#1{Eq.\ (\ref{#1})}
\def\mb#1{\mbox{\boldmath$#1$}}
\def\fig#1{Fig.\ \ref{#1}}

\begin{document}

\title{Geometry and linearly polarized cavity photon effects
on the charge and spin currents of spin-orbit interacting electrons in a quantum ring.}

\author{Thorsten Arnold}
\email{tla1@hi.is}
\affiliation{Science Institute, University of Iceland, Dunhaga 3,
        IS-107 Reykjavik, Iceland}
\author{Chi-Shung Tang}
\email{cstang@nuu.edu.tw}
\affiliation{Department of Mechanical Engineering,
        National United University,
        1, Lienda, Miaoli 36003, Taiwan}
\author{Andrei Manolescu}
\affiliation{School of Science and Engineering, Reykjavik University, 
        Menntavegur 1, IS-101 Reykjavik, Iceland}
\author{Vidar Gudmundsson}
\email{vidar@hi.is}
\affiliation{Science Institute, University of Iceland, Dunhaga 3,
        IS-107 Reykjavik, Iceland}
%

\begin{abstract}
 \ We calculate the persistent spin current inside a quantum ring
 as a function of the strength of the Rashba or Dresselhaus
 spin-orbit interaction.
 We provide analytical results for the spin current of a one-dimensional
 (1D) ring of non-interacting electrons for comparison.
 Furthermore, we calculate the time evolution in the transient regime
 of a two-dimensional (2D) quantum ring connected to electrically biased semi-infinite
 leads using a time-convolutionless
 non-Markovian generalized master equation. In the latter case, the electrons are  correlated via the Coulomb interaction
 and the ring can be embedded in a photon cavity
 with a single mode of linearly polarized photon field. 
 The electron-electron and electron-photon interactions are described by exact numerical diagonalization.
 The photon field can be polarized perpendicular or parallel to the
 charge transport.
 We find a pronounced charge current dip associated with many-electron level crossings at the Aharonov-Casher phase $\Delta\Phi=\pi$, which can be disguised by linearly polarized light.
 Qualitative agreement is found for the spin currents of the 1D and 2D ring. Quantatively, however, the spin currents are weaker in the more realistic
 2D ring, especially for weak spin-orbit interaction, but can be considerably enhanced 
 with the aid of a linearly polarized electromagnetic field.
 Specific spin current symmetries relating the Dresselhaus spin-orbit interaction case to
 the Rashba one are found to hold for the 2D ring in the photon cavity.
\end{abstract}


\maketitle
\section{Introduction}
Geometrical phases have captured much interest in the field of 
quantum transport. Electrons in a non-trivially connected region
like a quantum ring can show a variety of geometrical phases. 
An Aharonov-Bohm (AB) phase~\cite{PhysRev.115.485} is
acquired by a charged particle moving around a magnetic flux. 
An Aharonov-Casher (AC) phase~\cite{PhysRevLett.53.319} is acquired by a particle
with magnetic moment encircling, for example, a charged line. 
The Aharonov-Anadan (AA) phase~\cite{PhysRevLett.58.1593} is the remaining phase
of the AC phase
when subtracting the dynamical part.
The Berry phase~\cite{Berry_phase} is the adiabatic approximation of the AA phase.
Transport properties of magnetic-flux threaded rings~\cite{Szafran05:165301,Buchholz2010,Buttiker30.1982,Pichugin56:9662} 
have been investigated and the influence of a cavity photon mode
on the AB oscillations explored.~\cite{PhysRevB.87.035314}
Furthermore, the magnetic field leads to persistent charge currents~\cite{PhysRevB.37.6050}.
Both, the persistent current~\cite{Tan99:5626} and the conductance through the ring show characteristic oscillations 
with period $\Phi_0=hc/e$, the latter first having been measured in 1985.~\cite{Webb.54.2696}

The AC effect can be observed in the case of a more general electric field
than the one produced by a charged line, 
i.e.\ including the radial component and a component in $z$-direction.~\cite{PhysRevB.51.13441} 
Experimentally, it is
relatively simple to realize an electric field in $z$-direction, i.e.\ which is directed
perpendicular to the two-dimensional (2D) plane containing the quantum ring structure.
By changing the strength of the electric field, the spin-orbit interaction strength
of the Rashba effect~\cite{0022-3719-17-33-015} can be tuned.
The AC effect appears also for a Dresselhaus spin-orbit interaction~\cite{PhysRev.100.580},
which is typically stronger in GaAs.
Persistent equilibrium spin currents due to geometrical phases were addressed 
for the Zeeman interaction with an inhomogeneous, static magnetic field.~\cite{PhysRevLett.65.1655} 
Later, Balatsky and Altshuler studied persistent spin currents related to the AC phase~\cite{PhysRevLett.70.1678}.
Several authors addressed the persistent spin current oscillations as the strength
of the spin-orbit interaction is increased.~\cite{PhysRevB.51.13441,PhysRevB.76.125307,0295-5075-95-5-57008}
As opposed to the AB oscillations with the magnetic flux, the AC oscillations are not periodic
with the spin-orbit interaction strength. Optical control of the spin current 
can be achieved by a nonadiabatic, two-component laser pulse.~\cite{PhysRevB.83.155427}
Suggestions to measure persistent spin currents by the induced mechanical torque~\cite{PhysRevLett.99.266602} 
or the induced electric field~\cite{PhysRevB.77.035327} 
have been proposed. An analytical state-dependent expression for a 
specific spin polarization of the spin current has been stated in Ref.\ \onlinecite{PhysRevB.74.235315}.


Charge persistent currents in quantum rings can be produced by
two time-delayed light pulses with perpendicularly oriented,
linear polarization~\cite{PhysRevLett.94.166801}
and phase-locked laser pulses
based on the circular photon polarization
influencing the many-electron (ME) angular momentum.~\cite{PhysRevB.72.245331}
Moreover, energy splitting of degenerate states in interaction with
a monochromatic circularly polarized electromagnetic mode
and its vaccum fluctuations can lead to charge persistent currents.~\cite{PhysRevLett.107.106802,PhysRevB.87.245437}
Furthermore, the nonequilibrium dynamical response of the dipole moment
and spin polarization of a quantum ring
with spin-orbit interaction and magnetic field
under two linearly polarized electromagnetic pulses has been studied.~\cite{PhysRevB.77.235438}
Quantum systems embedded in an electromagnetic cavity have become one of the most promising
applications in quantum information processing devices. We are considering here the influence
of the cavity photons on the internal and external charge and spin transport
inside and into and out of the ring. 
We treat the electron-photon interaction by using exact numerical diagonalization
including many levels,~\cite{1367-2630-14-1-013036} 
i.e. beyond a two-level Jaynes-Cummings model
or the rotating wave approximation and higher order corrections of it.~\cite{1443594,PhysRevLett.98.013601,Sornborger04:052315}

Concentrating on the \textit{electronic} transport 
through a quantum ring connected to leads, which is embedded in a magnetic field,
several studies exist for only Rashba spin-orbit interaction~\cite{PhysRevB.69.235310,0256-307X-21-11-006},
only Dresselhaus spin-orbit interaction~\cite{PhysRevLett.70.343}
or both.~\cite{PhysRevB.55.10631,PhysRevB.72.165336}
Combining both, the light-matter interaction and the 
strong coupling of the quantum ring to leads, follow
even more involved questions,
especially when the leads have a bias, which
breaks additional transport symmetries.
The electronic transport through a quantum system in a strong system-lead coupling
regime was studied for longitudinally polarized
fields,~\cite{Tang99:1830,Zhou17:6663,Jung85:023420} or transversely
polarized fields~\cite{Tang00:127,Zhou68:155309} ---
though without taking into consideration spin-orbit effects.
For a weak coupling between the system and the leads,
the Markovian approximation,
which neglects memory effects in the system, can be used.~\cite{Spohn53:569,Gurvitz96:15932,Kampen01:00,Harbola06:235309}
To describe a stronger transient system-lead coupling,
we use a non-Markovian generalized master equation~\cite{PhysRevLett.72.1076,Braggio06:026805,Moldoveanu09:073019}
involving energy-dependent coupling elements.
The dynamics of the open system under
non-equilibrium conditions and realistic device geometries can be described with the
time-convolutionless generalized master equation,~\cite{PhysRevA.59.1633,PhysRevB.87.035314} 
which is suitable for higher system-lead coupling 
and allows for a controlled perturbative expansion
in the system-lead coupling strength.

The time-dependent transport of spin-orbit and Coulomb interacting
electrons through a topologically nontrivial broad ring geometry,
embedded in an electromagnetic cavity with a quantized photon mode, and connected to leads
has not yet been explored beyond the Markovian approximation.
One of the objectives of the present work,
is to present differences
between one-dimensional (1D) and 2D rings~\cite{PhysRevB.69.235310,0953-8984-23-33-335601}
focusing on the persistent spin current.
We derive the persistent spin current for arbitrary spin polarization
for the 1D ring with Rashba or Dresselhaus spin-orbit interaction
analytically giving us a robust tool to discern effects from
the 2D structure, Coulomb interaction between the electrons and
transient coupling to electrically biased leads.
For the 2D ring we performed numerical calculations
as analytical solutions are known only when neglecting spin-orbit interaction.~\cite{0268-1242-11-11-001}
Furthermore, we embed the 2D ring in a photon cavity
with $x$- or $y$-polarized photon field
to explore the influences of the photon field 
and its linear polarization on the current.
The comparisons are performed in the range of the 
Rashba or Dresselhaus interaction strength
almost up to an AC phase difference $\Delta\Phi \approx 3\pi$.  


The paper is organized as follows. In Sec.\ II, we provide a general description of the
central ring system
and its charge and spin currents, which applies to both
the 1D and 2D ring. In Sec.\ III, analytical expressions for the
charge and spin currents in a simpler 1D ring of spin-orbit interacting
electrons are given. Sec.\ IV describes our
dynamical model for the correlated electrons in the opened up 2D ring embedded
in a photon cavity. Sec.\ V shows the numerical transient results for the 2D ring and sets them in comparison with the analytical 1D results as a function of the Rashba spin-orbit interaction strength. 
The influence of the linearly polarized electromagnetic cavity field
on the spin currents is studied for different photon polarization. 
Furthermore, the differences between the Rashba and Dresselhaus interaction in a ring system are addressed. Conclusions will be drawn in Sec.\ VI. The time- and space-dependent spin photocurrents are provided as supplementary material.

\section{General description of the central ring system}

In this section, we give the most general Hamiltonian that we consider for the central ring system
including a homogeneous magnetic field in $z$-direction interacting with the electrons' spin,
Rashba and Dresselhaus spin-orbit interaction, Coulomb repulsion between the electrons
and a single cavity photon mode interacting with the electronic system. 
Furthermore, we use this general Hamiltonian to derive in two independent ways operators 
for the charge and spin density, charge and spin current density and spin source terms.
The spin source terms result from the fact that the spin transport is not satisfying a continuity equation
due to the spin-orbit coupling.

\subsection{Central system Hamiltonian}

The time-evolution operator of the closed system with respect to $t=0$,
\begin{equation}
 \hat{U}_{S}(t)=\exp\left(-\frac{i}{\hbar}\hat{H}^{S}t\right),
\end{equation}
is defined by a many-body (MB) system Hamiltonian
\begin{eqnarray}
\hat{H}^{S}&=&\int d^2 r\; \hat{\mathbf{\Psi}}^{\dagger}(\mathbf{r})\left[\left[\frac{\hat{\mathbf{p}}^2}{2m^{*}} +V_S(\mathbf{r})\right] + H_{Z}\nonumber \right. \\
 &&+\left. \hat{H}_{R}(\mathbf{r})+\hat{H}_{D}(\mathbf{r})\right]\hat{\mathbf{\Psi}}(\mathbf{r})+\hat{H}_{ee}+\hbar\omega \hat{a}^{\dagger}\hat{a}, \label{H^S}
\end{eqnarray}
with the two-component vector of field operators
\begin{equation}
 \hat{\mathbf{\Psi}}(\mathbf{r})=
 \begin{pmatrix} \hat{\Psi}(\uparrow,\mathbf{r}) \\ \hat{\Psi}(\downarrow,\mathbf{r}) \end{pmatrix},
\end{equation}
and
\begin{equation}
 \hat{\mathbf{\Psi}}^{\dagger}(\mathbf{r})=
 \begin{pmatrix} \hat{\Psi}^{\dagger}(\uparrow,\mathbf{r}), & \hat{\Psi}^{\dagger}(\downarrow,\mathbf{r}) \end{pmatrix}, \label{conj_FOS}
\end{equation}
where
\begin{equation}
 \hat{\Psi}(x)=\sum_{a}\psi_{a}^{S}(x)\hat{C}_{a}
\end{equation} 
is the field operator with $x\equiv \mathbf{r},\sigma$, $\sigma \in \{ \uparrow,\downarrow \}$ and the annihilation operator, $\hat{C}_{a}$,
for the single-electron state (SES) $\psi_a^S(x)$ in the central system, i.e.\ 
the eigenstate labeled by $a$
of the Hamiltonian $\hat{H}^{S}-\hat{H}_{ee}-\hbar\omega \hat{a}^{\dagger}\hat{a}$
for $\hat{\mathbf{A}}^{\mathrm{ph}}(\mathbf{r})=0$ (see \eq{mom}).
The momentum operator is
\begin{equation}
 \hat{\mathbf{p}}(\mathbf{r})=\begin{pmatrix} \hat{p}_x(\mathbf{r}) \\ \hat{p}_y(\mathbf{r}) \end{pmatrix}=\frac{\hbar}{i}\nabla +\frac{e}{c} \left[\mathbf{A}(\mathbf{r}) +
 \hat{\mathbf{A}}^{\mathrm{ph}}(\mathbf{r})\right].   \label{mom}
\end{equation}
The Hamiltonian in \eq{H^S} includes a kinetic part, a constant magnetic field $\mathbf{B} = B
\hat{\mb{z}}$, in Landau gauge being represented by 
$\mathbf{A}(\mathbf{r})= -By \mathbf{e}_x$ and a photon field.
Furthermore, in Eq. (\ref{H^S}), 
\begin{equation}
 H_{Z}=\frac{\mu_B g_S B}{2}\sigma_{z} \label{H_Z}
\end{equation}
describes the Zeeman interaction between the spin and the magnetic field, 
where $g_S$ is the electron spin g-factor and $\mu_B=e\hbar/ (2m_{e}c)$ is the Bohr magneton.
The interaction between the spin and the orbital motion is described by the Rashba part
\begin{equation}
 \hat{H}_{R}(\mathbf{r})=\frac{\alpha}{\hbar}\left( \sigma_{x} \hat{p}_y(\mathbf{r}) -\sigma_{y} \hat{p}_x(\mathbf{r}) \right) \label{H_R}
\end{equation}
with the Rashba coefficient $\alpha$ and the Dresselhaus part,
which here is restricted to the first-order term in the momentum,
\begin{equation}
 \hat{H}_{D}(\mathbf{r})=\frac{\beta}{\hbar}\left(\sigma_{x} \hat{p}_x(\mathbf{r}) - \sigma_{y}  \hat{p}_y(\mathbf{r})\right) \label{H_D}
\end{equation}
with the Dresselhaus coefficient $\beta$. In Eqs.\ (\ref{H_Z}-\ref{H_D}), $\sigma_x$, $\sigma_y$ and $\sigma_z$
represent the spin Pauli matrices.
Equation (\ref{H^S}) includes the exactly treated electron-electron interaction
\begin{equation}
 \hat{H}_{ee}=\frac{e^2}{2\kappa}\int dx' \; \int dx\; \frac{\hat{\Psi}^{\dagger}(x)\hat{\Psi}^{\dagger}(x')\hat{\Psi}(x')\hat{\Psi}(x)}{\sqrt{|\mathbf{r}-\mathbf{r'}|^2+\eta^2}} \label{Hee}
\end{equation}
with $e>0$ being the magnitude of the electron charge and the integral over $x$ being composed of
a continuous 2D space integral and a sum over the spin.
Only for numerical reasons, 
we include a small regularization parameter $\eta=0.2387$~nm in \eq{Hee}.
The last term in \eq{H^S} indicates the quantized photon field,
where $\hat{a}$ and $\hat{a}^{\dagger}$ are the
photon annihilation and creation operators, respectively,
and $\hbar\omega$ is the photon excitation energy. The photon field
interacts with the electron system via the vector potential
\begin{equation}
 \hat{\mathbf{A}}^{\mathrm{ph}}(\mathbf{r})=A(\mathbf{e}\hat{a}+\mathbf{e}^{*}\hat{a}^{\dagger})
\end{equation}
with
\begin{equation}
 \mathbf{e}=\begin{cases} \mathbf{e}_x, & \mathrm{TE}_{011} \\ \mathbf{e}_y, & \mathrm{TE}_{101}
\end{cases}
\end{equation}
for longitudinally-polarized ($x$-polarized) photon
field ($\mathrm{TE}_{011}$) and transversely-polarized ($y$-polarized)
photon field ($\mathrm{TE}_{101}$).
The electron-photon coupling
constant $g^{EM}=eA a_w \Omega_w/c$ scales with the amplitude $A$ of
the electromagnetic field. It is interesting to note that
the photon field couples directly to the spin via Eqs.\ (\ref{H_R}), (\ref{H_D}) and (\ref{mom}).
For reasons of comparison, we also
consider results without photons in the system. In this case,
$\hat{\mathbf{A}}^{\mathrm{ph}}(\mathbf{r})$ and $\hbar \omega
\hat{a}^{\dagger}\hat{a}$ drop out from the MB system Hamiltonian in
\eq{H^S}.

\subsection{Charge and spin operators}

The charge density satisfies the continuity equation
\begin{equation}
 \frac{\partial}{\partial t} n^{c}(\mathbf{r},t)+\nabla \mathbf{j}^{c}(\mathbf{r},t)=0 \label{chargecont}
\end{equation}
while the continuity equation for the spin density includes in general source terms
\begin{equation}
 \frac{\partial}{\partial t} n^{i}(\mathbf{r},t)+\nabla \mathbf{j}^{i}(\mathbf{r},t)=s^{i}(\mathbf{r},t) \label{spincont}
\end{equation}
for all spin polarizations $i=x,y,z$. Some controversy has been raised about spin currents and
their conservation and several conserved spin currents proposed.~\cite{PhysRevLett.96.076604, PhysRevB.80.012401}
Today, it is accepted that a redefinition of the Rashba expression~\cite{PhysRevB.68.241315}
is not necessary~\cite{PhysRevB.83.113307, PhysRevB.77.035327} as conservation laws cannot be restored in general~\cite{PhysRevB.85.235313, PhysRevB.77.035327}. 
We derived the expressions for all the corresponding operators from \eq{chargecont} and \eq{spincont}
by two independent ways, and come to the same conclusion, which is: 
\textit{though other definitions of the spin current are possible by a related compensation of the source, 
it is not possible to eliminate a spin source term for our Hamiltonian.}
First, we calculated the electron group velocity operator
\begin{equation}
\hat{\mathbf{v}}=\frac{1}{m^{*}i}\left(\hbar \nabla -\frac{ e}{ic}\hat{\mathbf{A}}(\mathbf{r})\right)+\frac{\alpha}{\hbar}\begin{pmatrix} -\sigma_y \\ \sigma_x  \end{pmatrix}+\frac{\beta}{\hbar}\begin{pmatrix} \sigma_x \\ -\sigma_y  \end{pmatrix}
\end{equation}
with the space-dependent vector potential
\begin{equation}
 \hat{\mathbf{A}}(\mathbf{r})=\mathbf{A}(\mathbf{r}) +
 \hat{\mathbf{A}}^{\mathrm{ph}}(\mathbf{r}).
\end{equation}
in first quantization for the standard expression, Eq. (6) in Ref. \onlinecite{PhysRevB.68.241315}. 
Second, we use the commutation relations for the field operators to 
derive expressions for the density, current density and source operators
in second quantization in the Heisenberg picture with the equation of motion,
\begin{equation}
i\hbar\frac{\partial}{\partial t}\hat{\Psi}(x,t)=[\hat{\Psi}(x,t),\hat{H}^S],
\end{equation}
starting from the continuity equation,
\begin{eqnarray}
&&\sum_{\sigma}\sum_{\sigma'}\frac{\partial}{\partial t}\left(\hat{\Psi}^{\dagger}(\mathbf{r},\sigma,t) \sigma_{\gamma}(\sigma, \sigma')\hat{\Psi}(\mathbf{r},\sigma',t)\right)\nonumber \\
&&=\sum_{\sigma}\sum_{\sigma'}\frac{1}{i\hbar}\left[ \hat{\Psi}^{\dagger}(\mathbf{r},\sigma,t)\sigma_{\gamma}(\sigma, \sigma')\hat{\Psi}(\mathbf{r},\sigma',t)\hat{H}^S\nonumber \right. \\
&&-\left. \hat{H}^S \hat{\Psi}^{\dagger}(\mathbf{r},\sigma,t)\sigma_{\gamma}(\sigma, \sigma')\hat{\Psi}(\mathbf{r},\sigma',t)\right] \label{sqder}
\end{eqnarray}
with $\sigma_{\gamma}(\sigma, \sigma')$ being proportional to the unity matrix coefficients if $\gamma=c$ (describing the charge),
\begin{equation}
 \sigma_{c}(\sigma, \sigma')=e\delta_{\sigma,\sigma'},
\end{equation}
or Pauli spin matrix coefficients if $\gamma=x,y,z$ (describing the spin polarization),
\begin{equation}
 \sigma_{x}(\sigma, \sigma')=\frac{\hbar}{2}(\delta_{\sigma,\uparrow}\delta_{\sigma',\downarrow}+\delta_{\sigma,\downarrow}\delta_{\sigma',\uparrow}), \label{xPauli}
\end{equation}
\begin{equation}
 \sigma_{y}(\sigma, \sigma')=\frac{i\hbar}{2}(-\delta_{\sigma,\uparrow}\delta_{\sigma',\downarrow}+\delta_{\sigma,\downarrow}\delta_{\sigma',\uparrow})
\end{equation}
and
\begin{equation}
 \sigma_{z}(\sigma, \sigma')=\frac{\hbar}{2}\delta_{\sigma,\sigma'}(\delta_{\sigma,\uparrow}-\delta_{\sigma,\downarrow}). \label{zPauli}
\end{equation}
In \eq{sqder}, the system Hamiltonian $\hat{H}^S$ from \eq{H^S} has to be written
with Heisenberg operators instead of the
Schr\"{o}dinger operators.
We attribute every contribution, 
which can be written in the form, $\nabla \mathbf{j}(\mathbf{r},t)$, to the current density operator,
thus aiming towards a minimal expression for the source operator.
Finally, we transform the operators into the Schr\"{o}dinger picture. 

The charge density operator
\begin{equation}
 \hat{n}^{c}(\mathbf{r})=e\mathbf{\Psi}^{\dagger}(\mathbf{r})\mathbf{\Psi}(\mathbf{r}) \label{ncq}
\end{equation}
and the spin density operator for spin polarization $S_i$
\begin{equation}
 \hat{n}^{i}(\mathbf{r})=\frac{\hbar}{2}\mathbf{\Psi}^{\dagger}(\mathbf{r})\sigma_{i}\mathbf{\Psi}(\mathbf{r}).
\end{equation}
The component labeled with $j\in \{x,y\}$ of the charge current density operator is given by
\begin{eqnarray}
\hat{j}^{c}_{j}(\mathbf{r})&=&\left[\frac{e\hbar}{2m^{*}i}\left[\hat{\mathbf{\Psi}}^{\dagger}(\mathbf{r})\nabla_{j} \hat{\mathbf{\Psi}}(\mathbf{r}) - \left[\nabla_{j} \hat{\mathbf{\Psi}}^{\dagger}(\mathbf{r})\right]\hat{\mathbf{\Psi}}(\mathbf{r}) \right] \right. \nonumber \\
&& +\left. \frac{e^2}{m^{*}c}\hat{A}_{j}(\mathbf{r}) \hat{\mathbf{\Psi}}^{\dagger}(\mathbf{r})\hat{\mathbf{\Psi}}(\mathbf{r})\right]\nonumber\\
&&+\frac{e}{\hbar}\hat{\mathbf{\Psi}}^{\dagger}(\mathbf{r}) (\beta \sigma_x-\alpha \sigma_y) \hat{\mathbf{\Psi}}(\mathbf{r}) \delta_{x,j}\nonumber \\
&&+\frac{e}{\hbar}\hat{\mathbf{\Psi}}^{\dagger}(\mathbf{r}) 
(\alpha \sigma_x- \beta \sigma_y)\hat{\mathbf{\Psi}}(\mathbf{r})\delta_{y,j}. \label{jcq}
\end{eqnarray}
The current density operator for the $j$-component and $S_x$ spin polarization
\begin{eqnarray}
\hat{j}^{x}_{j}(\mathbf{r})&=&\left[\frac{\hbar^2}{4m^{*}i}\left[\hat{\mathbf{\Psi}}^{\dagger}(\mathbf{r})\sigma_{x}\nabla_{j} \hat{\mathbf{\Psi}}(\mathbf{r}) - \left[\nabla_{j} \hat{\mathbf{\Psi}}^{\dagger}(\mathbf{r})\right]\sigma_{x}\hat{\mathbf{\Psi}}(\mathbf{r}) \right] \right. \nonumber \\
&& +\left. \frac{e \hbar}{2 m^{*}c}\hat{A}_{j}(\mathbf{r}) \hat{\mathbf{\Psi}}^{\dagger}(\mathbf{r})\sigma_x\hat{\mathbf{\Psi}}(\mathbf{r})\right]\nonumber\\
&&+\frac{\beta \delta_{x,j}+\alpha\delta_{y,j}}{2}\hat{\mathbf{\Psi}}^{\dagger}(\mathbf{r})\hat{\mathbf{\Psi}}(\mathbf{r}). \label{jsxq}
\end{eqnarray}\pagebreak
the current density operator for $S_y$ spin polarization
\begin{eqnarray}
\hat{j}^{y}_{j}(\mathbf{r})&=&\left[\frac{\hbar^2}{4m^{*}i}\left[\hat{\mathbf{\Psi}}^{\dagger}(\mathbf{r})\sigma_{y}\nabla_{j} \hat{\mathbf{\Psi}}(\mathbf{r}) - \left[\nabla_{j} \hat{\mathbf{\Psi}}^{\dagger}(\mathbf{r})\right]\sigma_{y}\hat{\mathbf{\Psi}}(\mathbf{r}) \right] \right. \nonumber \\
&& +\left. \frac{e \hbar}{2 m^{*}c}\hat{A}_{j}(\mathbf{r}) \hat{\mathbf{\Psi}}^{\dagger}(\mathbf{r})\sigma_y\hat{\mathbf{\Psi}}(\mathbf{r})\right]\nonumber\\
&&-\frac{\alpha \delta_{x,j}+\beta\delta_{y,j}}{2}\hat{\mathbf{\Psi}}^{\dagger}(\mathbf{r})\hat{\mathbf{\Psi}}(\mathbf{r}). \label{jsyq}
\end{eqnarray}
and $S_z$ spin polarization
\begin{eqnarray}
\hat{j}^{z}_{j}(\mathbf{r})&=&\left[\frac{\hbar^2}{4m^{*}i}\left[\hat{\mathbf{\Psi}}^{\dagger}(\mathbf{r})\sigma_{z}\nabla_{j} \hat{\mathbf{\Psi}}(\mathbf{r}) - \left[\nabla_{j} \hat{\mathbf{\Psi}}^{\dagger}(\mathbf{r})\right]\sigma_{z}\hat{\mathbf{\Psi}}(\mathbf{r}) \right] \right. \nonumber \\
&& +\left. \frac{e\hbar}{2m^{*}c}\hat{A}_{j}(\mathbf{r}) \hat{\mathbf{\Psi}}^{\dagger}(\mathbf{r})\sigma_{z}\hat{\mathbf{\Psi}}(\mathbf{r})\right]. \label{jszq}
\end{eqnarray}
The expressions for the source operators are given in appendix \ref{sourceop}. We note that our derivation agrees with the definition of the Rashba current when we limit ourselves to the
case without magnetic and photon field and without Dresselhaus spin-orbit interaction.~\cite{PhysRevB.68.241315, PhysRevB.76.033306}

\section{1D rings: exact expressions for the spin current}

In this section, we derive and describe analytical results for an ideal 1D ring, 
i.e.\ with infinitely narrow confinement, and 
with either Rashba or Dresselhaus spin-orbit interaction. 
Here, we will neglect the magnetic field, electron-electron interaction and the photons.
Accordingly, the general expressions for the Hamiltonian, \eq{H^S}, and the charge and spin operators, Eqs.\ (\ref{ncq}-\ref{jszq}) and the equations from appendix \ref{sourceop}, 
can be simplified for the purposes of this section.
Our aim is to clarify the role of the different parts of the central Hamiltonian \eq{H^S}
by comparing our numerical results to
the analytical results of this section.

\subsection{1D Rashba ring}

\begin{figure}[htbq] 
       \subfigure[]{
       \includegraphics[width=0.35\textwidth,angle=-90]{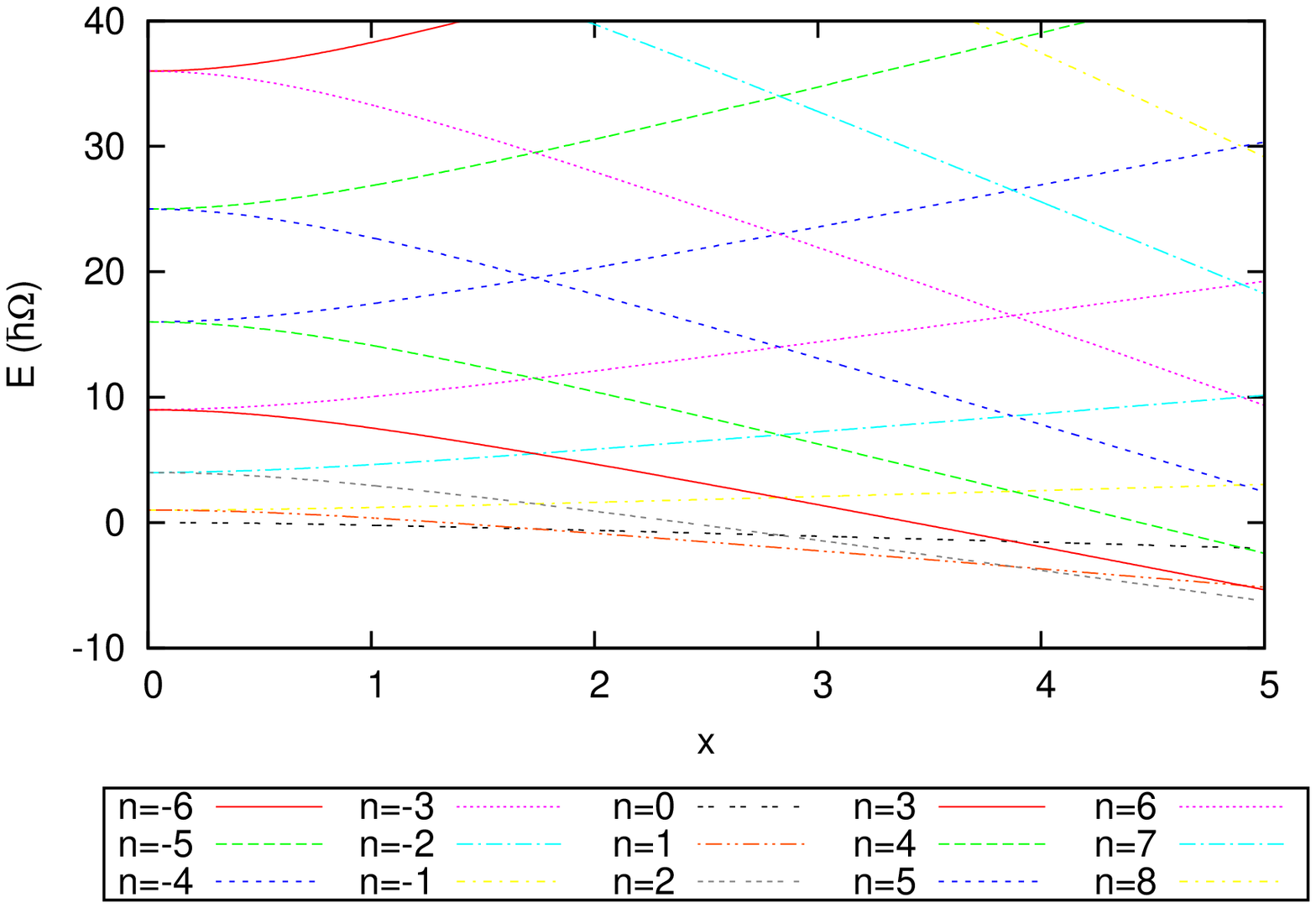}
       \label{spec_mu1}}
       \subfigure[]{
       \includegraphics[width=0.35\textwidth,angle=-90]{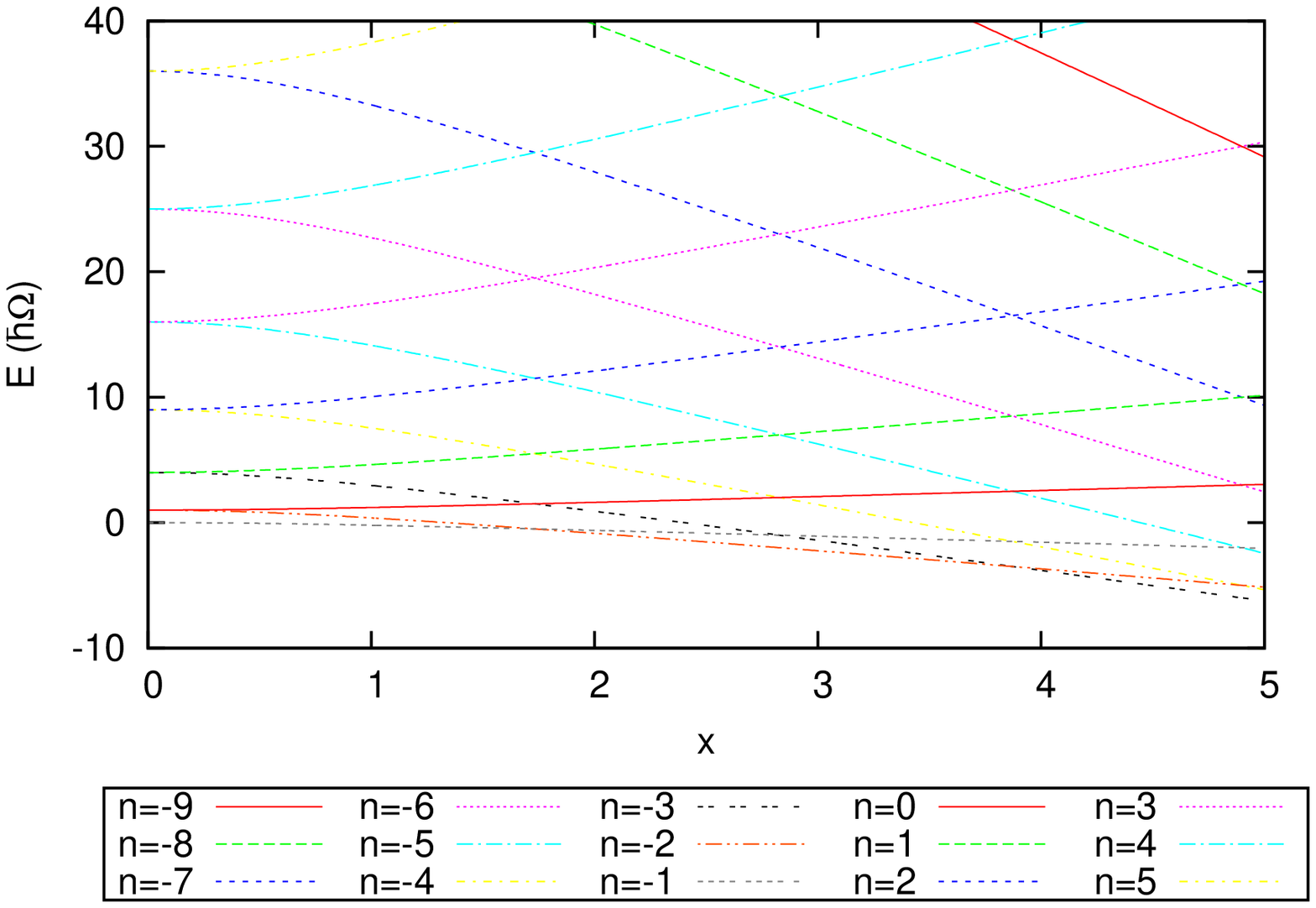}
       \label{spec_mu2}}
       \caption{(Color online) Spectrum from Eq.\ (\ref{spectrum}) as a function of $x=x_R$ or $x=x_D$ for \subref{spec_mu1} $\nu=-1$ and \subref{spec_mu2} $\nu=1$.}
       \label{spec}
\end{figure}

Our Hamiltonian containing the kinetic and the Rashba term
\begin{equation}
\hat{H}=-\frac{\hbar^2}{2m^{*}}\nabla^2+\frac{\alpha}{i}\left[\hat{\sigma}_x \frac{\partial}{\partial y} - \hat{\sigma}_y \frac{\partial}{\partial x}\right], \label{RashbaH}
\end{equation}
where $\alpha$ is the Rashba coefficient and $\hat{\sigma}_x$, $\hat{\sigma}_y$ and $\hat{\sigma}_z$
are the spin Pauli matrices, has the 1D ring limit:~\cite{PhysRevB.66.033107} 
\begin{eqnarray}
\hat{H}^{1D}&=&-\hbar\Omega \frac{\partial^2}{\partial \varphi^2}
-i\hbar\omega_{R}(\cos(\varphi)\hat{\sigma}_x+\sin(\varphi)\hat{\sigma}_y)\frac{\partial}{\partial \varphi}\nonumber \\
&-&i\frac{\hbar\omega_{R}}{2}(\cos(\varphi)\hat{\sigma}_y-\sin(\varphi)\hat{\sigma}_x). \label{corRas1D}
\end{eqnarray}
It is convenient to introduce the dimensionless Rashba parameter, $x_R$, which is independent
of the ring radius $a$ and scales linearly with the Rashba coefficient $\alpha$, given by
\begin{equation}
 x_{R}:=\frac{\omega_R}{\Omega} \label{defxR}
\end{equation}
with the Rashba frequency $\omega_{R}:=\alpha/(\hbar a)$ and kinetic frequency $\Omega:=\hbar/(2m^{*}a^2)$.
The eigenvalues of the Hamiltonian in Eq.\ (\ref{corRas1D}) are:~\cite{PhysRevB.69.155335}
\begin{equation}
 E_{\nu n}=\hbar\Omega\left[\left(n-\frac{\Phi^{\nu}}{2\pi}\right)^2-\frac{ x^{2}}{4}\right] \label{spectrum}
\end{equation}
with the Rashba AC phase
\begin{equation}
 \Phi^{\nu}=-\pi \left[1+\nu\sqrt{1+x^2}\right],
\end{equation}
where we call $n$ the angular momentum quantum number, $\nu=\pm 1$ the spin quantum number and $x=x_R$ 
in the Rashba ring case. The spectrum is shown in \fig{spec}. 
For zero temperature, $T=0$, the lowest $N_e/2$ states are occupied both for $\nu=-1$ and $\nu=1$. 
Occupation changes are possible at every other level crossing point.

The eigenfunctions are
\begin{eqnarray}
 \Psi_{\nu n}^{R}(\varphi)&=&\begin{pmatrix} \Psi_{\nu n}^{R}(\varphi,\uparrow)\\ \Psi_{\nu n}^{R}(\varphi,\downarrow)\end{pmatrix} \nonumber \\
 &=&\frac{\exp(in\varphi)}{\sqrt{2\pi a}}\begin{pmatrix} A_{\nu,1}^{R} \\ A_{\nu,2}^{R}\exp(i\varphi) \end{pmatrix} \label{Rwfc}
\end{eqnarray}
with the $2\times 2$ coefficient matrix
\begin{equation}
 A^{R}=\begin{pmatrix} A_{\nu,1}^{R} & A_{\nu,2}^{R} \end{pmatrix}=\begin{pmatrix} \cos\left(\frac{\theta_{R}}{2}\right) & \sin\left(\frac{\theta_{R}}{2}\right) \\
 \sin\left(\frac{\theta_{R}}{2}\right) & -\cos\left(\frac{\theta_{R}}{2}\right)
 \end{pmatrix} \label{Rcoef}
\end{equation}
and
\begin{equation}
 \tan\left(\frac{\theta_{R}}{2}\right)=\frac{1-\sqrt{1+x_R^2}}{x_R}. \label{Rtantheta}
\end{equation}

In our derivation of the exact analytical expressions for the spin currents given in the appendix \ref{xspincurder}, 
we assume that the number of electrons, $N_e$, is even, as this 
results in the same amount of states (distinguished by $n$) with $\nu=-1$ or $\nu=1$ to be occupied 
provided that $T=0$ (except possibly at the crossing points of the spectrum). 
Mathematically, we could phrase it that the cardinality (number of elements)
of the two sets of occupied states $N_{\pm}$ for $\nu=\pm 1$ 
is equal meaning that $|N_{-}|=|N_{+}|=N_e/2$.  
The charge density is given by
\begin{equation}
 n^{c}_{R}=\frac{eN_e}{2\pi a} \label{ncR1}
\end{equation}
and the charge current $j^{c}_{R}=0$. 
The spin densities are all vanishing:
\begin{equation}
 n^{x}_{R}(\varphi)=n^{y}_{R}(\varphi)=n^{z}_{R}(\varphi)=0.
\end{equation}
The spin current densities are given by
\begin{widetext}
\begin{equation}
j^{x}_{R}(\varphi)=\frac{\hbar j_{\Phi}\cos(\varphi)\left(x_R-x_R\sqrt{1+x_R^2}\right)}{N_e\left(2+2x_R^2-2\sqrt{1+x_R^2}\right)}\left[\sum_{n\in N_{-}}(2n+1)-\sum_{n\in N_{+}}(2n+1) \right] +\frac{x_R\hbar j_{\Phi}\cos(\varphi)}{2}, \label{jxR1}
\end{equation}
\begin{equation}
 j^{y}_{R}(\varphi)=\frac{\sin(\varphi)}{\cos(\varphi)}j^{x}_{R}(\varphi)
\end{equation}
and
\begin{eqnarray}
j^{z}_{R}&=&\frac{\hbar j_{\Phi}}{N_e\left(2+2x_R^2-2\sqrt{1+x_R^2}\right)}
\left[ \left(2+x_R^2-2\sqrt{1+x_R^2}\right)\left[\sum_{n\in N_{+}}n- \sum_{n\in N_{-}} (n+1) \right] \right. \nonumber \\
&&+ \left. x_R^2 \left[\sum_{n\in N_{-}}n-\sum_{n\in N_{+}}(n+1) \right]\right],
\end{eqnarray}
where $j_{\Phi}=N_e\hbar/(4\pi m^{*}a^2)$ is the maximum absolute value of the persistent charge current
in units of the electron charge $e$ as a function of the magnetic flux $\Phi$ for $\alpha=\beta=0$. Finally, the spin source terms
\begin{eqnarray}
s^{x}_{R}(\varphi)&=&-\frac{\hbar \sin(\varphi) j_{\Phi}}{N_e a \left(2+2x_R^2-2\sqrt{1+x_R^2}\right)} \left[\left(2x_R+x_R^3-2x_R\sqrt{1+x_R^2}\right) \nonumber \right. \\ 
&&\times \left. \left[ \sum_{n \in N_{-}} (n+1) - \sum_{n \in N_{+}} n \right] +  x_R^3 \left[ \sum_{n \in N_{+}} (n+1) - \sum_{n \in N_{-}} n \right] \right],
\end{eqnarray}
\end{widetext}
\begin{equation}
 s^{y}_{R}(\varphi)=-\frac{\cos(\varphi)}{\sin(\varphi)}s^{x}_{R}(\varphi) \label{syR1}
\end{equation}
and $s^{z}_{R}$ is vanishing as expected since $j^{z}_{R}$ depends not on $\varphi$.

To account properly for the rearrangements of the occupied states  $N_{-}$ and $N_{+}$,
we have to distinguish the case with the cardinalities, $|N_{-}|$ and $|N_{+}|=|N_{-}|$, to be even
and state rearrangements at $x^{\rm e}_{n}=\sqrt{(2n+1)^2-1}$, $n=0,1,\dots\;$
and the case with odd cardinalities
and state rearrangements at $x^{\rm o}_{n}=\sqrt{(2n+2)^2-1}$, $n=0,1,\dots\;$. 
In the even cardinality case, we define a multi-step function $\chi^{\rm e}=n$ 
for $x_{n}^{\rm e}<x<x_{n+1}^{\rm e}$, $n=0,1,\dots\;$.
In the odd cardinality case, $\chi^{\rm o}=0$ for $x<x_{0}^{\rm o}$ and
$\chi^{\rm o}=n+1$ for $x_{n}^{\rm o}<x<x_{n+1}^{\rm o}$, $n=0,1,\dots\;$.
Here, $x$ is the Rashba parameter $x_R$ or Dresselhaus parameter $x_D$ to be defined later.
Then, in the even cardinality case, we have
\begin{eqnarray}
 N_{-}^{\rm e}=\{-|N_{-}^{\rm e}|/2+\chi^{\rm e}+1, \nonumber \\
 -|N_{-}^{\rm e}|/2+\chi^{\rm e}+2, \dots, |N_{-}^{\rm e}|/2+\chi^{\rm e} \}
\end{eqnarray}
and
\begin{eqnarray}
 N_{+}^{\rm e}=\{-|N_{+}^{\rm e}|/2-\chi^{\rm e}-1, \nonumber \\
 -|N_{+}^{\rm e}|/2-\chi^{\rm e}, \dots, |N_{+}^{\rm e}|/2-\chi^{\rm e}-2 \}
\end{eqnarray}
while in the odd cardinality case, we have
\begin{eqnarray}
N_{-}^{\rm o}=\{-(|N_{-}^{\rm o}|-1)/2+\chi^{\rm o},   \nonumber \\
 -(|N_{-}^{\rm o}|-1)/2+\chi^{\rm o}+1, \dots, (|N_{-}^{\rm o}|-1)/2+\chi^{\rm o} \}
\end{eqnarray}
and
\begin{eqnarray}
 N_{+}^{\rm o}=\{-(|N_{+}^{\rm o}|-1)/2-\chi^{\rm o}-1,   \nonumber \\
 -(|N_{+}^{\rm o}|-1)/2-\chi^{\rm o}, \dots, (|N_{+}^{\rm o}|-1)/2-\chi^{\rm o}-1 \}.
\end{eqnarray}

The spin currents are
\begin{equation}
j^{x,\rm e/o}_{R}(\varphi)=\frac{\hbar j_{\Phi}}{2}\cos(\varphi)f^{\rm e/o}(x_R), \label{jxRres}
\end{equation}
\begin{equation}
j^{y,\rm e/o}_{R}(\varphi)=\frac{\hbar j_{\Phi}}{2}\sin(\varphi)f^{\rm e/o}(x_R)
\end{equation}
and
\begin{equation}
j_{R}^{z,\rm e/o}=\frac{\hbar j_{\Phi}}{2}g^{\rm e/o}(x_R).
\end{equation}
The non-vanishing source terms are
\begin{equation}
 s_{R}^{x,\rm e/o}(\varphi)=-\frac{\hbar j_{\Phi}}{2a} f^{\rm e/o}(x_R)\sin(\varphi)
\end{equation}
and
\begin{equation}
 s_{R}^{y,\rm e/o}(\varphi)=\frac{\hbar j_{\Phi}}{2a} f^{\rm e/o}(x_R)\cos(\varphi).
\end{equation}
The functions $f^{\rm e/o}(x)$ and $g^{\rm e/o}(x)$ describing the dependency
on the Rashba parameter $x=x_R$ or Dresselhaus parameter $x=x_D$
have to be distinguished according to their cardinality.

For even cardinality, we have
\begin{equation}
f^{\rm e}(x)=\frac{2x-2x\sqrt{1+x^2}}{2+2x^2-2\sqrt{1+x^2} }\left[2+2\chi^{\rm e}\right]+x
\end{equation}
and
\begin{eqnarray}
g^{\rm e}(x)&=&\frac{2}{{2+2x^2-2\sqrt{1+x^2}}}\left[ x^2 \left[\chi^{\rm e}+\frac{1}{2} \right] \nonumber \right. \\ 
&&-\left. \left(2+x^2-2\sqrt{1+x^2}\right) \left[\frac{3}{2}+ \chi^{\rm e} \right]\right].
\end{eqnarray}
For odd cardinality, they are
\begin{equation}
f^{\rm o}(x)=\frac{2x-2x\sqrt{1+x^2}}{2+2x^2-2\sqrt{1+x^2} }\left[1+2\chi^{\rm o}\right]+x
\end{equation}
and
\begin{eqnarray}
g^{\rm o}(x)&=&\frac{2}{{2+2x^2-2\sqrt{1+x^2}}}\left[ x^2 \chi^{\rm o} \right. \nonumber \\  &&-\left.\left(2+x^2-2\sqrt{1+x^2}\right) \left[1+ \chi^{\rm o} \right]\right]. \label{go}
\end{eqnarray}
Eqs. (\ref{jxRres}) to (\ref{go}) represent the main result of this section. In the following,
the properties of these Rashba spin currents and spin source terms will be described.  

\begin{figure}[htbq] 
       \includegraphics[width=0.48\textwidth,angle=0]{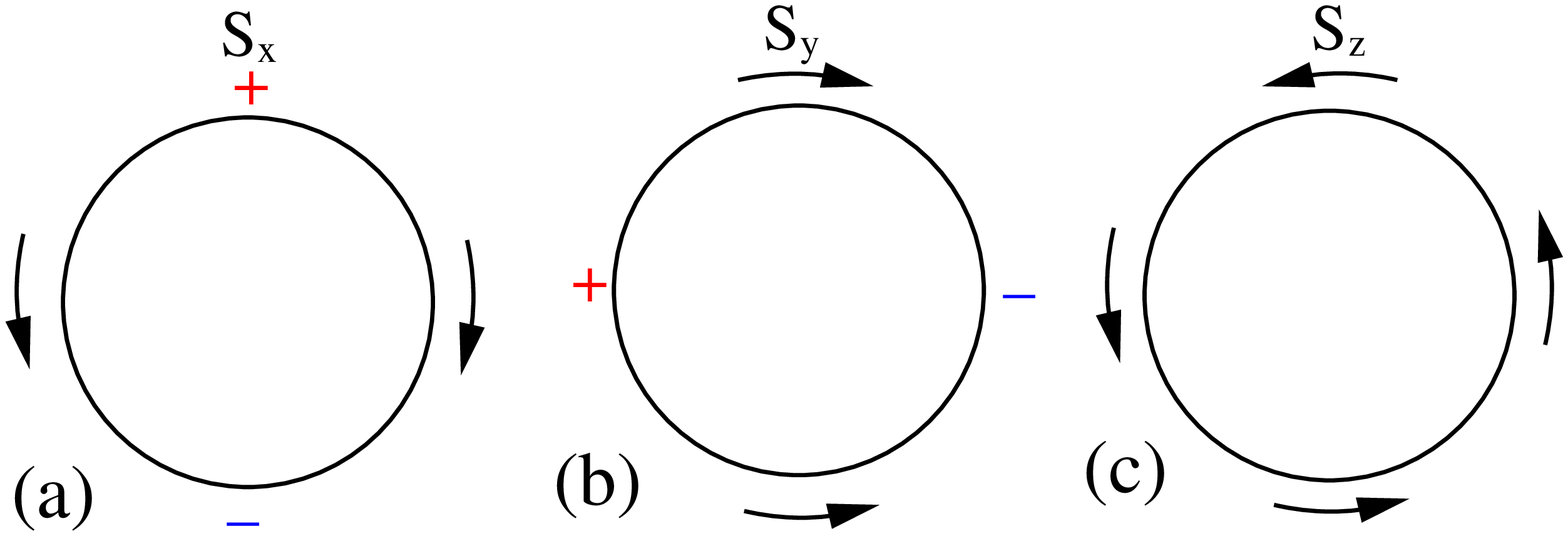}
       \caption{(Color online) Geometrical arrangement of (a) the source term $s_{R}^{x,\rm e/o}(\varphi)$ and spin current $j^{x,\rm e/o}_{R}(\varphi)$ of the $x$-component of the spin, (b) $s_{R}^{y,\rm e/o}(\varphi)$ and $j^{y,\rm e/o}_{R}(\varphi)$ of the $y$-component and (c) $j^{z,\rm e/o}_{R}$ for the $z$-component in the case that $B=\beta=0$. The spin current for the $z$-component of the spin is homogeneous in space due to the absence of the source term $s_{R}^{z,\rm e/o}(\varphi)$. The ``+''-sign and ``--''-sign indicate source and sink, respectively, in the case that $x_R$ is such that $f^{\rm e/o}(x_R))\leq 0$ or $g^{\rm e/o}(x_R)\geq 0$ and the arrows indicate the corresponding spin current direction and are shown (a) and (b) at the positions of maximum current magnitude and (c) at arbitrary positions for $j^{z,\rm e/o}_{R}$, which is homogeneous in space.}
       \label{Rashba_spin_currents}
       \includegraphics[width=0.48\textwidth,angle=0]{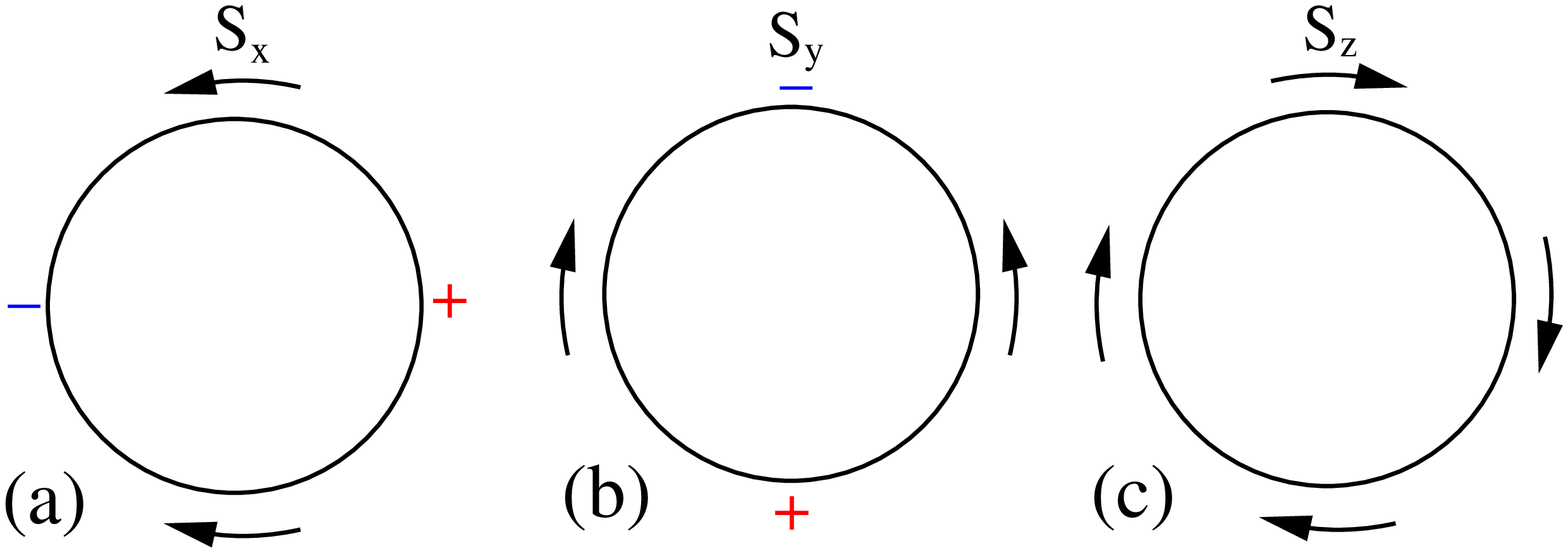}
       \caption{(Color online) Same as \fig{Rashba_spin_currents}, but for the Dresselhaus ring
       (the case $B=\alpha=0$).}
       \label{Dresselhaus_spin_currents}
\end{figure}

Figure \ref{Rashba_spin_currents} shows the geometrical arrangement of the sources and spin currents.
For the $x$- and $y$-component of the spin, \fig{Rashba_spin_currents} (a) and (b), respectively, source and sink term are largest on opposite sites of the ring.
Correspondingly, a non-homogeneous spin current is flowing from the source to the sink with maxima at the intermediate positions, where the source term is zero. The only difference between the spin components $S_x$ and $S_y$ is a rotation by $\pi/2$.
The source can interchange its position with the sink if we allow for variations in the Rashba parameter $x_R$.
The ranges of $x_R$, where the case of \fig{Rashba_spin_currents} applies is dependent on the cardinality 
and can be alternatively summarized by the either condition, $f^{\rm e/o}(x_R))\leq 0$ or $g^{\rm e/o}(x_R)\geq 0$.
The current for $S_z$ spin polarization in \fig{Rashba_spin_currents} (c) is equally large everywhere and circulating around the ring 
similar to the persistent current invoked by a magnetic flux.~\cite{PhysRevLett.7.46, Viefers20041} The $z$-component of the spin is therefore source-free. This might be understood in the following way: 
similarly to the magnetic field acting on the spin via the Zeeman term, one can define an effective magnetic field for the Rashba spin-orbit interaction, $\mathbf{B}_{R}=-\hat{\mathbf{p}}\times \mathbf{E}/(m^* c)$, due to the electronic motion inside the electric field $\mathbf{E}=E\mathbf{e}_z$ causing the Rashba interaction. The effective magnetic field is perpendicular to the 
electric field in $z$-direction and the effective momentum of the electrons in $\mathbf{e}_{\varphi}$-direction. Consequently, the effective magnetic field is always perpendicular
to the $S_z$ spin polarization suggesting that $\frac{\partial}{\partial t}n^{z}(\mathbf{r},t)=0$. 
In a local interpretation of the spin continuity equation \eq{spincont}, this corresponds to the case
that the driving mechanism, i.e.\ the source term $s_{R}^{z,\rm e/o}(\varphi)=0$.
We note that the geometrical aspects of the spin current flow could be summarized
by stating that the spin current for spin polarization $S_i$ flows freely 
in the plane perpendicular to the unit vector 
$\mathbf{e}_i$. In our case, the spin current is confined to a 1D system along $\mathbf{e}_{\varphi}$.
The spin current magnitude along the ring is then given by the projection $\mathbf{e}_{\varphi}$
onto the plane perpendicular to $\mathbf{e}_i$. For $S_z$ spin polarization, $\mathbf{e}_{\varphi}$
is inside this plane and therefore the spin current is space independent.
As the spin currents are not dependent on time, we will call them persistent spin currents, not 
distinguishing whether
they are homogeneous in space or not.

\begin{figure}[htbq] 
       \subfigure[]{
       \includegraphics[width=0.35\textwidth,angle=-90]{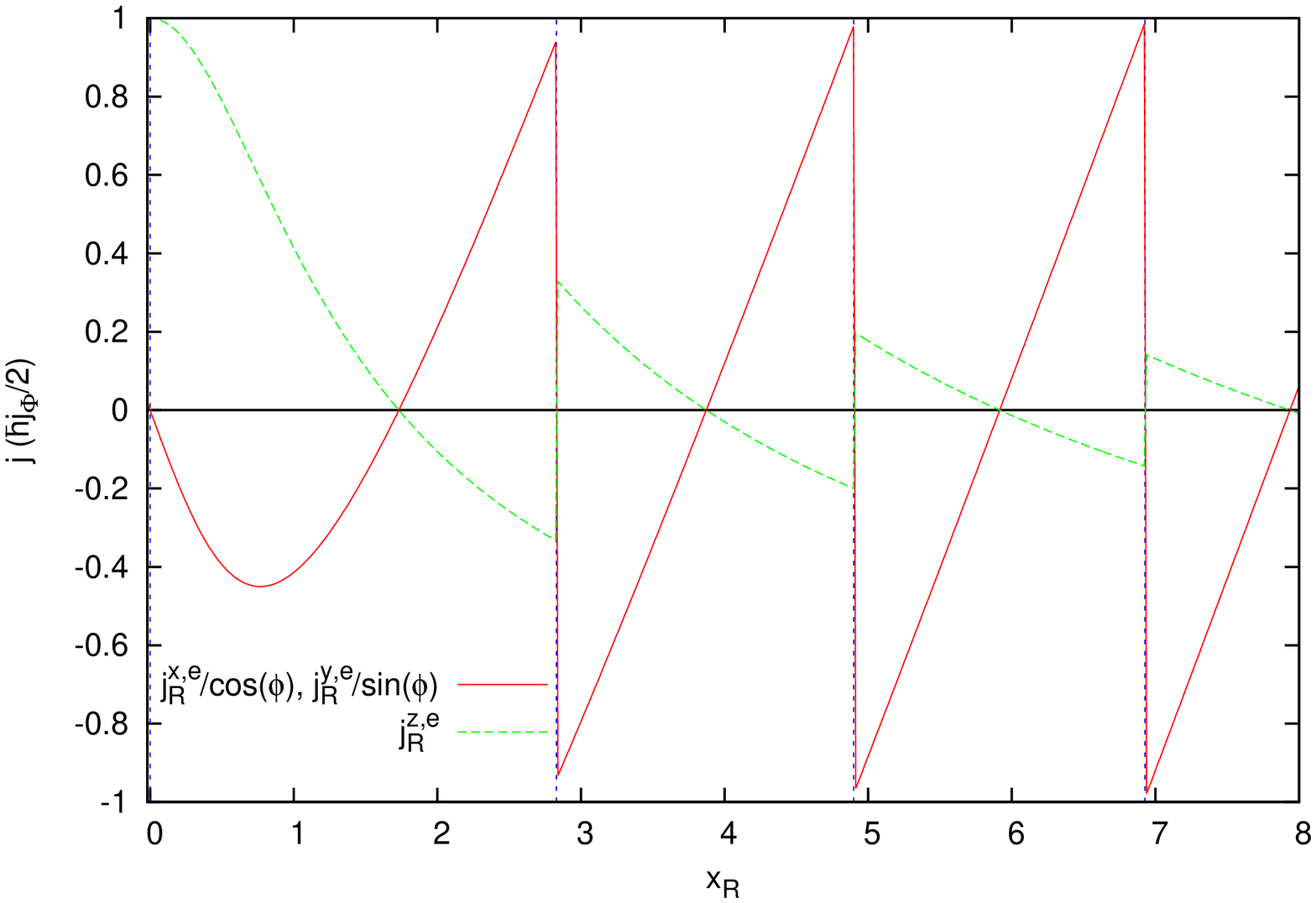}
       \label{pers_cur_alpha_even}}
       \subfigure[]{
       \includegraphics[width=0.35\textwidth,angle=-90]{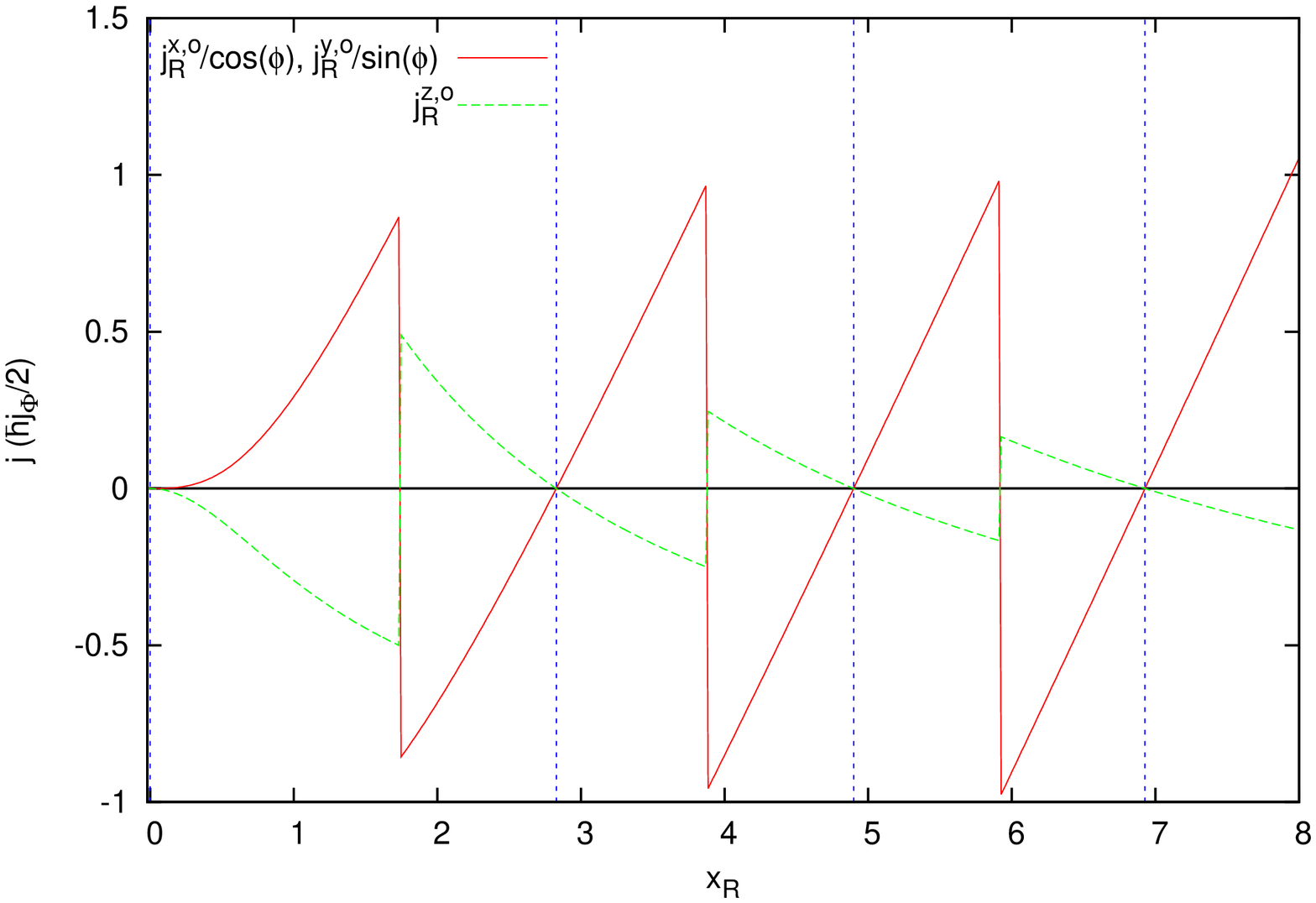}
       \label{pers_cur_alpha_odd}}
       \caption{(Color online) Rashba parameter dependency of $j^{x,\rm e/o}_{R}(x_R)/\cos(\varphi)$, $j^{y,\rm e/o}_{R}(x_R)/\sin(\varphi)$ and $j^{z,\rm e/o}_{R}(x_R)$ for $B=\beta=0$ and \subref{pers_cur_alpha_even} $N_e/2$ even, and \subref{pers_cur_alpha_odd} $N_e/2$ odd. The positions of constructive AC interference satisfying $\Phi^{\nu}=2\pi n$, $n=0,1,\dots$ are indicated by the blue vertical lines.}
       \label{pers_cur_alpha}
\end{figure}

Figure \ref{pers_cur_alpha} shows the spin currents as a function of the Rashba parameter $x_R$.
As opposed to the magnetic flux dependency of the charge current, the Rashba parameter dependency of the spin currents
is not exactly periodic, in particular for small $x_R$. At the zero points of all the even cardinality spin currents, 
the odd cardinality spin currents are largest, changing discontinuously by sign due to sudden reoccupations
among states of the same spin quantum number $\nu$. 
Likewise, at the discontinuities of the even cardinality spin currents, the odd cardinality spin currents are zero. 
It is interesting to note that the $z$-component
of the spin current is commonly larger for small $x_R$ and, in particular, that
an infinitesimal small Rashba coefficient should lead to the relatively large
spin current $j_{R}^{z,\rm e}=\frac{\hbar j_{\Phi}}{2}$ 
provided that the total electron number $N_e$ is divisible by $4$. 
This way, an infinitesimal small effective electric field is enough 
to generate a considerable persistent AC current 
provided the system can be cooled down and ME interactions neglected.
We note that for $x_R$ exactly equal to zero,
all spin currents are vanishing as $j_{R}^{z,\rm e}$ changes discontinuously at $x_R=0$.

\subsection{1D Dresselhaus ring}

Here, we consider the case that spin and orbital momentum couple via the Dresselhaus instead of
the Rashba interaction. The corresponding Hamiltonian containing the kinetic and the Dresselhaus term,
\begin{equation}
\hat{H}=-\frac{\hbar^2}{2m^{*}}\nabla^2+\frac{\beta}{i}\left[\hat{\sigma}_x \frac{\partial}{\partial x} - \hat{\sigma}_y \frac{\partial}{\partial y}\right], \label{DresselhausH}
\end{equation}
where $\beta$ is the Dresselhaus coefficient. In analogy to the Rashba parameter $x_R$, 
it is convenient to introduce the dimensionless Dresselhaus parameter, $x_D$, which is independent
of the ring radius and scales linearly with the Dresselhaus coefficient $\beta$, given by
\begin{equation}
 x_{D}:=\frac{\omega_D}{\Omega}
\end{equation}
with the Dresselhaus frequency $\omega_{D}:=\beta/(\hbar a)$.
The Dresselhaus eigenvalues and coefficient matrix are derived in appendix \ref{Deigenvalder}. The Dresselhaus and Rashba spectrum are identical and shown in \fig{spec}.
The charge density is constant as in the Rashba case,
\begin{equation}
 n^{c}_{D}=\frac{eN_e}{2\pi a}, \label{ncD1}
\end{equation}
and the charge current $j^{c}_{D}=0$.

The spectrum, charge density and charge current are the same for the Rashba and 
Dresselhaus ring. We have calculated also the spin densities, spin currents and
spin source terms in analogy to the Rashba case. For the Dresselhaus ring,
we will present the results by
a comparison to the Rashba case and give an explanation of our findings by a comparison of the two Hamiltonians.
The Dresselhaus Hamiltonian \eq{DresselhausH} is invariant to the
Rashba Hamiltonian \eq{RashbaH}, if the replacement
\begin{equation}
\begin{pmatrix}
 \hat{\sigma}_x \\ \hat{\sigma}_y \\ \beta
\end{pmatrix} \to
\begin{pmatrix}
 -\hat{\sigma}_y \\ -\hat{\sigma}_x \\ \alpha
\end{pmatrix}.
\end{equation}
is performed. Moreover, using the commutation relation $\hat{\sigma}_z=[\hat{\sigma}_x,\hat{\sigma}_y]/(2i)$,
the $z$-spin Pauli matrix transforms according to $\hat{\sigma}_z \to -\hat{\sigma}_z$.
This suggests the following relations for the Dresselhaus spin densites for $x_D=x_R$:
\begin{eqnarray}
\begin{pmatrix} n^{x}_{D} \\ n^{y}_{D} \\ n_{D}^{z} \end{pmatrix} 
= \begin{pmatrix} -n^{y}_{R} \\ -n^{x}_{R} \\ -n_{R}^{z} \end{pmatrix}. 
 \label{RDrelspinden}
\end{eqnarray}
As a consequence, also in the Dresselhaus case, all spin densities are vanishing:
\begin{equation}
 n^{x}_{D}(\varphi)=n^{y}_{D}(\varphi)=n^{z}_{D}(\varphi)=0. \label{Dspinden}
\end{equation}
Furthermore, the Dresselhaus spin currents and spin sources are related to the Rashba ones for $x_D=x_R$:
\begin{eqnarray}
\begin{pmatrix} j^{x,\rm e/o}_{D} \\ j^{y,\rm e/o}_{D} \\ j_{D}^{z,\rm e/o} \end{pmatrix} 
= \begin{pmatrix} -j^{y,\rm e/o}_{R} \\ -j^{x,\rm e/o}_{R} \\ -j_{R}^{z,\rm e/o} \end{pmatrix}, 
\begin{pmatrix}
 s_{D}^{x,\rm e/o} \\ s_{D}^{y,\rm e/o} \\ s_{D}^{z,\rm e/o}
\end{pmatrix}=
\begin{pmatrix}
 -s_{R}^{y,\rm e/o} \\ -s_{R}^{x,\rm e/o} \\ -s_{R}^{z,\rm e/o}
\end{pmatrix}. \label{RDrel}
\end{eqnarray}
Figure \ref{Dresselhaus_spin_currents} shows the geometrical arrangement of the sources and spin currents. The differences to the Rashba ring can be stated as follows:
\begin{enumerate}
 \item The transport pattern for the $x$-component of the spin is rotated by $-\pi/2$.
 \item The transport pattern for the $y$-component of the spin is rotated by $\pi/2$.
 \item The $\varphi$-independent current for the $z$-component of the spin flows in the opposite direction.
\end{enumerate}

\section{Model and theory for the 2D ring coupled to external leads}

In this section, we describe the central system potential $V_S$
for the broad quantum ring and its connection to the leads.
The electronic ring system is embedded in an electromagnetic cavity by
coupling a many-level electron system with photons using the full
photon energy spectrum of a single cavity mode. The central ring
system is described by a MB system Hamiltonian
$\hat{H}_{S}$ with a uniform perpendicular magnetic
field, in which the electron-electron interaction and the
electron-photon coupling to the $x$- or $y$-polarized photon field is
explicitly taken into account. We employ the TCL-GME approach to
explore the non-equilibrium electronic transport when the system is
coupled to leads by a transient switching potential.

\subsection{Quantum ring potential}

\begin{figure}[htbq] 
       \includegraphics[width=0.45\textwidth,angle=0, bb= 69 89 466 310]{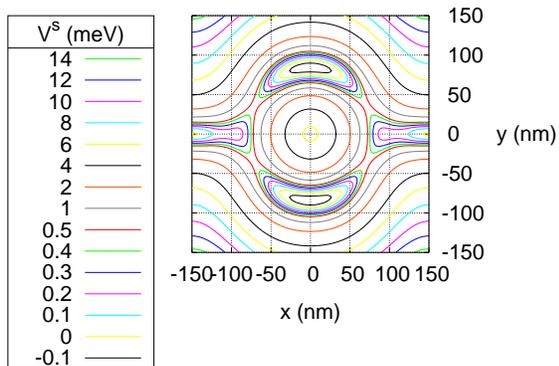}
       \caption{(Color online) Equipotential lines in the central ring  system connected
       to the left and right leads.
       Note that the isolines are refined close to the bottom of the ring structure.}
       \label{ext_pot}
\end{figure}

The quantum ring is embedded in the central system of length $L_x = 300$~\textrm{nm}
situated between two contact areas that will be coupled to the
external leads, as is depicted in \fig{ext_pot}. The system
potential is described by
\begin{eqnarray}
 V_S(\mathbf{r})&=& \sum_{i=1}^{6}V_{i}\exp
 \left[
 -\left(\beta_{xi}(x-x_{0i})\right)^2 - \left(\beta_{yi}y\right)^2
 \right] \nonumber \\
 &&+\frac{1}{2}m^* \Omega_{0}^2y^2, \label{V_S}
\end{eqnarray}
with the parameters from table \ref{table:ringpot}, 
which are selected such that the potential is a bit higher at the
contact regions (the place where electrons tend otherwise to accumulate)
than at the ring arms to guarantee a uniform density distribution along the ring.
$x_{03}=\epsilon$ is a small numerical symmetry breaking parameter and
$|\epsilon|=10^{-5}$~nm is enough for numerical stability. 
In \eq{V_S}, $\hbar\Omega_0 = 1.0$~meV is the characteristic energy of the confinement and
$m^* = 0.067 m_e$ is the effective mass of an electron in GaAs-based material.
The ring radius $a\approx 80$~nm, 
which means that $\alpha \approx x_R \times 7.1$~meV\,nm.
\begin{table}
 \caption{Parameters of the central region ring potential.}
 \centering
 \begin{tabular}{c    |    c    |    c    |    c    |    c}
\hline\hline
\\ [-1.4ex]
$i$ &  $V_{i}$ in meV &  $\beta_{xi}$ in $\frac{1}{\mathrm{nm}}$
    &  $x_{0i}$ in nm &  $\beta_{yi}$ in $\frac{1}{\mathrm{nm}}$ \\ [0.5ex]
\hline
1 & 10 & 0.013 & 150 & 0 \\
2 & 10 & 0.013 & -150 & 0 \\
3 & 11.1 & 0.0165 & $\epsilon$ & 0.0165 \\
4 & -4.7 & 0.02 & 149 & 0.02 \\
5 & -4.7 & 0.02 & -149 & 0.02 \\
6 & -5.33 & 0 & 0 & 0 \\[0.5ex]
\hline\hline
\end{tabular}
\label{table:ringpot}
\end{table}

\subsection{Lead Hamiltonian}

The Hamiltonian for the semi-infinite lead $l\in\{L,R\}$ (left or right lead),
\begin{eqnarray}
\hat{H}^{l}&=&\int d^2 r\; \int d^2 r'\; \hat{\mathbf{\Psi}}_{l}^{\dagger}(\mathbf{r}')\delta(\mathbf{r}'-\mathbf{r})\left[\left[\frac{\hat{\mathbf{p}}_l^2}{2m^{*}} +V_l(\mathbf{r})\right]\nonumber \right. \\
&&+\left.H_{Z}+\hat{H}_{R}(\mathbf{r})+ \hat{H}_{D}(\mathbf{r})\right]\hat{\mathbf{\Psi}}_{l}(\mathbf{r}), \label{H^l}
\end{eqnarray}
with the momentum operator containing only the kinetic momentum 
and the vector potential coming from the magnetic field (i.e.\ no photon field)
\begin{equation}
 \hat{\mathbf{p}}_l(\mathbf{r})=\frac{\hbar}{i}\nabla +\frac{e}{c} \mathbf{A}(\mathbf{r}).   \label{mom_l}
\end{equation}
We remind the reader that the Rashba part, $\hat{H}_{R}(\mathbf{r})$, (\eq{H_R}) 
and Dresselhaus part, $\hat{H}_{D}(\mathbf{r})$, (\eq{H_D})
of the spin-orbit interaction are momentum dependent and it is the momentum from \eq{mom_l},
which is used for these terms in \eq{H^l}. Equation (\ref{H^l}) contains the lead field operator
\begin{equation}
 \hat{\Psi}_{l}(x)=\sum_{q}\psi_{ql}(x)\hat{C}_{ql} \label{FOl}
\end{equation}
in the two-component vector
\begin{equation}
 \hat{\mathbf{\Psi}}_{l}(\mathbf{r})=\begin{pmatrix} \hat{\Psi}_{l}(\uparrow,\mathbf{r}) \\ \hat{\Psi}_{l}(\downarrow,\mathbf{r}) \end{pmatrix} 
\end{equation}
and a corresponding definition of the hermitian conjugate to \eq{conj_FOS}.
In \eq{FOl}, $\psi_{ql}(x)$ is a SES in the lead $l$ (eigenstate with quantum number $q$ of Hamiltonian \eq{H^l})
and $\hat{C}_{ql}$ is the associated electron annihilation operator.
The lead potential
\begin{equation}
 V_{l}(\mathbf{r})=\frac{1}{2} m^{*} \Omega_{l}^2 y^2
\end{equation}
confines the electrons parabolically in $y$-direction.
We use a relatively strong confinement, 
$\hbar\Omega_l = 2.0$~meV, to reduce the number of subbands
in the leads and thereby the computational effort 
for our total time-dependent quantum system.

\subsection{Time-convolutionless generalized master equation approach}

We use the time-convolutionless generalized master equation~\cite{PhysRevA.59.1633} TCL-GME,
which is a non-Markovian master equation that is local in time.
This master equation satisfies the positivity conditions~\cite{Whitney08:175304}
for the MB state occupation probabilities in the RDO usually
to a higher system-lead coupling strength~\cite{PhysRevB.87.035314}.
We assume, the
initial total statistical density matrix can be written as 
a product of the system and leads density matrices, before
switching on the coupling to the leads,
\begin{equation}
 \hat{W}(0)=\hat{\rho}_L \otimes \hat{\rho}_R \otimes \hat{\rho}_S(0),
\end{equation}
with $\rho_l$, $l\in \{L,R\}$, being the normalized density matrices
of the leads. The coupling Hamiltonian between the central system
and the leads reads
\begin{equation}
 \hat{H}_{T}(t)= \sum_{l=L,R}\int dq\;\chi^{l}(t)\left[\hat{\mathfrak{T}}^{l}(q)\hat{C}_{ql}
 +\hat{C}_{ql}^{\dagger} \hat{\mathfrak{T}}^{l\dagger}(q) \right]\,
 .
\end{equation}
The coupling is switched on at $t=0$ via the switching function
\begin{equation}
 \chi^{l}(t)=1-\frac{2}{e^{\alpha^{l}t}+1}
\end{equation}
with switching parameter $\alpha^l$ and
\begin{equation}
 \hat{\mathfrak{T}}^{l}(q)
 = \sum_{\alpha \beta}\ketmes{\alpha}\brames{\beta}\sum_{a}T_{qa}^{l}
 \brames{\alpha} \hat{C}_{a}^{\dagger}\ketmes{\beta}.
 \label{mathfrakT}
\end{equation}
Equation (\ref{mathfrakT}) is written in
the system Hamiltonian MB eigenbasis $\{\ketmes{\alpha}\}$.  The
coupling tensor \cite{Gudmundsson85:075306}
\begin{eqnarray}
 T_{qa}^{l}&=&\sum_{\sigma}\sum_{\sigma'}\int_{\Omega^l} d^2r\; \int_{\Omega_S^l} d^2r'\;
 \psi_{ql}^{*}(\mathbf{r},\sigma) \nonumber \\
 && \times g_{aq}^{l}(\mathbf{r},\mathbf{r'},\sigma,\sigma')\psi_a^S(\mathbf{r}',\sigma')
\end{eqnarray}
couples the lead SES
$\{\psi_{ql}(\mathbf{r},\sigma)\}$ with energy spectrum $\{\epsilon^l(q)\}$
to the system SES $\{\psi_a^S(\mathbf{r},\sigma)\}$ with energy spectrum
$\{E_a\}$ that reach into the contact regions,
\cite{1367-2630-11-11-113007} $\Omega_S^l$ and $\Omega_l$, of system
and lead $l$, respectively, and
\begin{eqnarray}
 g_{aq}^l(\mathbf{r},\mathbf{r'},\sigma,\sigma')&=&g_0^l\delta_{\sigma,\sigma'}\exp\left[-\delta_x^l(x-x')^2-\delta_y^l(y-y')^2 \right]\nonumber \\
&&\times \exp{
\left(
 -\frac{|E_a-\epsilon^l(q)|}{\Delta^l_E}
\right)} \label{gaql}
\end{eqnarray}
includes the same-spin coupling condition. Note that the meaning of $x$ in \eq{gaql}
is $\mathbf{r}=(x,y)$ and not $x=\mathbf{r},\sigma$.
In \eq{gaql}, $g_0^l$ is the lead coupling strength. In addition, $\delta^l_x$
and $\delta^l_y$ are the contact region parameters for lead $l$ in $x$-
and $y$-direction, respectively. Moreover, $\Delta^l_E$ denotes the
affinity constant between the central system SES energy levels $\{E_a\}$
and the lead energy levels $\{\epsilon^l(q)\}$.

The reduced density
operator (RDO) of the system,
\begin{equation}
 \hat{\rho}_S(t)= \mathrm{Tr}_{L}\mathrm{Tr}_{R}[\hat{W}(t)],
\end{equation}
propagated with the TCL-GME~\cite{PhysRevA.59.1633,PhysRevB.87.035314} in the
Schr\"{o}dinger picture evolves to second order in the lead coupling strength in time via
\begin{eqnarray}
\dot{\hat{\rho}}_{S}(t)&=&-\frac{i}{\hbar}[\hat{H}_{S},\hat{\rho}_{S}(t)]-
\Bigg[\sum_{l=L,R} \int dq\;\Big[\hat{\mathfrak{T}}^{l}(q),
\hat{\Omega}^{l}(q,t)\hat{\rho}_{S}(t) \nonumber \\ &&-
f(\epsilon^{l}(q)) \left\{ \hat{\rho}_{S}(t),\hat{\Omega}^{l}(q,t)
\right\} \Big] +\mathrm{H.c.}\Bigg]
\end{eqnarray}
with
\begin{eqnarray}
 \hat{\Omega}^{l}(q,t) &=& \frac{1}{\hbar^2}\chi^{l}(t)\exp\left(-\frac{i}{\hbar}t\epsilon^{l}(q)\right) \nonumber \\
 && \times \hat{U}_{S}(t) \hat{\Pi}^{l}(q,t)  \hat{U}_{S}^{\dagger}(t),
\end{eqnarray}
\begin{eqnarray}
 \hat{\Pi}^{l}(q,t)&=& \int_{0}^{t}dt'\; \left[ \exp\left(\frac{i}{\hbar}t'\epsilon^{l}(q)\right) \chi^{l}(t') \right. \nonumber \\
 && \times \left. \hat{U}_{S}^{\dagger}(t') \hat{\mathfrak{T}}^{l\dagger}(q) \hat{U}_{S}(t') \right]
\end{eqnarray}
and $f(E)$ being the Fermi distribution function.

\section{Non-equilibrium transport properties for a 2D ring connected to leads}

We investigate the non-equilibrium electron
transport properties through a quantum ring system, 
which is situated in a photon
cavity and weakly coupled to leads. We assume GaAs-based
material with electron effective mass $m^*=0.067m_e$ and background
relative dielectric constant $\kappa = 12.4$. We consider a single cavity mode
with fixed photon excitation energy $\hbar\omega = 0.4$~meV. The
electron-photon coupling constant in the central system is $g^{EM} =
0.1$~meV.  Before switching on the coupling, we assume the central
system to be in the pure initial state with electron occupation
number $N_{e,\rm{init}}=0$ and and --- unless otherwise stated --- photon occupation
number $N_{ph,\rm{init}}=1$ of the electromagnetic field.

A small external perpendicular uniform magnetic field $B=10^{-5}$~T is applied through
the central ring system and the lead reservoirs to lift the spin degeneracy.  The area of the
central ring system is $A =\pi a^2\approx 2\times 10^{4}$~$\mathrm{nm}^2$ leading to
the magnetic field $B_0=\Phi_0/A \approx 0.2$~T corresponding to one flux quantum
$\Phi_0=hc/e$. The applied magnetic field $B<<B_0$ is therefore order of magnitudes
outside the AB regime.  
The temperature of
the reservoirs is assumed to be $T = 0.5$~K. The chemical potentials
in the leads are $\mu_L=1.55$~meV and $\mu_R=0.7$~meV leading to a
source-drain bias window $\Delta \mu = 0.85$~meV. We let the affinity
constant $\Delta_E^{l} = 0.25$~meV to be close to the characteristic electronic excitation energy in $x$-direction. In
addition, we let the contact region parameters for lead $l\in\{L,R\}$ in $x$-
and $y$-direction be $\delta^l_x = \delta^l_y= 4.39\times
10^{-4}$~$\rm{nm}^{-2}$.  The system-lead coupling strength $g_0^l =
1.371\times 10^{-3}$~${\rm meV}/{\rm nm}^{3/2}$.

There are several relevant length and time scales that should be
mentioned.  The 2D magnetic length is $l =
[c\hbar/(eB)]^{1/2} = 8.12$~$\mu$m.  The ring system
is parabolically confined in the $y$-direction with characteristic
energy $\hbar\Omega_0 = 1.0$~meV leading to a much shorter magnetic
length scale
\begin{align}
a_w &= \left(\frac{\hbar}{m^*\Omega_0}\right)^{1/2}
    \frac{1}{\sqrt[4]{1+[eB/(m^*c\Omega_0)]^2}}\nonumber\\
    &= 33.74\ {\rm nm}.
\end{align}
%
The time-scale for the switching on of the system-lead coupling is
$(\alpha^{l})^{-1}=3.291$~\textrm{ps}, the single-electron state
(1ES) charging time-scale $\tau_{\rm 1ES}\approx 30$~\textrm{ps},
and the two-electron state (2ES) charging time-scale $\tau_{\rm 2ES}
\gg 200$~\textrm{ps} described in the sequential tunneling regime. 
We study the transport properties for $0\leq t<\tau_{\rm 2ES}$,
when the system has not yet reached a steady state.

To get more insight into the local current flow in the ring system,
we define the top local charge ($\gamma=c$) and spin
($\gamma=x,y,z$) 
current through the upper arm ($y>0$) of the ring
\begin{equation}
 I_{\mathrm{top}}^{\gamma}(t)=\int_{0}^{\infty}dy\;j_x^{\gamma}(x=0,y,t)
\end{equation}
and the bottom local charge and spin current through the lower arm ($y<0$) of
the ring
\begin{equation}
 I_{\mathrm{bottom}}^{\gamma}(t)=\int_{-\infty}^{0}dy\;j_x^{\gamma}(x=0,y,t)\, .
\end{equation}
Here, the charge and spin current density,
\begin{equation}
 \mathbf{j}^{\gamma}(\mathbf{r},t)
 =\begin{pmatrix} j_x^{\gamma}(\mathbf{r},t)\\j_y^{\gamma}(\mathbf{r},t) \end{pmatrix}
 =\mathrm{Tr}[\hat{\rho}_{S}(t)\hat{\mathbf{j}}^{\gamma}(\mathbf{r})],
\end{equation}
is given by the expectation value of the charge and spin current density
operator, \eq{jcq}, \eq{jsxq}, \eq{jsyq} and \eq{jszq}.
Furthermore, to distinguish better the type and driving schemes
of the dynamical transport features,
we define the total local (TL) charge or spin current
\begin{equation}
 I_{\rm tl}^{\gamma}(t)=I_{\rm top}^{\gamma}(t)+I_{\rm bottom}^{\gamma}(t)
\end{equation}
and circular local (CL) charge or spin current
\begin{equation}
 I_{\rm cl}^{\gamma}(t)=\frac{1}{2}\left[I_{\rm bottom}^{\gamma}(t)-I_{\rm top}^{\gamma}(t)\right],
\end{equation}
which is positive if the electrons move counter-clockwise in the ring.
The TL charge current is usually bias driven while the CL charge current
could be driven by a magnetic field or circularly polarized photon field.
The TL spin current is usually related to non-vanishing sources
while a CL spin current can exist without sources.
In the supplemental material, we present
the spin photocurrent densities
\begin{equation}
 \mathbf{j}_{\rm ph}^{\gamma,p}(\mathbf{r},t)=\mathbf{j}^{\gamma,p}(\mathbf{r},t)-\mathbf{j}^{\gamma,0}(\mathbf{r},t),
\end{equation}
which are given by the difference of the associated local spin current densities 
with ($\mathbf{j}^{\gamma,p}(\mathbf{r},t)$) and without ($\mathbf{j}^{\gamma,0}(\mathbf{r},t)$) photons, where
$p=x,y$ denotes the polarization of the photon field
($x$: $x$-polarization, $y$: $y$-polarization) and $\gamma\in\{x,y,z\}$.
Below, we shall explore the influence of the Rashba and Dresselhaus parameter and the
photon field polarization on the non-equilibrium
quantum transport in terms of the above time-dependent
currents in the broad quantum ring system connected to leads.

\subsection{2D Rashba ring}

Here, we will describe our numerical results (ME spectrum and charge and spin currents)
for the finite-width ring with only Rashba spin-orbit interaction and  
and compare them to the analytical results for the 1D ring.

\subsubsection{Local charge current}

\begin{figure}[htbq] 
       \subfigure[]{
       \includegraphics[width=0.34\textwidth,angle=-90]{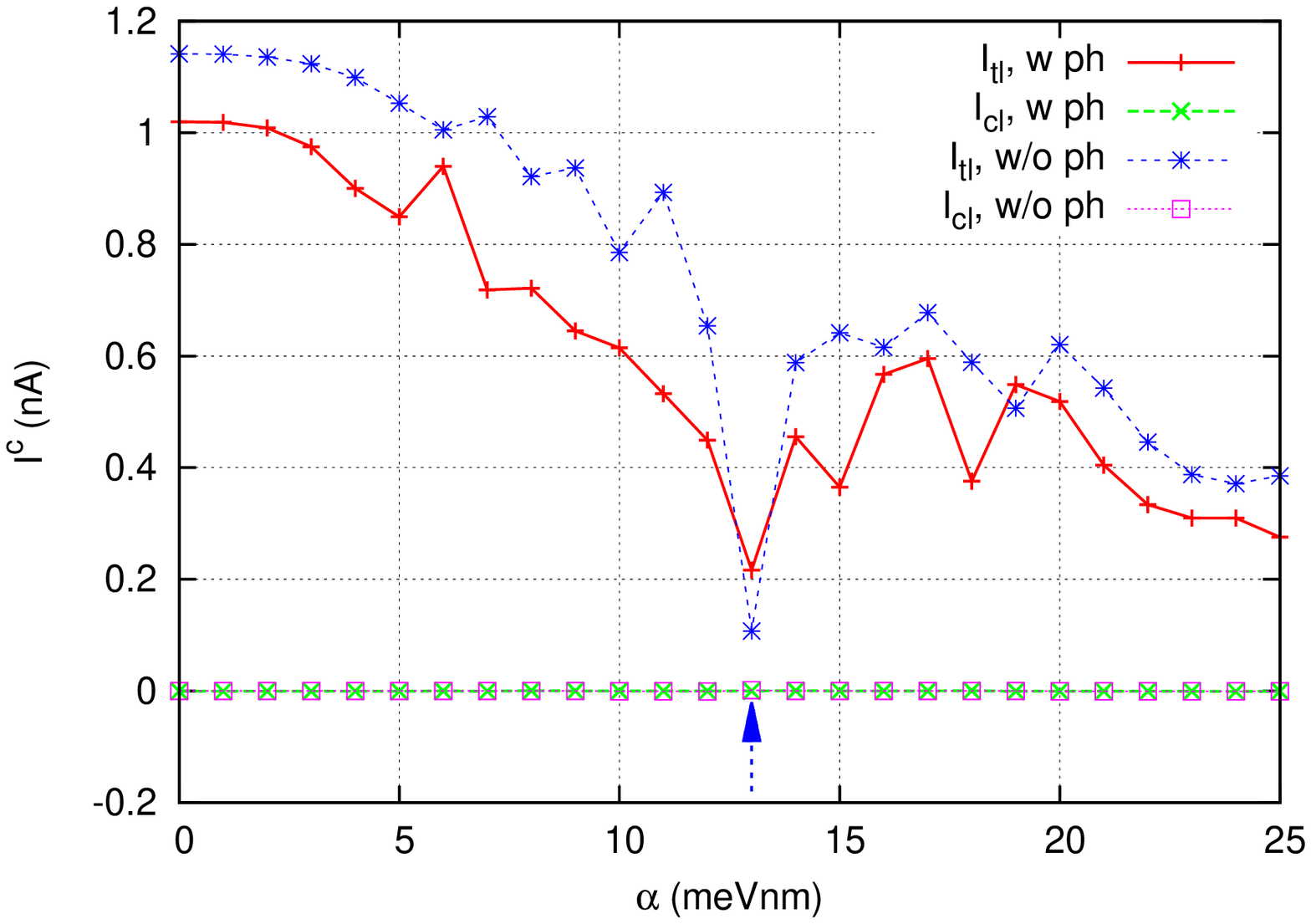}
       \label{current_arm_x_alpha_200}}
       \subfigure[]{
       \includegraphics[width=0.34\textwidth,angle=-90]{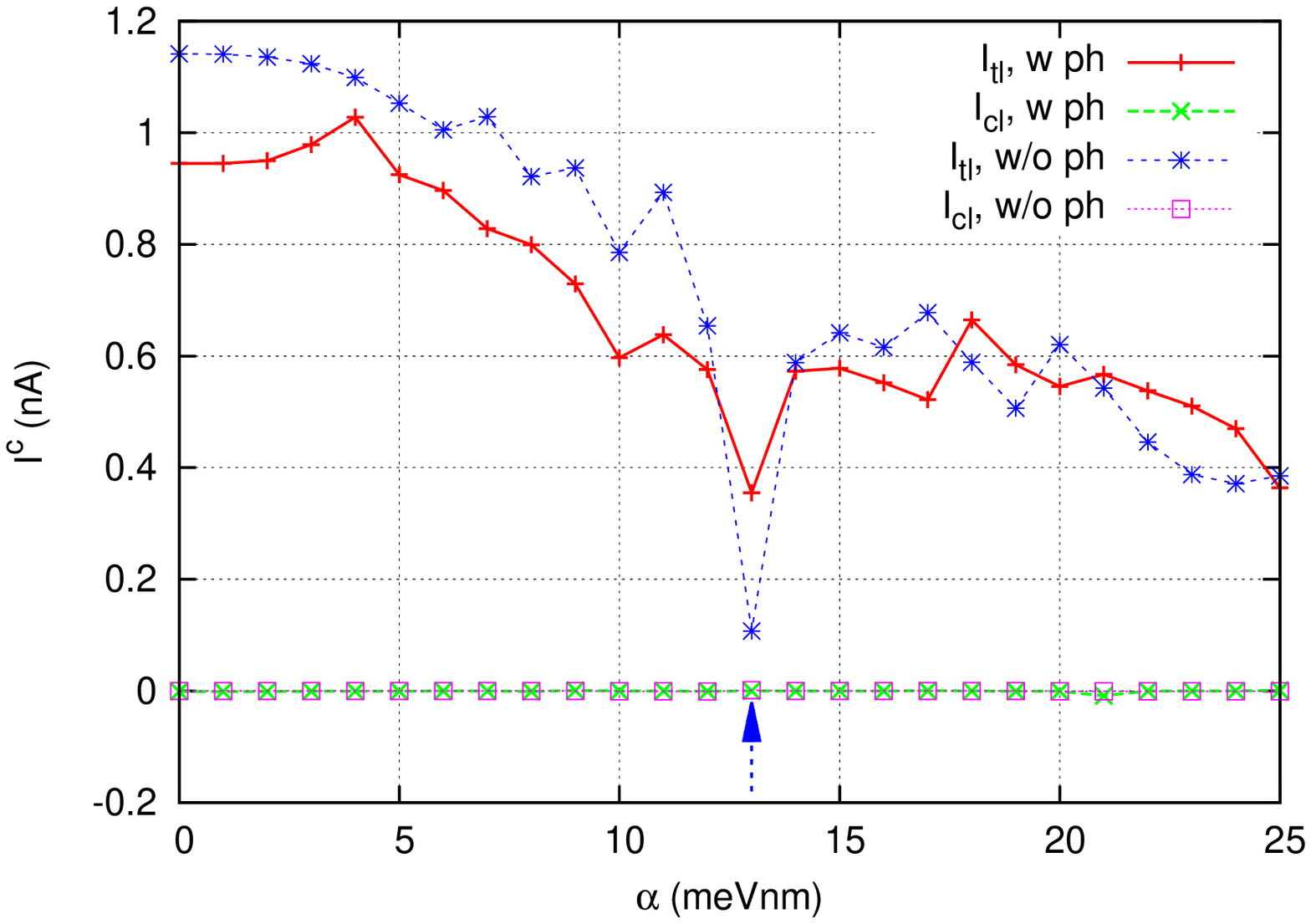}
       \label{current_arm_y_alpha_200}}
       \caption{(Color online) Total local current
       ($I_{\rm tl}^{c}$) and CL current
       ($I_{\rm cl}^{c}$) versus the Rashba coefficient and
       averaged over the time interval $[180,220]$~ps to 
       give a more representative picture in the transient regime 
       with (w) \subref{current_arm_x_alpha_200} $x$-polarized photon field 
       and \subref{current_arm_y_alpha_200} $y$-polarized photon field
       or without (w/o) photon cavity.
       The Dresselhaus coefficient $\beta=0$. The blue arrow denotes the
       position of the first AC destructive phase.}
       \label{current_arm_alpha_200}
\end{figure}

Figure \ref{current_arm_alpha_200} shows the local charge currents as a function
of the Rashba coefficient. The CL charge current is close to zero 
as the linearly polarized photon field and negligible magnetic field promote
no circular charge motion. This is in agreement with the
exact result of the 1D closed (i.e.\ not connected to electron reservoirs) Rashba ring, 
where the charge current vanishes. The non-vanishing TL charge current
is therefore solely induced by the bias between the leads.
Around $\alpha^c\approx 13$~meV\,nm (blue arrow) the TL charge current
has a pronounced minimum coming from the AC destructive phase interference
at $x_R^c=x_0^o=\sqrt{3} \approx 1.73$ or
$\alpha^c \approx x_R^c \times 7.1 \approx 12.3$~meV\,nm.
The linearly polarized photons tend in general to suppress the local charge current
as the increasing number of possible MB states tends to constrict them
to smaller energy differences in the MB spectrum. 
However, especially for the $y$-polarized photon field \fig{current_arm_alpha_200}\subref{current_arm_y_alpha_200}, 
the AC minimum appears weaker, and for large values $\alpha \geq 18$ 
the TL current can be enhanced.

\begin{figure}[htbq] 
       \includegraphics[width=0.34\textwidth,angle=-90]{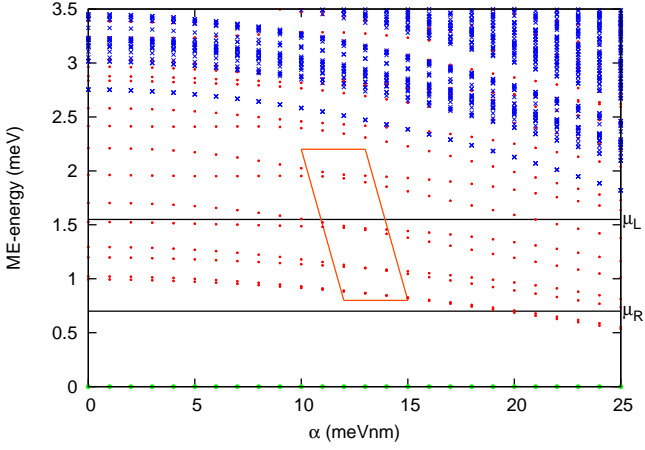}
       \caption{(Color online) ME energy spectrum of the system Hamiltonian \eq{H^S}
       versus the Rashba coefficient $\alpha$ without photon cavity.
       The states are differentiated according to their electron content $N_e$:
       zero-electron state ($N_e=0$, 0ES, green dot), single electron states ($N_e=1$, SES, red dots)
       and two electron states ($N_e=2$, 2ES, blue crosses). The Dresselhaus coefficient $\beta=0$.
       The bias window $[\mu_R,\mu_L]$ is depicted by solid black lines. The orange parallelogram
       indicates the location of SES crossings.}
       \label{MB_spec_valpha_0}
\end{figure}

To investigate the charge current minimum (blue arrow in \fig{current_arm_alpha_200}) further, 
we have a look at the ME spectrum as a function of the Rashba coefficient,
\fig{MB_spec_valpha_0}, where the zero-electron state is marked in green color, the SESs in red color
and the two electron states in blue color.
Around $\alpha\approx 13$~meV\,nm, we observe crossings of the SESs (inside the orange parallelogram in \fig{MB_spec_valpha_0}), which
correspond to the AC destructive phase interference at $x_R^c$. 
We see clearly that the phase relation and the TL charge current behavior are linked
due to the appearance of a current-suppressing ME degeneracy.~\cite{PhysRevB.87.035314}  
It is also interesting to notice that the critical Rashba coeffient describing the location of the crossing point is the smaller the higher a selected SES lies in energy.
As the spin-orbit wavefunctions of the higher-in-energy SESs are more extended 
the associated effective 1D ring radius $a$ increases.
Now, since $\alpha_0^o=\frac{\hbar^2 x_0^o}{2m^* a}$ obtained from \eq{defxR}, 
the first crossing point value  $\alpha_0^o$ is located at a smaller $\alpha$-value. 
It is important, however, to be aware that mainly the SES
around the bias window $[\mu_R,\mu_L]$ are contributing to the transport properties.

\subsubsection{Local spin current and 1D comparison}

\begin{figure}[htbq]
       \subfigure[]{
       \includegraphics[width=0.41\textwidth,angle=-90]{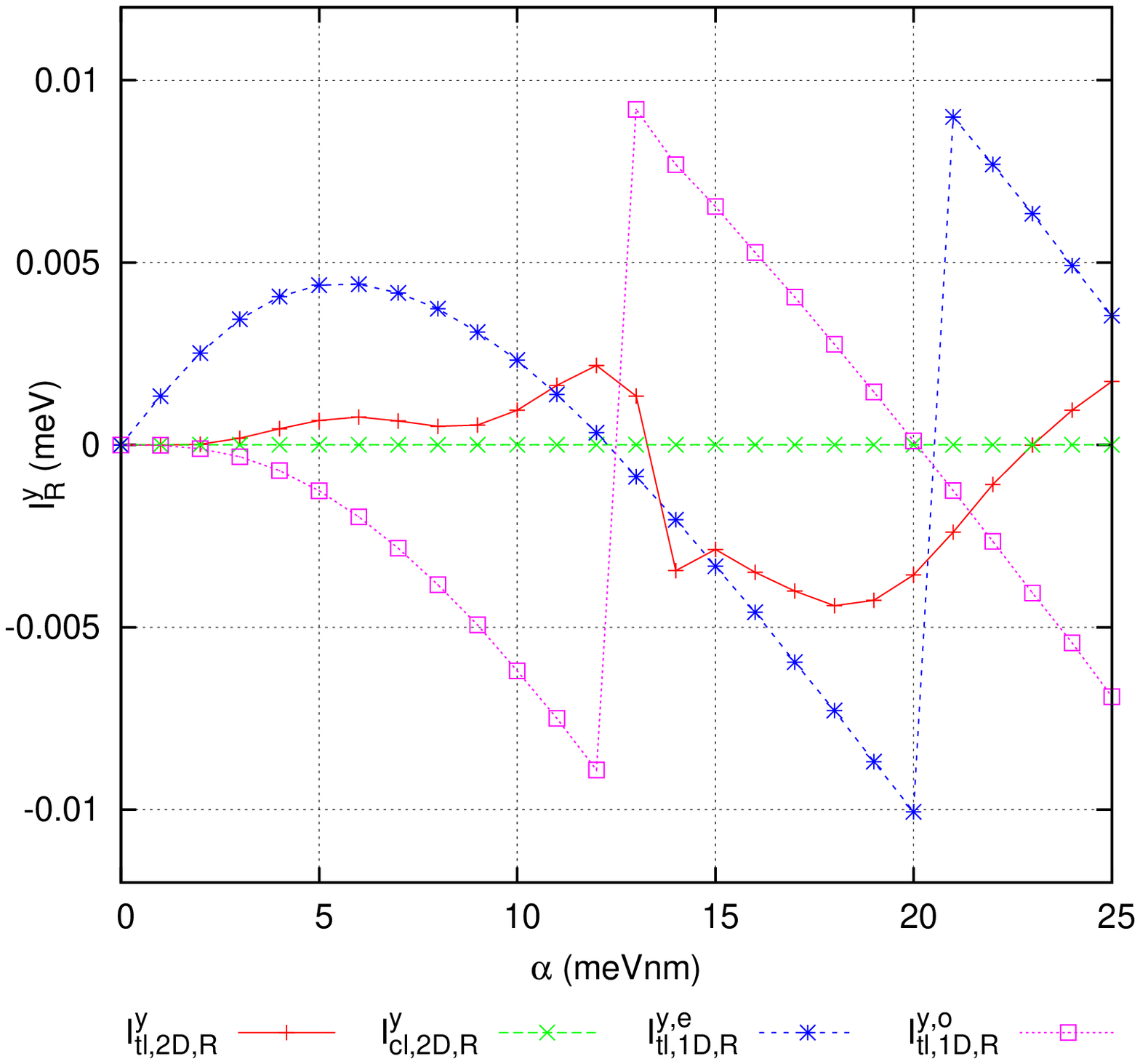}
       \label{current_arm_x_sy_alpha_200_comp2}}
       \subfigure[]{
       \includegraphics[width=0.41\textwidth,angle=-90]{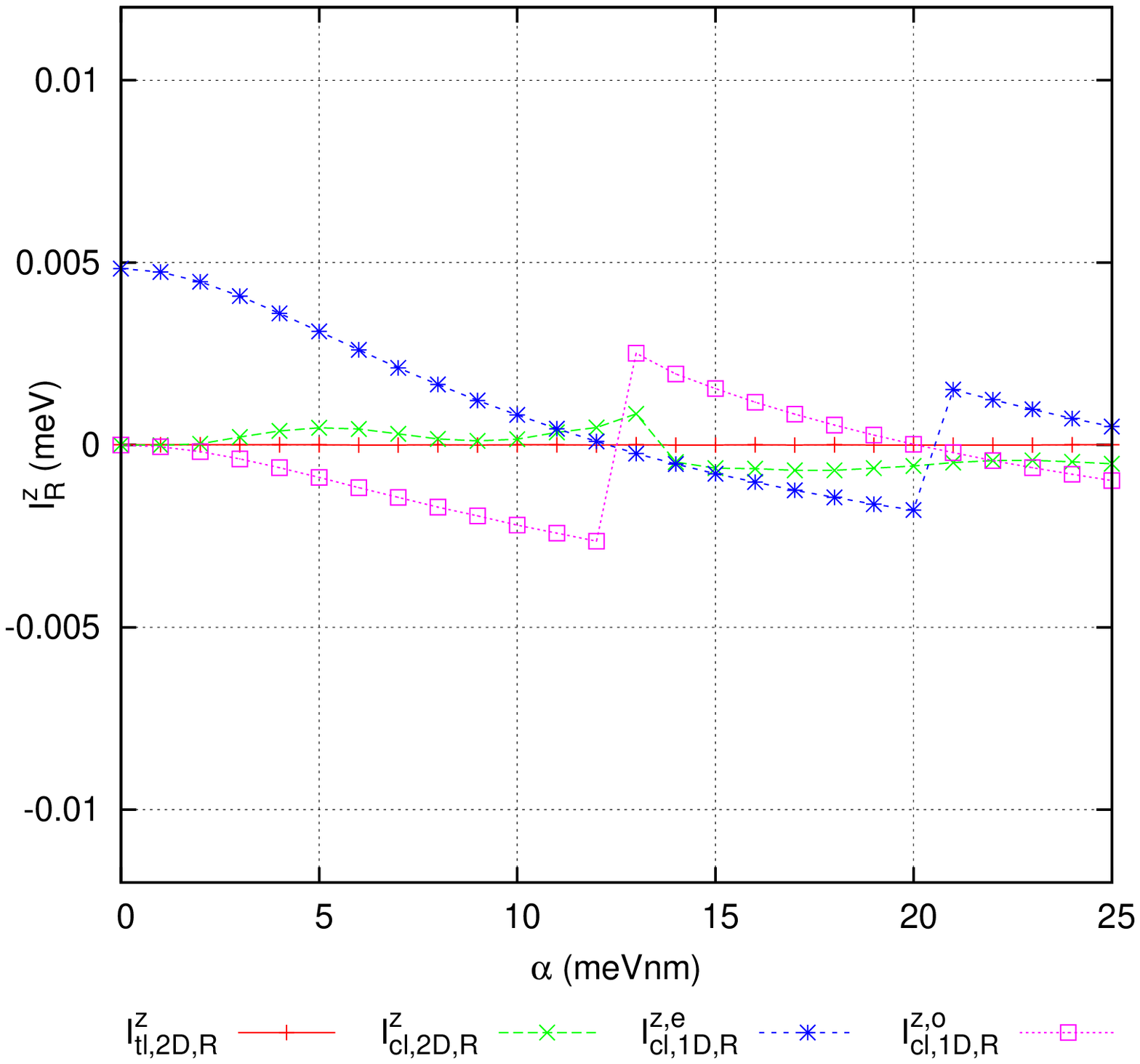}
       \label{current_arm_x_sz_alpha_200_comp2}}
       \caption{(Color online) 2D TL Rashba spin current $I_{tl}^{i}$
       or 2D CL Rashba spin current $I_{cl}^{i}$ 
       averaged over the time interval $[180,220]$~ps 
       without photon cavity in comparison
       with the 1D TL Rashba spin current $I_{tl,1D,R}^{i,e/o}$ or
       1D CL Rashba spin current $I_{cl,1D,R}^{i,e/o}$
       for even or odd cardinalities and
       with the electron number $N_e$ taken from the 2D case. 
       The Rashba spin currents are shown for
       \subref{current_arm_x_sy_alpha_200_comp2} $S_y$ spin polarization
       and \subref{current_arm_x_sz_alpha_200_comp2} $S_z$ spin polarization versus the
       Rashba coefficient $\alpha$. The Dresselhaus coefficient $\beta=0$ and the ring radius $a=80$~nm. 
       The 1D TL and CL spin currents, which are equal to zero,
       $I_{cl,1D,R}^{y,e/o}=I_{tl,1D,R}^{z,e/o}=0$, 
       are not shown.}
       \label{current_arm_x_alpha_200_comp}
\end{figure}


In \fig{current_arm_x_alpha_200_comp}, we compare the 2D local Rashba spin currents $I_{tl/cl,2D,R}^{i}$ without photon field with the analogously defined 1D TL Rashba spin current
\begin{equation}
 I_{tl,1D,R}^{i,e/o}=-j_{R}^{i,e/o}(\frac{\pi}{2})+j_{R}^{i,e/o}(-\frac{\pi}{2})
\end{equation}
and 1D CL Rashba spin current 
\begin{equation}
 I_{cl,1D,R}^{i,e/o}=\frac{1}{2}\left(j_{R}^{i,e/o}(\frac{\pi}{2})+j_{R}^{i,e/o}(-\frac{\pi}{2})\right).
\end{equation}
For the electron number $N_e$, which $j_{1D,R}^{i,e/o}$ depends on, we have
chosen the corresponding value of the 2D Rashba ring without photon cavity
averaged over the time interval $[180,220]$~ps.
Neither is $N_e$ an integer number in general, 
nor are the state occupancies in the central system following
a Fermi distribution due to the geometry- and energy-dependent 
coupling to the biased leads and electron correlations
suggesting to compare to the cases
of both even and odd cardinalities for the 1D Rashba spin current $j_{1D,R}^{i,e/o}$.
For $S_x$ spin polarization, in a plot of the same scale as 
\fig{current_arm_x_alpha_200_comp}, the 2D TL and CL Rashba spin currents,
$I_{tl}^{x}$ and $I_{cl}^{x}$, can not be distinguished from a zero line.
The corresponding 1D TL and CL Rashba spin currents are zero: $I_{tl,1D,R}^{x,e/o}=I_{cl,1D,R}^{x,e/o}=0$.

Figure \ref{current_arm_x_alpha_200_comp} shows that the 2D spin currents are in general smaller
than the 1D spin currents (often in between the 1D Rashba spin currents for even and odd cardinality, $I_{1D,R}^{i,e}$ and $I_{1D,R}^{i,o}$, respectively). This is because many ME states are contributing, which are only fractionally occupied.
Furthermore, the 2D structure smoothens the discontinuities with respect to $\alpha$ thus reducing further
the peaks in the 1D currents.
Nonetheless, some similarities can be found for the $\alpha$-values regarding the position of the zero transitions.
In particular considering the zero transitions, it seems that the even cardinality case is the more appropriate 
case to describe the spin currents of the finite-width ring. 
Furthermore, there is strong agreement in the spin currents, which are supposed to be zero. 
The 1D local Rashba spin currents for spin polarization $S_x$, 
$I_{tl,1D,R}^{x,e/o}$ and $I_{cl,1D,R}^{x,e/o}$, are zero.
The same is true for the 1D CL Rashba spin current for spin polarization $S_y$, $I_{cl,1D,R}^{y,e/o}$, 
and the 1D TL Rashba spin current for spin polarization $S_z$, $I_{tl,1D,R}^{z,e/o}$.
When looking at the associated 2D local Rashba spin currents, we find that these currents are also close to zero.

\begin{figure}[htbq] 
       \subfigure[]{
       \includegraphics[width=0.34\textwidth,angle=-90]{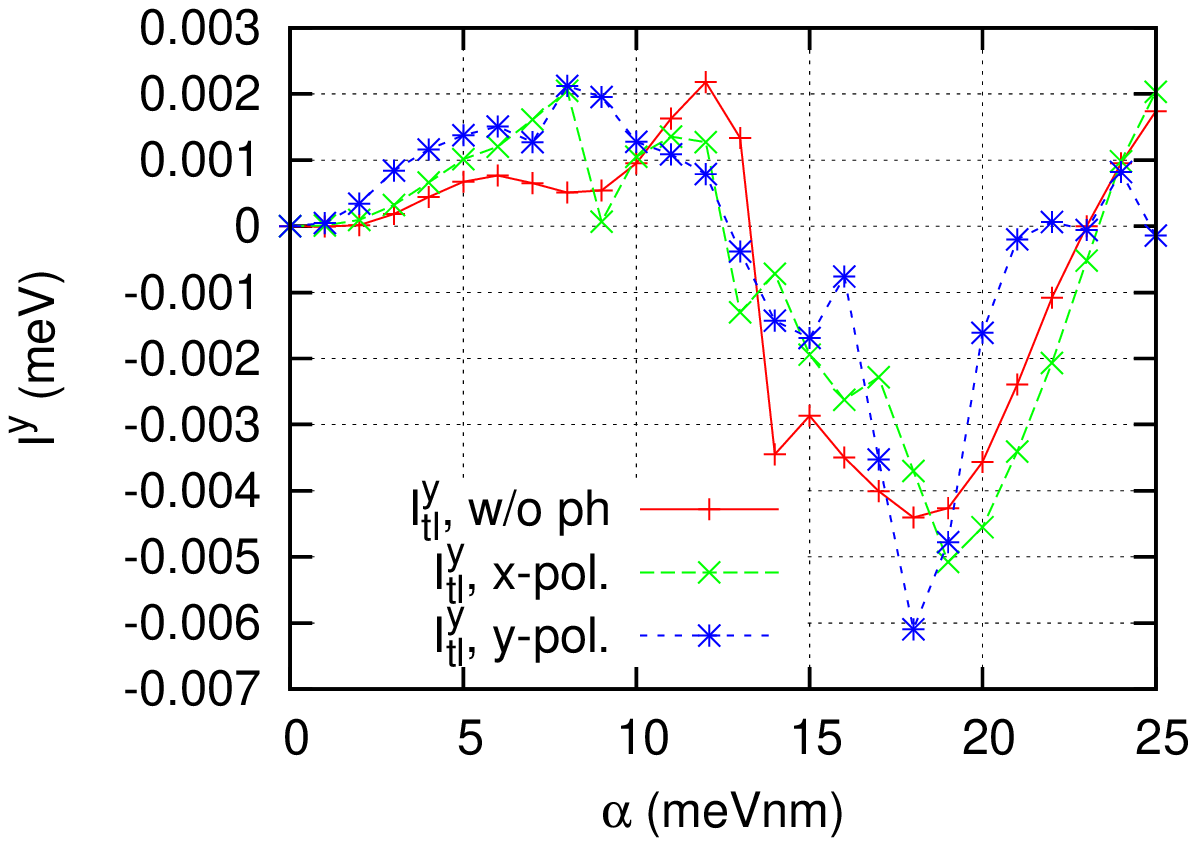}
       \label{current_arm_0xy_sy_alpha_200}}
       \subfigure[]{
       \includegraphics[width=0.34\textwidth,angle=-90]{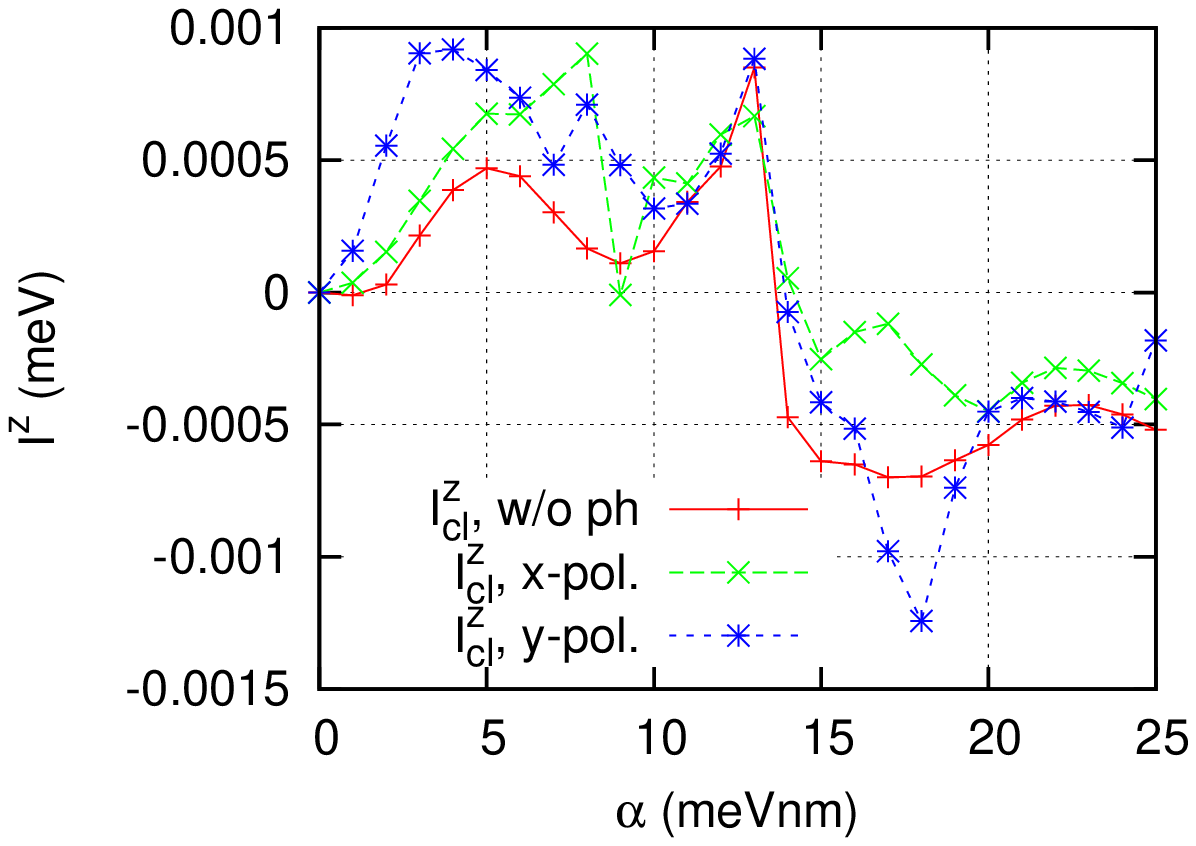}
       \label{current_arm_0xy_sz_alpha_200}}
       \caption{(Color online) \subref{current_arm_0xy_sy_alpha_200} 
       Total local Rashba spin current $I_{tl}^{y}$ for $S_y$ spin polarization
       and \subref{current_arm_0xy_sz_alpha_200} 
       CL Rashba spin current $I_{cl}^{z}$ for $S_z$ spin polarization
       averaged over the time interval $[180,220]$~ps without (w/o) photon cavity,
       $x$-polarized photon field and $y$-polarized photon field versus the
       Rashba coefficient $\alpha$ for the ring of finite width.
       The Dresselhaus coefficient $\beta=0$.}
       \label{current_arm_0xy_s_alpha_200}
\end{figure}

Figure \ref{current_arm_0xy_s_alpha_200} shows the Rashba local spin currents, which are far from zero
for $x$- or $y$-polarized photon field (and without photon field for comparison). 
The other spin currents remain close to zero even with photon field.
For $\alpha \leq 8$ the photon cavity field enhances the spin currents for both polarizations as opposed to the local charge current. In general, the modifications of the $y$-polarized photon field are a bit stronger due to the closer agreement of the characteristic electronic excitation energy
in $y$-direction with the photon mode energy $\hbar \omega =0.4$~meV.

 \begin{figure*}[htbq]
       \includegraphics[width=0.4\textwidth,angle=0,bb=89 59 252 219]{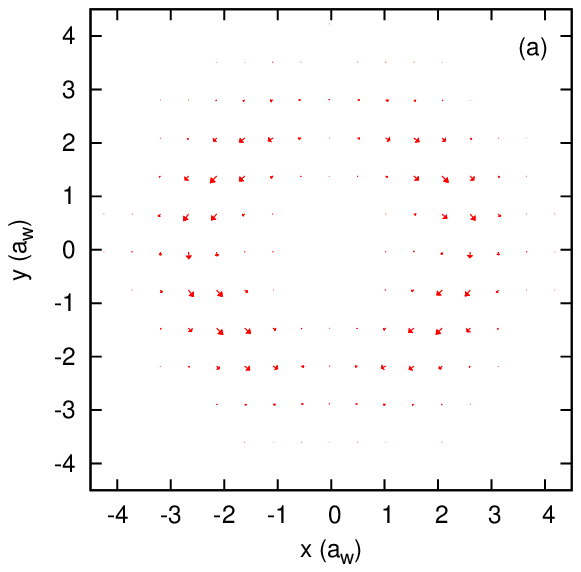}
       \includegraphics[width=0.4\textwidth,angle=0,bb=89 59 252 219]{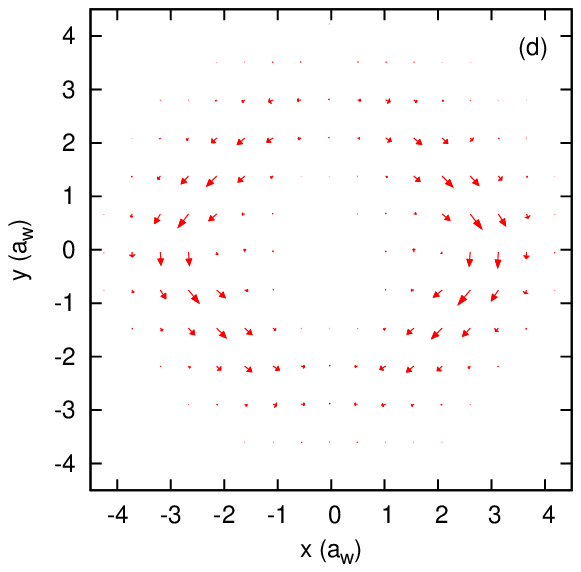}\\
       \includegraphics[width=0.4\textwidth,angle=0,bb=89 59 252 219]{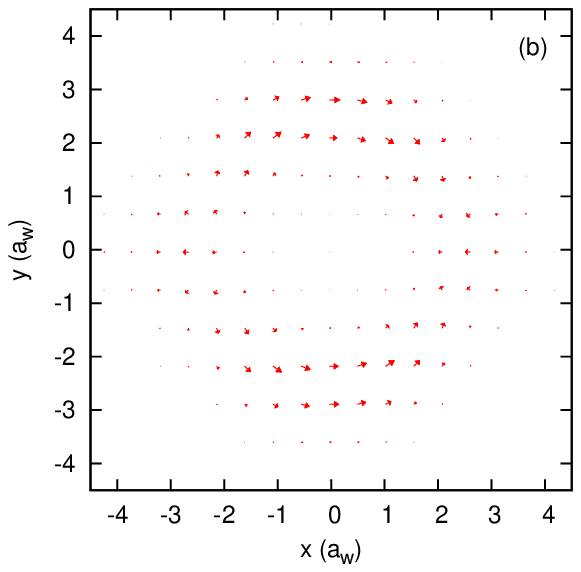}
       \includegraphics[width=0.4\textwidth,angle=0,bb=89 59 252 219]{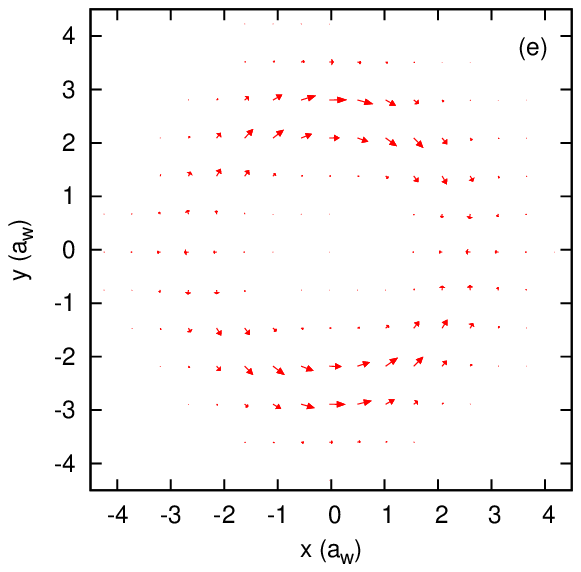}\\
       \includegraphics[width=0.4\textwidth,angle=0,bb=89 59 252 219]{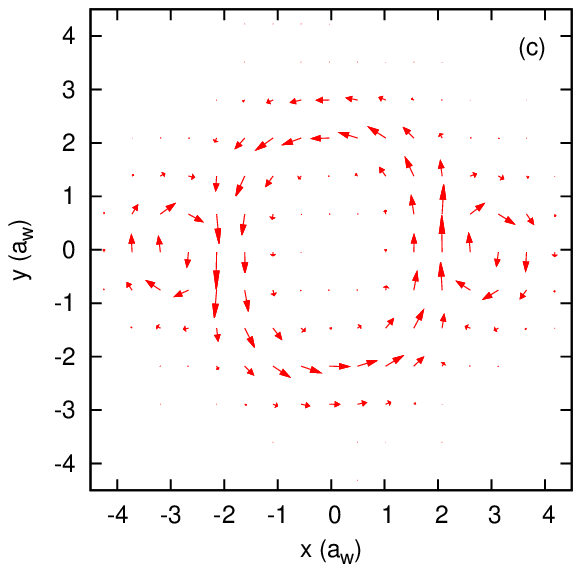}
       \includegraphics[width=0.4\textwidth,angle=0,bb=89 59 252 219]{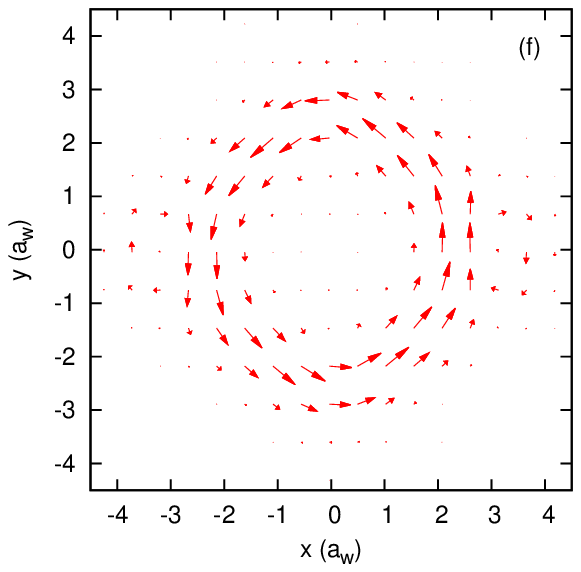}      
       \caption{(Color online) Spin current densities $\mathbf{j}^{i}(x,y)$, $i=x,y,z$ at $t=200$~ps 
       (a)-(c) without photon field or (d)-(e) with $x$-polarized photon field
       for (a) and (d) $S_x$ spin polarization, (b) and (e) $S_y$ spin polarization or
       (e) and (f) $S_z$ spin polarization. The Rashba coefficient $\alpha=5$~meV\,nm and
       the Dresselhaus coefficient $\beta=0$. 
       A spin current density vector of length $a_w$ corresponds to $1.25\times 10^{-3} \rm{meV}/a_w$.}
       \label{cur_den_alpha5_beta0_B000001_200_lowres}
\end{figure*}


Figure \ref{cur_den_alpha5_beta0_B000001_200_lowres} shows the spin current densities $\mathbf{j}^{x}(x,y)$ (top panels),
$\mathbf{j}^{y}(x,y)$ (middle panels) and $\mathbf{j}^{z}(x,y)$ (bottom panels). 
The photon field is switched off (left panels) or it is $x$-polarized (right panels).
The spin current densities are depicted for a Rashba coefficient $\alpha=5$~meV\,nm 
below the first destructive AC interference. 
We note that the spin densities show increasingly vortex structures for larger $\alpha$.
The results without photon cavity (left panels) have numerous similarities to the 1D ring:
the $S_x$ spin current density is maximal at $\varphi=0,\pi$ (\fig{cur_den_alpha5_beta0_B000001_200_lowres}(a)) and 
the $S_y$ spin current density is maximal at $\varphi=-\pi/2,\pi/2$ (\fig{cur_den_alpha5_beta0_B000001_200_lowres}(b)). 
Furthermore, the spin flow is along the $\varphi$-direction.
Also, the $S_z$ spin current density is almost homogeneous in $\varphi$ (\fig{cur_den_alpha5_beta0_B000001_200_lowres}(c)).
Furthermore, the relative directions of the spin flow are in agreement with the 1D case, 
when the flow directions are compared for the different spin polarizations.
A difference is the vortices around charge density maxima at the contact regions of the 2D ring
for $S_z$ spin polarization.

Next, we want to study the influence of linearly polarized photons on the spin current density distributions
(right versus left panels in \fig{cur_den_alpha5_beta0_B000001_200_lowres} for $x$-polarized photons).
All spin current densities are a bit larger for $x$-polarized photons except the vortices at the contact regions due to a redistribution of the charge density (i.e.\ by the density of potential spin carriers) from the contact regions to the ring arms.
The influence of the $x$-polarized photons is still considerably time-dependent in the time regime
shown in \fig{cur_den_alpha5_beta0_B000001_200_lowres}. The same is true for
$y$-polarization and the spin current densites without photon cavity (left panels in \fig{cur_den_alpha5_beta0_B000001_200_lowres}) reminding us about the non-equilibrium situation.
For supplemental material for the time dependency of 
the spin current densities
in the form of movies, we refer the reader to Ref. \onlinecite{supplmat}.

\subsection{2D Dresselhaus ring and comparison to the 2D Rashba ring}

\begin{figure}[htbq]       
       \subfigure[]{
       \includegraphics[width=0.218\textwidth,angle=-90,bb=255 53 551 358]{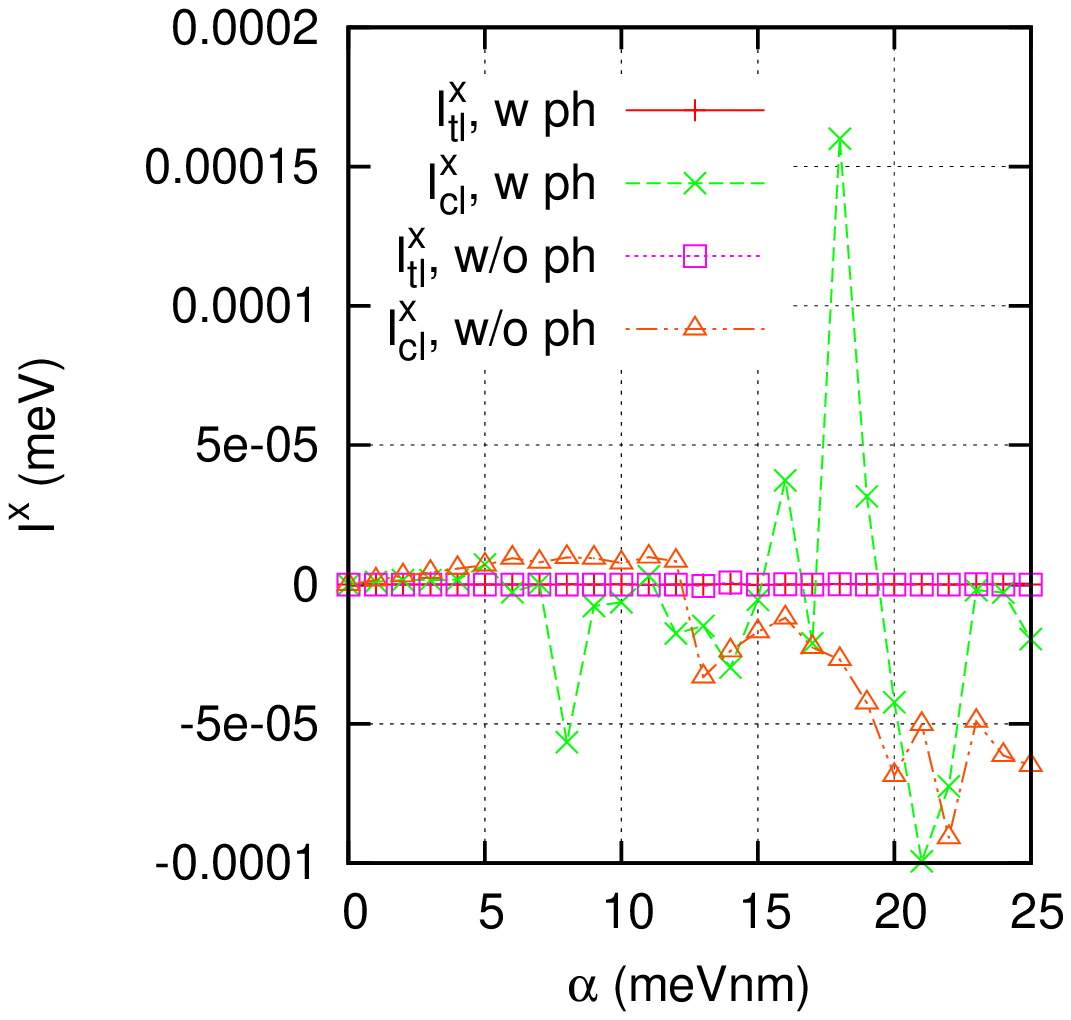} 
       \label{current_arm_x_sx_alpha_200}}
       \subfigure[]{
       \includegraphics[width=0.218\textwidth,angle=-90,bb=255 53 551 358]{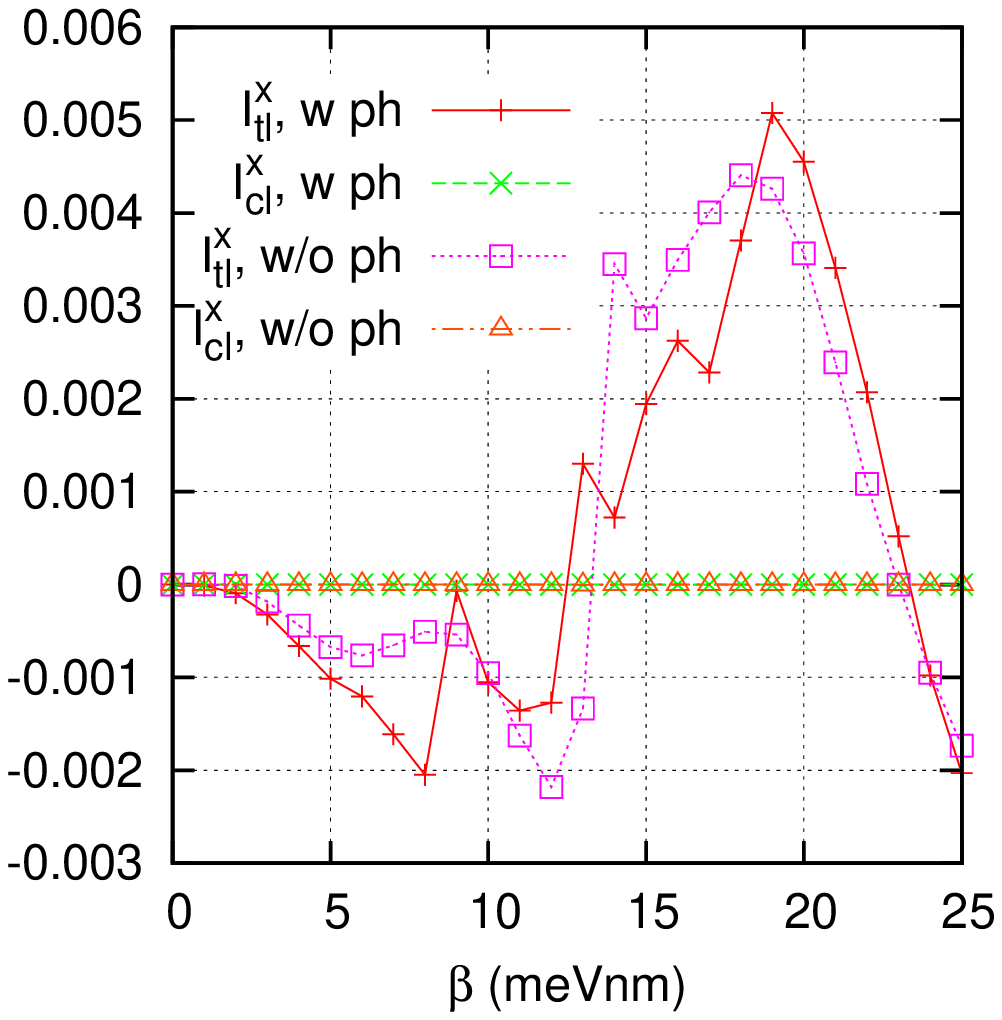}
       \label{current_arm_x_sx_beta_200}}
       \subfigure[]{
       \includegraphics[width=0.218\textwidth,angle=-90,bb=255 53 551 358]{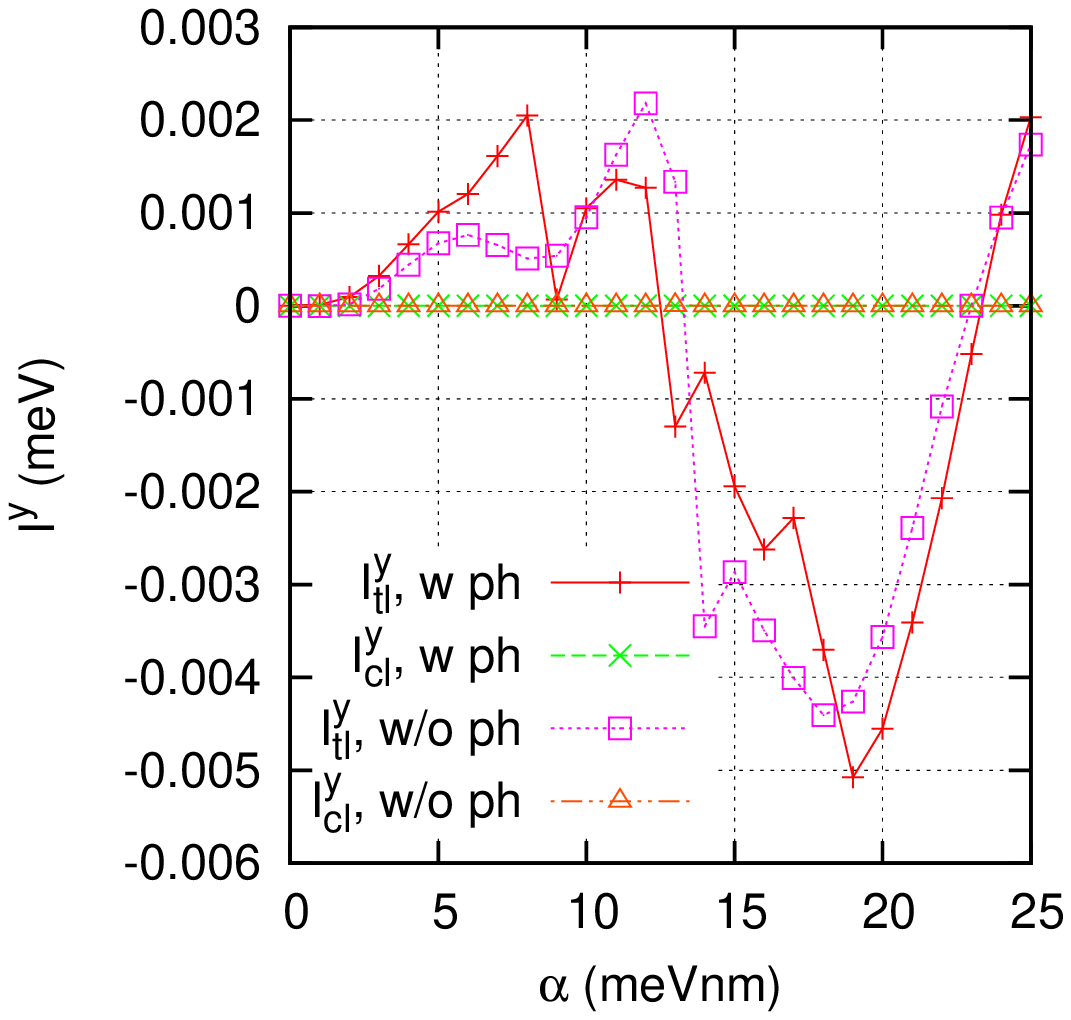} 
       \label{current_arm_x_sy_alpha_200}}
       \subfigure[]{
       \includegraphics[width=0.218\textwidth,angle=-90,bb=255 53 551 358]{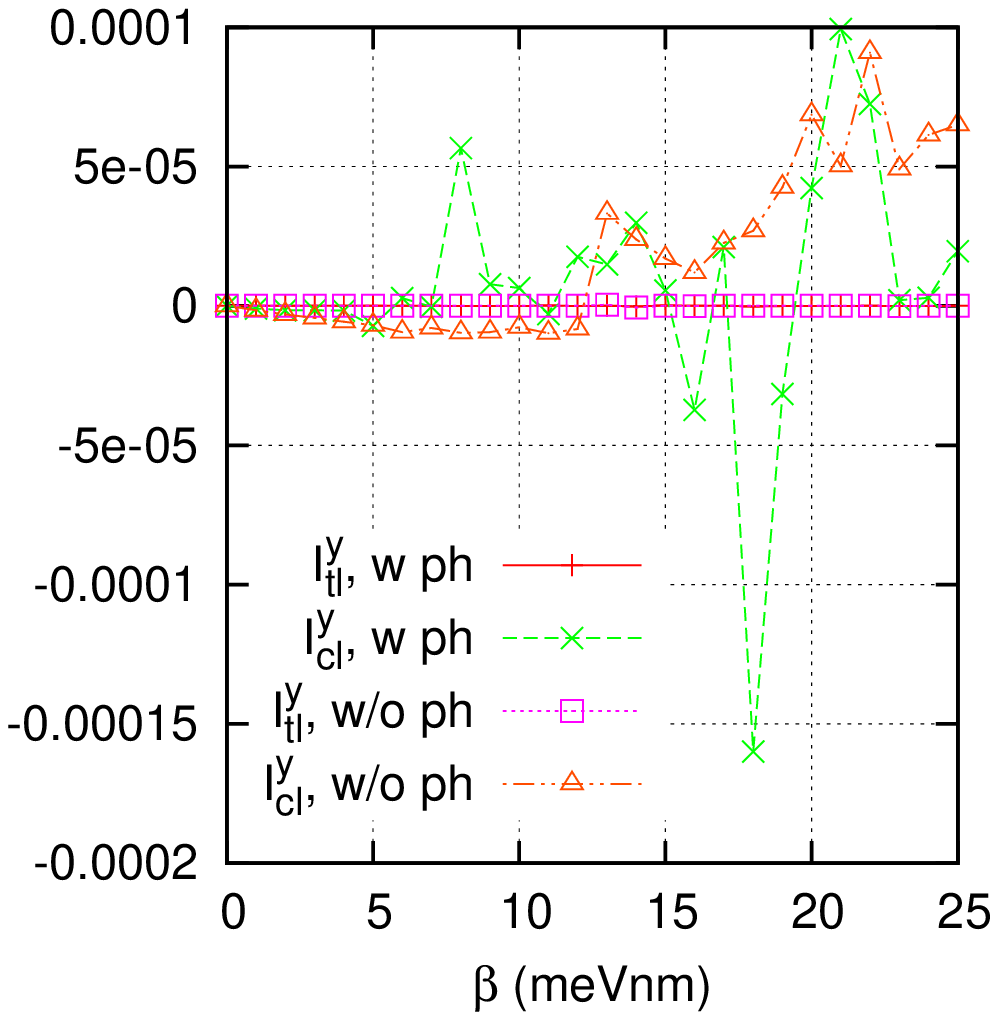}
       \label{current_arm_x_sy_beta_200}}
       \subfigure[]{
       \includegraphics[width=0.218\textwidth,angle=-90,bb=255 53 551 358]{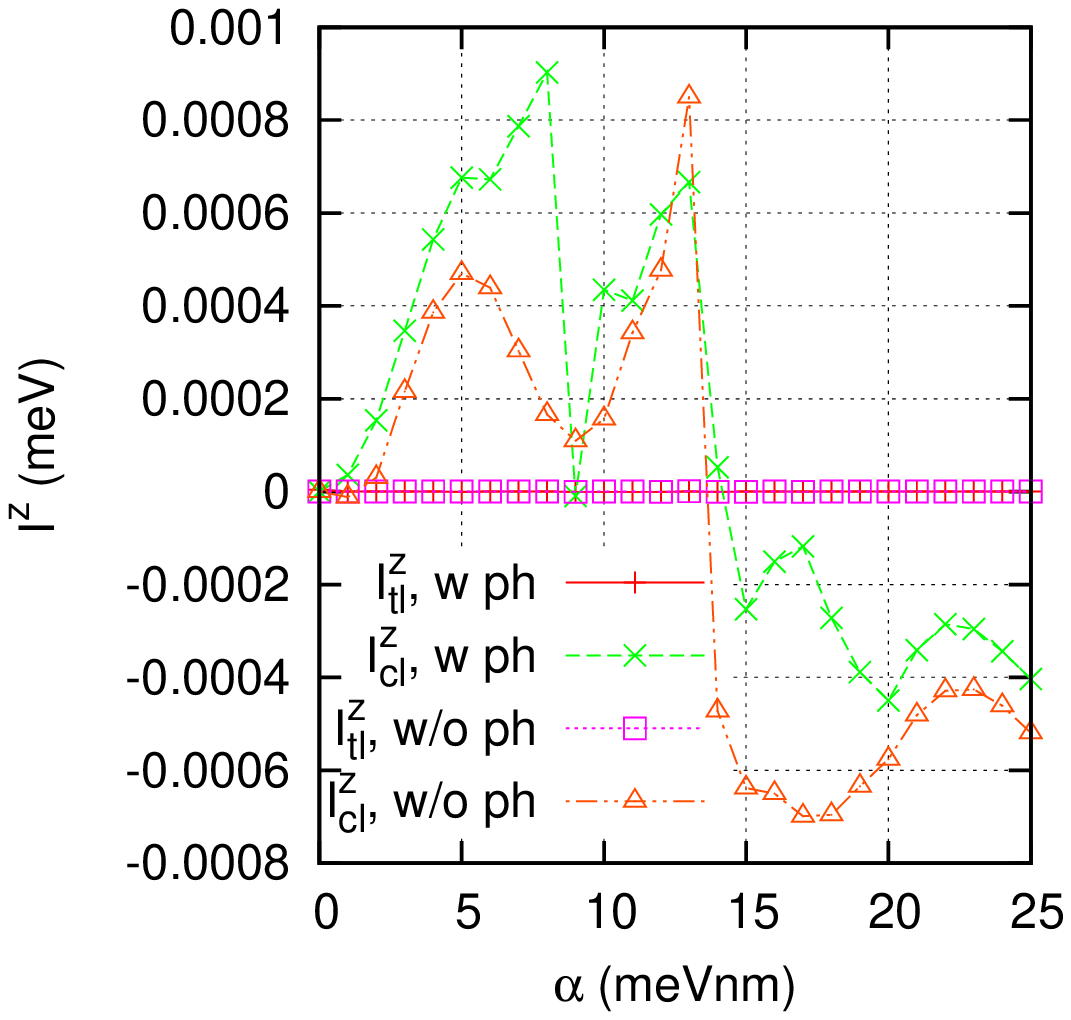} 
       \label{current_arm_x_sz_alpha_200}}
       \subfigure[]{
       \includegraphics[width=0.218\textwidth,angle=-90,bb=255 53 551 358]{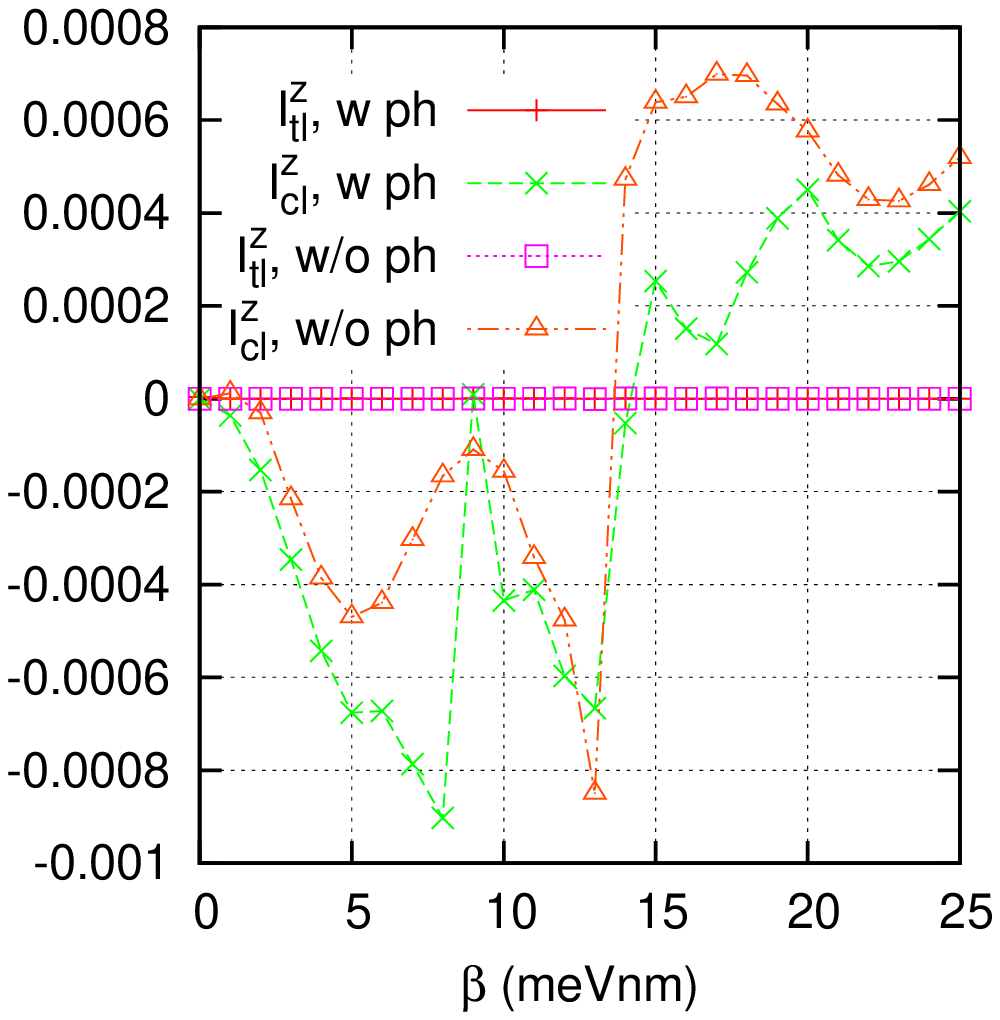}
       \label{current_arm_x_sz_beta_200}}
       \caption{(Color online) Total local and CL 2D spin current,
       $I_{\rm tl}^{i}$ and $I_{\rm cl}^{i}$, respectively, averaged over the time interval $[180,220]$~ps
       with (w) $x$-polarized photon field or without (w/o) photon cavity
       for \subref{current_arm_x_sx_alpha_200} $S_x$ spin polarization and Rashba interaction ($\beta=0$),
       \subref{current_arm_x_sx_beta_200} $S_x$ spin polarization and Dresselhaus interaction ($\alpha=0$),
       \subref{current_arm_x_sy_alpha_200} $S_y$ spin polarization and Rashba interaction ($\beta=0$),
       \subref{current_arm_x_sy_beta_200} $S_y$ spin polarization and Dresselhaus interaction ($\alpha=0$),
       \subref{current_arm_x_sz_alpha_200} $S_z$ spin polarization and Rashba interaction ($\beta=0$) and
       \subref{current_arm_x_sz_beta_200} $S_z$ spin polarization and Dresselhaus interaction ($\alpha=0$). Note that the scale for the ordinate may differ dramatically between the subfigures.}
       \label{current_arm_x_200_RD}
\end{figure}

 \begin{figure*}[htbq]
       \includegraphics[width=0.4\textwidth,angle=0,bb=89 59 252 219]{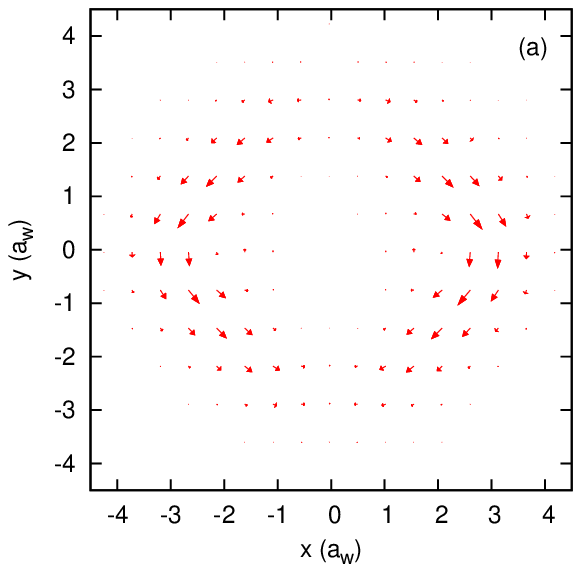}
       \includegraphics[width=0.4\textwidth,angle=0,bb=89 59 252 219]{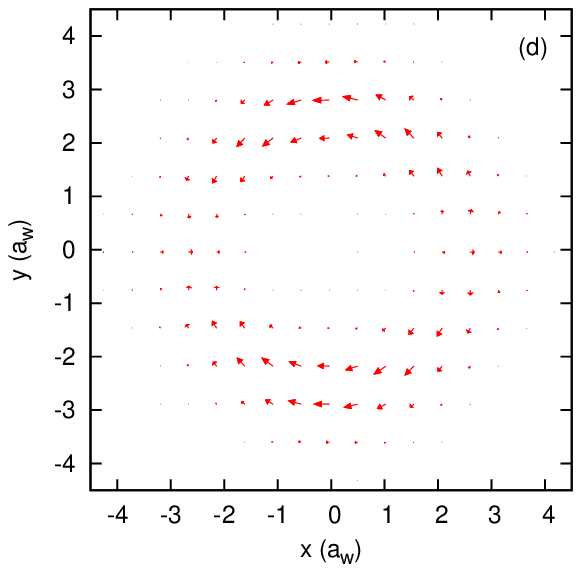}\\
       \includegraphics[width=0.4\textwidth,angle=0,bb=89 59 252 219]{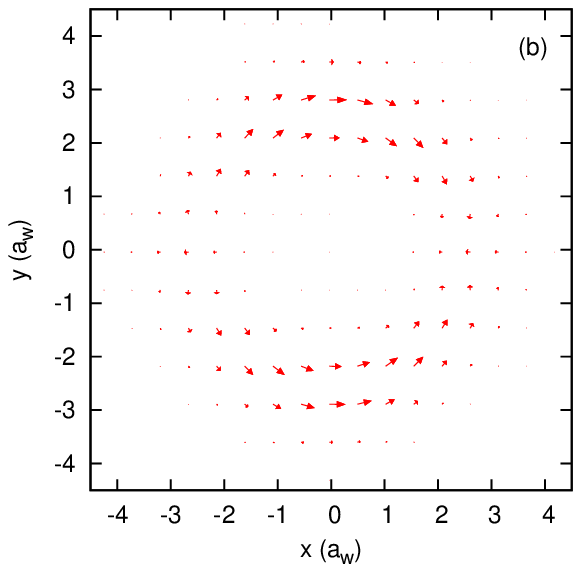}
       \includegraphics[width=0.4\textwidth,angle=0,bb=89 59 252 219]{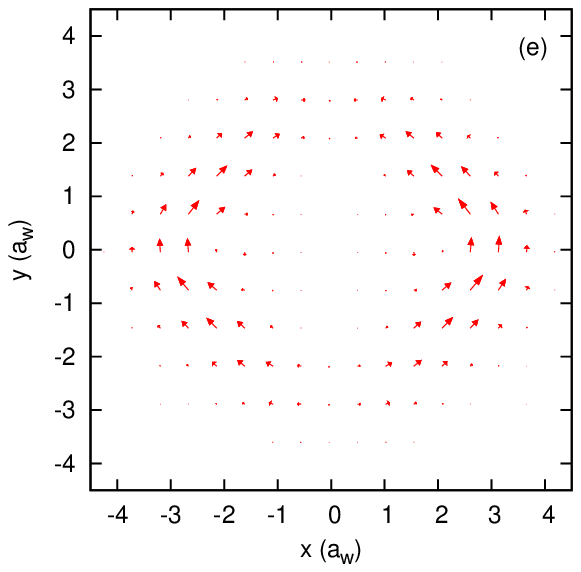}\\
       \includegraphics[width=0.4\textwidth,angle=0,bb=89 59 252 219]{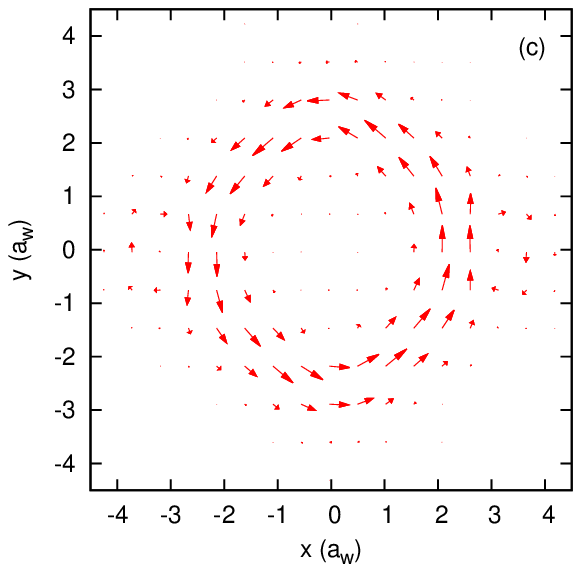}
       \includegraphics[width=0.4\textwidth,angle=0,bb=89 59 252 219]{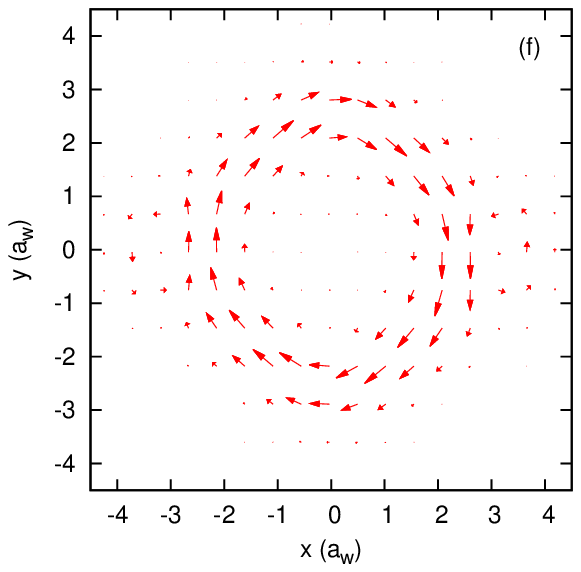}      
       \caption{(Color online) Spin current densities $\mathbf{j}^{i}(x,y)$, $i=x,y,z$
       at $t=200$~ps for
       (a)-(c) only Rashba spin-orbit interaction ($\alpha=5$~meV\,nm and $\beta=0$)
       or (d)-(e) only Dresselhaus spin-orbit interaction ($\alpha=0$ and $\beta=5$~meV\,nm)
       and for (a) and (d) $S_x$ spin polarization, (b) and (e) $S_y$ spin polarization or
       (e) and (f) $S_z$ spin polarization. The photon field is $x$-polarized.
       A spin current density vector of length $a_w$ corresponds to $1.25\times 10^{-3} \rm{meV}/a_w$.}
       \label{cur_den_alpha5_beta5_B000001x_200_lowres}
\end{figure*}

Here, we will compare the 2D results of the Rashba ring to the 2D results of the Dresselhaus ring.
The charge currents from \fig{current_arm_alpha_200} remain the same in the Dresselhaus case.
How the spin currents shown in Figs. \ref{current_arm_x_alpha_200_comp} and \ref{current_arm_0xy_s_alpha_200} look like in the Dresselhaus case 
can be deduced from \fig{current_arm_x_200_RD}, 
which shows the TL and CL 2D spin currents 
with and without $x$-polarized photon field. 
The left panels correspond to the situation of only Rashba spin-orbit interaction,
the right panels to the situation of only Dresselhaus spin-orbit interaction.
It becomes clear from these figures that the symmtries between the Rashba and Dresselhaus ring, \eq{RDrel},
apply to the non-equilibrium situation of a 2D ring of interacting electrons, which is connected to leads. This is because neither the Coulomb interaction nor the 2D ring potential depend on the spin. The leads include spin-orbit interaction and
the contact regions allow for tunneling of electron between same-spin states of the central system and leads. It is therefore, that the symmetries, \eq{RDrel}, are conserved. 
To support this, we note in passing that also the non-local spin currents, which describe the
spin transport from the left lead into the system or from the system to the right lead,
satisfy Eqs.\ (\ref{RDrel}).
However, if we would allow for spin-flip
coupling between system and lead states, Eqs.\ (\ref{RDrel}) might be violated.
Furthermore, the ring may be embedded in a photon cavity with linearly polarized photons without
breaking the symmetry relations \eq{RDrel} (the symmetries are conserved also for $y$-polarization,
not shown in \fig{current_arm_x_200_RD}).
It can be easily understood that the photon field
does not break this symmetry as the photonic part of the vector potential operator $\hat{A}^{\rm ph}(\mathbf{r})$
enters the Rashba Hamiltonian \eq{H_R} and Dresselhaus Hamiltonian \ref{H_D} in the same way
as the momentum operator.
Figure \ref{cur_den_alpha5_beta5_B000001x_200_lowres}  
shows the spin current densities $\mathbf{j}^{x}(x,y)$ (top panels),
$\mathbf{j}^{y}(x,y)$ (middle panels) and $\mathbf{j}^{z}(x,y)$ (bottom panels)
for the 2D Rashba (left panels) and 2D Dresselhaus ring (right panels) in comparison.
It confirms that the Eqs.\ (\ref{RDrel}) are valid at any location in the central system.
Finally, we note that the Zeeman term \eq{H_Z} breaks the symmetry relations.
The intricate effect from the magnetic field can be recognized for $B\geq 0.1$~T.

\vspace{0.7cm}

\section{Conclusions}

The transport of electrons can be controlled 
by various interference phenomena and geometric phases.
In this work, we turned our focus to the AC phase, which can be influenced
by the strength of the spin-orbit coupling, device geometry and cavity photons.
We have presented the charge and spin currents inside a quantum ring, 
in which the electrons' spin interacts with the orbital motion 
via the Rashba or Dresselhaus interaction. First, for a 1D ring,
we presented analytical results for the currents. 
For zero temperature and divisibility of the electron number by $4$, 
we predict a finite spin current of non-interacting electrons in the
limit of the electric field causing the Rashba effect approaching zero.
The current for the $S_z$ spin polarization 
is flowing homogeneously around the ring, but 
the currents for the other spin polarizations, flows from a local source
to a local sink.
Second, 
for a finite-width ring connected to leads, where the electrons are correlated
by Coulomb interaction 
we calculated numerically
the transient currents before the equilibrium
situation is reached using a time-convolutionless generalized master equation
formalism. We included spin-orbit coupling, but excluded Coulomb interaction
in the electrically biased leads. In addition, 
we allow the system electrons to interact with
a single cavity photon mode of $x$- or $y$-polarized photons.
The broad ring geometry together with the spin degree of freedom
required a substantial computational effort on state of the art machines.

A pronounced AC charge current dip can be recognized in the TL current
flowing from the higher-biased lead through the ring to the lower-biased lead
at the predicted position of the Rashba coefficient derived from the 1D model. 
The dip structure is linked to crossings in the ME spectrum and can be removed partly
by embedding the ring system in a photon cavity of preferably $y$-polarized photons.
The spin currents of the 1D and 2D rings agree qualitatively in their kind (TL or CL) and spin polarization ($S_x$, $S_y$ or $S_z$), the position of sign changes
with respect to the Rashba parameter and the geometric shape of the current flow distribution.
Quantatively, we can conclude that it is preferable to choose a narrow ring of weakly correlated electrons to obtain a strong spin current. 
The linearly polarized photon field interacting with the electrons suppresses in general
the charge current but enhances the spin current in the small Rashba coefficient regime.
Therefore, the linearly polarized photon field might be used to restore to some extent the
strong spin current for $S_z$ spin polarization in the small Rashba coefficient regime, which is suppressed 
for the broad ring with electron correlations and coupling to the leads.
The local spin current and spin photocurrent are subjected to stronger changes in time
than non-local quantities as the total charge in the system, 
emphasizing the non-equilibrium situation.
We established symmetry relations of the spin currents between the Rashba and Dresselhaus ring. 
We have shown that they remain valid for a finite-width ring of correlated electrons connected 
to electrically biased leads via a spin-conserving coupling tensor. 
Furthermore, switching on the cavity photon field does not destroy the symmetry relations. 
The sign of the spin current for $S_z$ spin polarization
could be used to distinguish the Rashba and Dresselhaus spin-orbit interactions
provided that they are sufficiently weak ($x<x^c:=x_R^c$) and care is taken that the induced magnetic
field of a charge current through the ring is not too large.
The conceived quantum ring system in a photon cavity 
with adjustable spin-orbit interaction and photon field polarization
could be used for future applications as
an elementary optoelectronic quantum device 
for quantum information processing.

\begin{acknowledgments}
      The authors acknowledge discussions with Tomas Orn Rosdahl.
      This work was financially supported by the Icelandic Research
      and Instruments Funds, the Research Fund of the University of Iceland, and the
      National Science Council of Taiwan under contract
      No.\ NSC100-2112-M-239-001-MY3. We acknowledge also support 
      from the computational facilities of the Nordic High Performance Computing (NHPC).
\end{acknowledgments}

\appendix
\section{source operators} \label{sourceop}
Here, we give the expressions for the spin source operators:

\begin{eqnarray}
 \hat{s}^{x}(\mathbf{r})&=&-\frac{\mu_B g_S B}{2} \left[\hat{\mathbf{\Psi}}^{\dagger}(\mathbf{r}) \sigma_{y}\hat{\mathbf{\Psi}}(\mathbf{r})\right]\nonumber \\
 &&-\frac{i\alpha}{2} \left[\frac{\partial}{\partial x}\left[\hat{\mathbf{\Psi}}^{\dagger}(\mathbf{r})\right] \sigma_{z}\hat{\mathbf{\Psi}}(\mathbf{r})
 -\hat{\mathbf{\Psi}}^{\dagger}(\mathbf{r})\sigma_z\frac{\partial}{\partial x}\hat{\mathbf{\Psi}}(\mathbf{r})\right]\nonumber \\
 &&-\frac{i\beta}{2} \left[\frac{\partial}{\partial y}\left[\hat{\mathbf{\Psi}}^{\dagger}(\mathbf{r})\right]\sigma_z\hat{\mathbf{\Psi}}(\mathbf{r})-
 \hat{\mathbf{\Psi}}^{\dagger}(\mathbf{r})\sigma_z\frac{\partial}{\partial y}\hat{\mathbf{\Psi}}(\mathbf{r})\right]\nonumber \\
 &&-\frac{e\alpha}{c\hbar}\hat{A}_x(\mathbf{r}) \hat{\mathbf{\Psi}}^{\dagger}(\mathbf{r})\sigma_z\hat{\mathbf{\Psi}}(\mathbf{r})\nonumber \\
 &&-\frac{e\beta}{c\hbar}\hat{A}_y(\mathbf{r}) \hat{\mathbf{\Psi}}^{\dagger}(\mathbf{r})\sigma_z\hat{\mathbf{\Psi}}(\mathbf{r}).
\end{eqnarray}

\begin{eqnarray}
 \hat{s}^{y}(\mathbf{r})&=&\frac{\mu_B g_S B}{2} \left[\hat{\mathbf{\Psi}}^{\dagger}(\mathbf{r}) \sigma_{x}\hat{\mathbf{\Psi}}(\mathbf{r})\right]\nonumber \\
 &&-\frac{i\beta}{2} \left[\frac{\partial}{\partial x}\left[\hat{\mathbf{\Psi}}^{\dagger}(\mathbf{r})\right] \sigma_{z}\hat{\mathbf{\Psi}}(\mathbf{r})
 -\hat{\mathbf{\Psi}}^{\dagger}(\mathbf{r})\sigma_z\frac{\partial}{\partial x}\hat{\mathbf{\Psi}}(\mathbf{r})\right]\nonumber \\
 &&-\frac{i\alpha}{2} \left[\frac{\partial}{\partial y}\left[\hat{\mathbf{\Psi}}^{\dagger}(\mathbf{r})\right]\sigma_z\hat{\mathbf{\Psi}}(\mathbf{r})-
 \hat{\mathbf{\Psi}}^{\dagger}(\mathbf{r})\sigma_z\frac{\partial}{\partial y}\hat{\mathbf{\Psi}}(\mathbf{r})\right]\nonumber \\
 &&-\frac{e\beta}{c\hbar}\hat{A}_x(\mathbf{r}) \hat{\mathbf{\Psi}}^{\dagger}(\mathbf{r})\sigma_z\hat{\mathbf{\Psi}}(\mathbf{r})\nonumber \\
 &&-\frac{e\alpha}{c\hbar}\hat{A}_y(\mathbf{r}) \hat{\mathbf{\Psi}}^{\dagger}(\mathbf{r})\sigma_z\hat{\mathbf{\Psi}}(\mathbf{r}).
\end{eqnarray}

\begin{eqnarray}
\hat{s}^{z}(\mathbf{r})&=&
\frac{1}{2} \left[\frac{\partial}{\partial x}\left[\hat{\mathbf{\Psi}}^{\dagger}(\mathbf{r})\right]\left(i\alpha\sigma_x+i\beta\sigma_y\right)\hat{\mathbf{\Psi}}(\mathbf{r})\nonumber \right. \\
&& -\left. \hat{\mathbf{\Psi}}^{\dagger}(\mathbf{r})\left(i\alpha\sigma_x+i\beta\sigma_y\right)\frac{\partial}{\partial x}\hat{\mathbf{\Psi}}(\mathbf{r}) \right]\nonumber \\
&&\frac{1}{2} \left[\frac{\partial}{\partial y}\left[\hat{\mathbf{\Psi}}^{\dagger}(\mathbf{r})\right]\left(i\beta\sigma_x+i\alpha\sigma_y\right)\hat{\mathbf{\Psi}}(\mathbf{r})\nonumber \right. \\
&& -\left. \hat{\mathbf{\Psi}}^{\dagger}(\mathbf{r})\left(i\beta\sigma_x+i\alpha\sigma_y\right)\frac{\partial}{\partial y}\hat{\mathbf{\Psi}}(\mathbf{r}) \right]\nonumber \\
&&+\frac{e}{ic\hbar}\hat{A}_x(\mathbf{r}) \hat{\mathbf{\Psi}}^{\dagger}(\mathbf{r})\left(i\alpha\sigma_x+i\beta\sigma_y\right)\hat{\mathbf{\Psi}}(\mathbf{r})\nonumber \\
&&+\frac{e}{ic\hbar}\hat{A}_y(\mathbf{r}) \hat{\mathbf{\Psi}}^{\dagger}(\mathbf{r})\left(i\beta\sigma_x+i\alpha\sigma_y\right)\hat{\mathbf{\Psi}}(\mathbf{r}).
\end{eqnarray}

\begin{widetext}
\section{derivation of Eq. (\ref{jxR1})} \label{xspincurder}
Here, we show only the derivation of Eq. (\ref{jxR1}) in detail. 
All other Rashba or Dresselhaus charge or spin densities, currents or source terms (Eqs.\ (\ref{ncR1}) to (\ref{syR1}) and, in the Dresselhaus case, \eq{ncD1}, \eq{Dspinden} and the corresponding expressions, which can be infered from \eq{RDrel} can be derived in analogy.
For a 1D ring geometry without magnetic and photon field, Eq. (\ref{jsxq}) can be simplified
and the $\varphi$-component of the current density along the ring is given in first quantization:
\begin{eqnarray}
j^{x}(\varphi)&=&\frac{\hbar^2}{4im^{*}a}\sum_{\nu=\pm 1}\sum_{n\in N_{\nu}}\left[\Psi_{\nu n}^{*}(\varphi,\uparrow)\frac{\partial}{\partial \varphi} \Psi_{\nu n}(\varphi,\downarrow) - \Psi_{\nu n}(\varphi,\downarrow) \frac{\partial}{\partial \varphi} \Psi_{\nu n}^{*}(\varphi,\uparrow)+
\Psi_{\nu n}^{*}(\varphi,\downarrow)\frac{\partial}{\partial \varphi} \Psi_{\nu n}(\varphi,\uparrow) \nonumber \right. \\ 
&&-\left. \Psi_{\nu n}(\varphi,\uparrow) \frac{\partial}{\partial \varphi} \Psi_{\nu n}^{*}(\varphi,\downarrow)\right]+ 
\frac{\alpha \cos(\varphi)}{2}\sum_{\nu=\pm 1}\sum_{n \in N_{\nu}}\left[\Psi_{\nu n}^{*}(\varphi,\uparrow)\Psi_{\nu n}(\varphi,\uparrow)+\Psi_{\nu n}^{*}(\varphi,\downarrow)\Psi_{\nu n}(\varphi,\downarrow)\right] \label{jxfq}
\end{eqnarray}
Now, we introduce the eigenfunctions, Eq. (\ref{Rwfc}), into Eq. (\ref{jxfq}) making already use of the fact that the coefficients $A_{\nu,\sigma}^{R}$ from Eq. (\ref{Rcoef}) are real:
\begin{eqnarray}
j^{x}_{R}(\varphi)&=&\frac{\hbar^2}{8\pi m^{*}a^2}\left[A_{1,1}^{R} A_{1,2}^{R} \exp(i\varphi)\sum_{n \in N_{-}}(n+1)+A_{2,1}^{R} A_{2,2}^{R}\exp(i\varphi)\sum_{n \in N_{+}}(n+1)+
A_{1,1}^{R} A_{1,2}^{R} \exp(i\varphi)\sum_{n \in N_{-}}n\nonumber \right. \\ 
&&+\left. A_{2,1}^{R} A_{2,2}^{R}\exp(i\varphi)\sum_{n \in N_{+}}n+
A_{1,1}^{R} A_{1,2}^{R} \exp(-i\varphi)\sum_{n \in N_{-}}n+A_{2,1}^{R} A_{2,2}^{R}\exp(-i\varphi)\sum_{n \in N_{+}}n+\nonumber \right. \\ 
&&+ \left.A_{1,1}^{R} A_{1,2}^{R} \exp(-i\varphi)\sum_{n \in N_{-}}(n+1)+A_{2,1}^{R} A_{2,2}^{R}\exp(-i\varphi)\sum_{n \in N_{+}}(n_2+1)\right] \nonumber \\
&&+\frac{\alpha \cos(\varphi)\left[(A_{1,1}^{R})^2|N_{-}|+(A_{2,1}^{R})^2|N_{+}|+(A_{1,2}^{R})^2|N_{-}|+(A_{2,2}^{R})^2|N_{+}|\right]}{4\pi a}.
\end{eqnarray}
This can be further simplified and the coefficients from Eq. (\ref{Rcoef}) introduced 
yielding for even electron number $N_e$:
\begin{equation}
j^{x}_{R}(\varphi)=\frac{\hbar j_{\Phi}\cos(\varphi)\cos(\frac{\theta_{R}}{2})\sin(\frac{\theta_{R}}{2})}{N_e}\left[\sum_{n \in N_{-}}(n+1)-\sum_{n \in N_{+}}(n+1)+\sum_{n \in N_{-}}n-\sum_{n \in N_{+}}n\right]+
\frac{\alpha \cos(\varphi)N_e}{4\pi a}. \label{jxR0}
\end{equation}
With the aid of Eq. (\ref{Rtantheta}), a relation
\begin{equation}
\cos\left(\frac{\theta_{R}}{2}\right)\sin\left(\frac{\theta_{R}}{2}\right)=\frac{x_R-x_R\sqrt{1+x_R^2}}{2+2x_R^2-2\sqrt{1+x_R^2}}
\end{equation}
can be established and introduced in Eq. (\ref{jxR0}) together
with the definition of the Rashba parameter, Eq. (\ref{defxR}), to get Eq. (\ref{jxR1}).
\end{widetext}

\section{dresselhaus eigenvalues and coefficient matrix} \label{Deigenvalder}

The Hamiltonian \eq{DresselhausH} has the 1D ring (strong confinement) limit, 
\begin{eqnarray}
\hat{H}^{1D}&=&-\hbar\Omega \frac{\partial^2}{\partial \varphi^2}
+i\hbar\omega_{D}(\cos(\varphi)\hat{\sigma}_y+\sin(\varphi)\hat{\sigma}_x)\frac{\partial}{\partial \varphi}\nonumber \\
&+&i\frac{\hbar\omega_{D}}{2}(\cos(\varphi)\hat{\sigma}_x-\sin(\varphi)\hat{\sigma}_y). \label{corDre1D}
\end{eqnarray} 
which can be reformulated
\begin{equation}
\hat{H}^{\rm 1D}(\varphi)=\hbar\Omega \left[\left(-i\frac{\partial}{\partial \varphi}-\frac{x_{D}}{2}\hat{R}(\varphi)\right)^2-\frac{x_{D}^{2}}{4}\right] \label{corDre1Dref}
\end{equation}
with
\begin{equation}
 R(\varphi)=\begin{pmatrix} 0 & \exp(i(\varphi-\frac{\pi}{2})) \\
\exp(-i(\varphi-\frac{\pi}{2})) & 0 \end{pmatrix}.
\end{equation}
Using the ansatz
\begin{eqnarray}
 \Psi_{\nu n}^{D}(\varphi)&=&\begin{pmatrix} \Psi_{\nu n}^{D}(\varphi,\uparrow)\\ \Psi_{\nu n}^{D}(\varphi,\downarrow)\end{pmatrix} \nonumber \\
 &=&\frac{\exp(in\varphi)}{\sqrt{2\pi a}}\begin{pmatrix} A_{\nu,1}^{D}\exp(i\varphi) \\ A_{\nu,2}^{D} \end{pmatrix} \label{Dwfc}
\end{eqnarray}
leads to the eigenvalue problem 
\begin{equation}
\begin{pmatrix} 1 & i\frac{x_D}{2} \\  -i\frac{x_D}{2} & 0  \end{pmatrix}\begin{pmatrix} A^{D}_{\nu,1} \\A^{D}_{\nu,2} \end{pmatrix} = \left(\Lambda_{\nu n}-n \right)\begin{pmatrix} A^{D}_{\nu,1} \\A^{D}_{\nu,2} \end{pmatrix},
\end{equation}
where $\Lambda_{\nu n}$ is related to the
Dresselhaus eigenvalues $E_{\nu n}^{D}$ of Hamiltonian Eq.\ (\ref{corDre1Dref}) by
\begin{equation}
 E_{\nu n}^{D}=\hbar\Omega\left(\Lambda_{\nu n}^2-\frac{x_D^2}{4}\right).
\end{equation}
The resulting Dresselhaus eigenvalues are identical with the Rashba eigenvalues (Eq. (\ref{spectrum})).
The complex Dresselhaus coefficient matrix is given by
\begin{equation}
 A^{D}=\begin{pmatrix} A_{\nu,1}^{D} & A_{\nu,2}^{D} \end{pmatrix}=\begin{pmatrix} -i\cos\left(\frac{\theta_{D}}{2}\right) & \sin\left(\frac{\theta_{D}}{2}\right) \\
 -i\sin\left(\frac{\theta_{D}}{2}\right) & -\cos\left(\frac{\theta_{D}}{2}\right)
 \end{pmatrix} \label{Dcoef}
\end{equation}
and
\begin{equation}
 \tan\left(\frac{\theta_{D}}{2}\right)=\frac{1+\sqrt{1+x_D^2}}{x_D}. \label{Dtantheta}
\end{equation}

%
%
%
\bibliographystyle{apsrev4-1}

\begin{thebibliography}{66}%
\makeatletter
\providecommand \@ifxundefined [1]{%
 \@ifx{#1\undefined}
}%
\providecommand \@ifnum [1]{%
 \ifnum #1\expandafter \@firstoftwo
 \else \expandafter \@secondoftwo
 \fi
}%
\providecommand \@ifx [1]{%
 \ifx #1\expandafter \@firstoftwo
 \else \expandafter \@secondoftwo
 \fi
}%
\providecommand \natexlab [1]{#1}%
\providecommand \enquote  [1]{``#1''}%
\providecommand \bibnamefont  [1]{#1}%
\providecommand \bibfnamefont [1]{#1}%
\providecommand \citenamefont [1]{#1}%
\providecommand \href@noop [0]{\@secondoftwo}%
\providecommand \href [0]{\begingroup \@sanitize@url \@href}%
\providecommand \@href[1]{\@@startlink{#1}\@@href}%
\providecommand \@@href[1]{\endgroup#1\@@endlink}%
\providecommand \@sanitize@url [0]{\catcode `\\12\catcode `\$12\catcode
  `\&12\catcode `\#12\catcode `\^12\catcode `\_12\catcode `\%12\relax}%
\providecommand \@@startlink[1]{}%
\providecommand \@@endlink[0]{}%
\providecommand \url  [0]{\begingroup\@sanitize@url \@url }%
\providecommand \@url [1]{\endgroup\@href {#1}{\urlprefix }}%
\providecommand \urlprefix  [0]{URL }%
\providecommand \Eprint [0]{\href }%
\providecommand \doibase [0]{http://dx.doi.org/}%
\providecommand \selectlanguage [0]{\@gobble}%
\providecommand \bibinfo  [0]{\@secondoftwo}%
\providecommand \bibfield  [0]{\@secondoftwo}%
\providecommand \translation [1]{[#1]}%
\providecommand \BibitemOpen [0]{}%
\providecommand \bibitemStop [0]{}%
\providecommand \bibitemNoStop [0]{.\EOS\space}%
\providecommand \EOS [0]{\spacefactor3000\relax}%
\providecommand \BibitemShut  [1]{\csname bibitem#1\endcsname}%
\let\auto@bib@innerbib\@empty
\bibitem [{\citenamefont {Aharonov}\ and\ \citenamefont
  {Bohm}(1959)}]{PhysRev.115.485}%
  \BibitemOpen
  \bibfield  {author} {\bibinfo {author} {\bibfnamefont {Y.}~\bibnamefont
  {Aharonov}}\ and\ \bibinfo {author} {\bibfnamefont {D.}~\bibnamefont
  {Bohm}},\ }\href {\doibase 10.1103/PhysRev.115.485} {\bibfield  {journal}
  {\bibinfo  {journal} {Phys. Rev.}\ }\textbf {\bibinfo {volume} {115}},\
  \bibinfo {pages} {485} (\bibinfo {year} {1959})}\BibitemShut {NoStop}%
\bibitem [{\citenamefont {Aharonov}\ and\ \citenamefont
  {Casher}(1984)}]{PhysRevLett.53.319}%
  \BibitemOpen
  \bibfield  {author} {\bibinfo {author} {\bibfnamefont {Y.}~\bibnamefont
  {Aharonov}}\ and\ \bibinfo {author} {\bibfnamefont {A.}~\bibnamefont
  {Casher}},\ }\href {\doibase 10.1103/PhysRevLett.53.319} {\bibfield
  {journal} {\bibinfo  {journal} {Phys. Rev. Lett.}\ }\textbf {\bibinfo
  {volume} {53}},\ \bibinfo {pages} {319} (\bibinfo {year} {1984})}\BibitemShut
  {NoStop}%
\bibitem [{\citenamefont {Aharonov}\ and\ \citenamefont
  {Anandan}(1987)}]{PhysRevLett.58.1593}%
  \BibitemOpen
  \bibfield  {author} {\bibinfo {author} {\bibfnamefont {Y.}~\bibnamefont
  {Aharonov}}\ and\ \bibinfo {author} {\bibfnamefont {J.}~\bibnamefont
  {Anandan}},\ }\href {\doibase 10.1103/PhysRevLett.58.1593} {\bibfield
  {journal} {\bibinfo  {journal} {Phys. Rev. Lett.}\ }\textbf {\bibinfo
  {volume} {58}},\ \bibinfo {pages} {1593} (\bibinfo {year}
  {1987})}\BibitemShut {NoStop}%
\bibitem [{\citenamefont {Berry}(1984)}]{Berry_phase}%
  \BibitemOpen
  \bibfield  {author} {\bibinfo {author} {\bibfnamefont {M.~V.}\ \bibnamefont
  {Berry}},\ }\href@noop {} {\bibfield  {journal} {\bibinfo  {journal} {Proc.
  R. Soc. Lond. A}\ }\textbf {\bibinfo {volume} {392}},\ \bibinfo {pages} {45}
  (\bibinfo {year} {1984})}\BibitemShut {NoStop}%
\bibitem [{\citenamefont {Szafran}\ and\ \citenamefont
  {Peeters}(2005)}]{Szafran05:165301}%
  \BibitemOpen
  \bibfield  {author} {\bibinfo {author} {\bibfnamefont {B.}~\bibnamefont
  {Szafran}}\ and\ \bibinfo {author} {\bibfnamefont {F.~M.}\ \bibnamefont
  {Peeters}},\ }\href {\doibase 10.1103/PhysRevB.72.165301} {\bibfield
  {journal} {\bibinfo  {journal} {Phys. Rev. B}\ }\textbf {\bibinfo {volume}
  {72}},\ \bibinfo {pages} {165301} (\bibinfo {year} {2005})}\BibitemShut
  {NoStop}%
\bibitem [{\citenamefont {Buchholz}\ \emph {et~al.}(2010)\citenamefont
  {Buchholz}, \citenamefont {Fischer}, \citenamefont {Kunze}, \citenamefont
  {Bell}, \citenamefont {Reuter},\ and\ \citenamefont {Wieck}}]{Buchholz2010}%
  \BibitemOpen
  \bibfield  {author} {\bibinfo {author} {\bibfnamefont {S.~S.}\ \bibnamefont
  {Buchholz}}, \bibinfo {author} {\bibfnamefont {S.~F.}\ \bibnamefont
  {Fischer}}, \bibinfo {author} {\bibfnamefont {U.}~\bibnamefont {Kunze}},
  \bibinfo {author} {\bibfnamefont {M.}~\bibnamefont {Bell}}, \bibinfo {author}
  {\bibfnamefont {D.}~\bibnamefont {Reuter}}, \ and\ \bibinfo {author}
  {\bibfnamefont {A.~D.}\ \bibnamefont {Wieck}},\ }\href@noop {} {\bibfield
  {journal} {\bibinfo  {journal} {Phys. Rev. B}\ }\textbf {\bibinfo {volume}
  {82}},\ \bibinfo {pages} {045432} (\bibinfo {year} {2010})}\BibitemShut
  {NoStop}%
\bibitem [{\citenamefont {B\"uttiker}\ \emph {et~al.}(1984)\citenamefont
  {B\"uttiker}, \citenamefont {Imry},\ and\ \citenamefont
  {Azbel}}]{Buttiker30.1982}%
  \BibitemOpen
  \bibfield  {author} {\bibinfo {author} {\bibfnamefont {M.}~\bibnamefont
  {B\"uttiker}}, \bibinfo {author} {\bibfnamefont {Y.}~\bibnamefont {Imry}}, \
  and\ \bibinfo {author} {\bibfnamefont {M.~Y.}\ \bibnamefont {Azbel}},\ }\href
  {\doibase 10.1103/PhysRevA.30.1982} {\bibfield  {journal} {\bibinfo
  {journal} {Phys. Rev. A}\ }\textbf {\bibinfo {volume} {30}},\ \bibinfo
  {pages} {1982} (\bibinfo {year} {1984})}\BibitemShut {NoStop}%
\bibitem [{\citenamefont {Pichugin}\ and\ \citenamefont
  {Sadreev}(1997)}]{Pichugin56:9662}%
  \BibitemOpen
  \bibfield  {author} {\bibinfo {author} {\bibfnamefont {K.~N.}\ \bibnamefont
  {Pichugin}}\ and\ \bibinfo {author} {\bibfnamefont {A.~F.}\ \bibnamefont
  {Sadreev}},\ }\href@noop {} {\bibfield  {journal} {\bibinfo  {journal} {Phys
  Rev. B}\ }\textbf {\bibinfo {volume} {56}},\ \bibinfo {pages} {9662}
  (\bibinfo {year} {1997})}\BibitemShut {NoStop}%
\bibitem [{\citenamefont {Arnold}\ \emph {et~al.}(2013)\citenamefont {Arnold},
  \citenamefont {Tang}, \citenamefont {Manolescu},\ and\ \citenamefont
  {Gudmundsson}}]{PhysRevB.87.035314}%
  \BibitemOpen
  \bibfield  {author} {\bibinfo {author} {\bibfnamefont {T.}~\bibnamefont
  {Arnold}}, \bibinfo {author} {\bibfnamefont {C.-S.}\ \bibnamefont {Tang}},
  \bibinfo {author} {\bibfnamefont {A.}~\bibnamefont {Manolescu}}, \ and\
  \bibinfo {author} {\bibfnamefont {V.}~\bibnamefont {Gudmundsson}},\ }\href
  {\doibase 10.1103/PhysRevB.87.035314} {\bibfield  {journal} {\bibinfo
  {journal} {Phys. Rev. B}\ }\textbf {\bibinfo {volume} {87}},\ \bibinfo
  {pages} {035314} (\bibinfo {year} {2013})}\BibitemShut {NoStop}%
\bibitem [{\citenamefont {Cheung}\ \emph {et~al.}(1988)\citenamefont {Cheung},
  \citenamefont {Gefen}, \citenamefont {Riedel},\ and\ \citenamefont
  {Shih}}]{PhysRevB.37.6050}%
  \BibitemOpen
  \bibfield  {author} {\bibinfo {author} {\bibfnamefont {H.-F.}\ \bibnamefont
  {Cheung}}, \bibinfo {author} {\bibfnamefont {Y.}~\bibnamefont {Gefen}},
  \bibinfo {author} {\bibfnamefont {E.~K.}\ \bibnamefont {Riedel}}, \ and\
  \bibinfo {author} {\bibfnamefont {W.-H.}\ \bibnamefont {Shih}},\ }\href
  {\doibase 10.1103/PhysRevB.37.6050} {\bibfield  {journal} {\bibinfo
  {journal} {Phys. Rev. B}\ }\textbf {\bibinfo {volume} {37}},\ \bibinfo
  {pages} {6050} (\bibinfo {year} {1988})}\BibitemShut {NoStop}%
\bibitem [{\citenamefont {Tan}\ and\ \citenamefont
  {Inkson}(1999)}]{Tan99:5626}%
  \BibitemOpen
  \bibfield  {author} {\bibinfo {author} {\bibfnamefont {W.-C.}\ \bibnamefont
  {Tan}}\ and\ \bibinfo {author} {\bibfnamefont {J.~C.}\ \bibnamefont
  {Inkson}},\ }\href@noop {} {\bibfield  {journal} {\bibinfo  {journal} {Phys.
  Rev. B}\ }\textbf {\bibinfo {volume} {60}},\ \bibinfo {pages} {5626}
  (\bibinfo {year} {1999})}\BibitemShut {NoStop}%
\bibitem [{\citenamefont {Webb}\ \emph {et~al.}(1985)\citenamefont {Webb},
  \citenamefont {Washburn}, \citenamefont {Umbach},\ and\ \citenamefont
  {Laibowitz}}]{Webb.54.2696}%
  \BibitemOpen
  \bibfield  {author} {\bibinfo {author} {\bibfnamefont {R.~A.}\ \bibnamefont
  {Webb}}, \bibinfo {author} {\bibfnamefont {S.}~\bibnamefont {Washburn}},
  \bibinfo {author} {\bibfnamefont {C.~P.}\ \bibnamefont {Umbach}}, \ and\
  \bibinfo {author} {\bibfnamefont {R.~B.}\ \bibnamefont {Laibowitz}},\ }\href
  {\doibase 10.1103/PhysRevLett.54.2696} {\bibfield  {journal} {\bibinfo
  {journal} {Phys. Rev. Lett.}\ }\textbf {\bibinfo {volume} {54}},\ \bibinfo
  {pages} {2696} (\bibinfo {year} {1985})}\BibitemShut {NoStop}%
\bibitem [{\citenamefont {Oh}\ and\ \citenamefont
  {Ryu}(1995)}]{PhysRevB.51.13441}%
  \BibitemOpen
  \bibfield  {author} {\bibinfo {author} {\bibfnamefont {S.}~\bibnamefont
  {Oh}}\ and\ \bibinfo {author} {\bibfnamefont {C.-M.}\ \bibnamefont {Ryu}},\
  }\href {\doibase 10.1103/PhysRevB.51.13441} {\bibfield  {journal} {\bibinfo
  {journal} {Phys. Rev. B}\ }\textbf {\bibinfo {volume} {51}},\ \bibinfo
  {pages} {13441} (\bibinfo {year} {1995})}\BibitemShut {NoStop}%
\bibitem [{\citenamefont {Bychkov}\ and\ \citenamefont
  {Rashba}(1984)}]{0022-3719-17-33-015}%
  \BibitemOpen
  \bibfield  {author} {\bibinfo {author} {\bibfnamefont {Y.~A.}\ \bibnamefont
  {Bychkov}}\ and\ \bibinfo {author} {\bibfnamefont {E.~I.}\ \bibnamefont
  {Rashba}},\ }\href {http://stacks.iop.org/0022-3719/17/i=33/a=015} {\bibfield
   {journal} {\bibinfo  {journal} {Journal of Physics C: Solid State Physics}\
  }\textbf {\bibinfo {volume} {17}},\ \bibinfo {pages} {6039} (\bibinfo {year}
  {1984})}\BibitemShut {NoStop}%
\bibitem [{\citenamefont {Dresselhaus}(1955)}]{PhysRev.100.580}%
  \BibitemOpen
  \bibfield  {author} {\bibinfo {author} {\bibfnamefont {G.}~\bibnamefont
  {Dresselhaus}},\ }\href {\doibase 10.1103/PhysRev.100.580} {\bibfield
  {journal} {\bibinfo  {journal} {Phys. Rev.}\ }\textbf {\bibinfo {volume}
  {100}},\ \bibinfo {pages} {580} (\bibinfo {year} {1955})}\BibitemShut
  {NoStop}%
\bibitem [{\citenamefont {Loss}\ \emph {et~al.}(1990)\citenamefont {Loss},
  \citenamefont {Goldbart},\ and\ \citenamefont
  {Balatsky}}]{PhysRevLett.65.1655}%
  \BibitemOpen
  \bibfield  {author} {\bibinfo {author} {\bibfnamefont {D.}~\bibnamefont
  {Loss}}, \bibinfo {author} {\bibfnamefont {P.}~\bibnamefont {Goldbart}}, \
  and\ \bibinfo {author} {\bibfnamefont {A.~V.}\ \bibnamefont {Balatsky}},\
  }\href {\doibase 10.1103/PhysRevLett.65.1655} {\bibfield  {journal} {\bibinfo
   {journal} {Phys. Rev. Lett.}\ }\textbf {\bibinfo {volume} {65}},\ \bibinfo
  {pages} {1655} (\bibinfo {year} {1990})}\BibitemShut {NoStop}%
\bibitem [{\citenamefont {Balatsky}\ and\ \citenamefont
  {Altshuler}(1993)}]{PhysRevLett.70.1678}%
  \BibitemOpen
  \bibfield  {author} {\bibinfo {author} {\bibfnamefont {A.~V.}\ \bibnamefont
  {Balatsky}}\ and\ \bibinfo {author} {\bibfnamefont {B.~L.}\ \bibnamefont
  {Altshuler}},\ }\href {\doibase 10.1103/PhysRevLett.70.1678} {\bibfield
  {journal} {\bibinfo  {journal} {Phys. Rev. Lett.}\ }\textbf {\bibinfo
  {volume} {70}},\ \bibinfo {pages} {1678} (\bibinfo {year}
  {1993})}\BibitemShut {NoStop}%
\bibitem [{\citenamefont {Kovalev}\ \emph {et~al.}(2007)\citenamefont
  {Kovalev}, \citenamefont {Borunda}, \citenamefont {Jungwirth}, \citenamefont
  {Molenkamp},\ and\ \citenamefont {Sinova}}]{PhysRevB.76.125307}%
  \BibitemOpen
  \bibfield  {author} {\bibinfo {author} {\bibfnamefont {A.~A.}\ \bibnamefont
  {Kovalev}}, \bibinfo {author} {\bibfnamefont {M.~F.}\ \bibnamefont
  {Borunda}}, \bibinfo {author} {\bibfnamefont {T.}~\bibnamefont {Jungwirth}},
  \bibinfo {author} {\bibfnamefont {L.~W.}\ \bibnamefont {Molenkamp}}, \ and\
  \bibinfo {author} {\bibfnamefont {J.}~\bibnamefont {Sinova}},\ }\href
  {\doibase 10.1103/PhysRevB.76.125307} {\bibfield  {journal} {\bibinfo
  {journal} {Phys. Rev. B}\ }\textbf {\bibinfo {volume} {76}},\ \bibinfo
  {pages} {125307} (\bibinfo {year} {2007})}\BibitemShut {NoStop}%
\bibitem [{\citenamefont {Maiti}\ \emph {et~al.}(2011)\citenamefont {Maiti},
  \citenamefont {Dey}, \citenamefont {Sil}, \citenamefont {Chakrabarti},\ and\
  \citenamefont {Karmakar}}]{0295-5075-95-5-57008}%
  \BibitemOpen
  \bibfield  {author} {\bibinfo {author} {\bibfnamefont {S.~K.}\ \bibnamefont
  {Maiti}}, \bibinfo {author} {\bibfnamefont {M.}~\bibnamefont {Dey}}, \bibinfo
  {author} {\bibfnamefont {S.}~\bibnamefont {Sil}}, \bibinfo {author}
  {\bibfnamefont {A.}~\bibnamefont {Chakrabarti}}, \ and\ \bibinfo {author}
  {\bibfnamefont {S.~N.}\ \bibnamefont {Karmakar}},\ }\href
  {http://stacks.iop.org/0295-5075/95/i=5/a=57008} {\bibfield  {journal}
  {\bibinfo  {journal} {EPL (Europhysics Letters)}\ }\textbf {\bibinfo {volume}
  {95}},\ \bibinfo {pages} {57008} (\bibinfo {year} {2011})}\BibitemShut
  {NoStop}%
\bibitem [{\citenamefont {Nita}\ \emph {et~al.}(2011)\citenamefont {Nita},
  \citenamefont {Marinescu}, \citenamefont {Manolescu},\ and\ \citenamefont
  {Gudmundsson}}]{PhysRevB.83.155427}%
  \BibitemOpen
  \bibfield  {author} {\bibinfo {author} {\bibfnamefont {M.}~\bibnamefont
  {Nita}}, \bibinfo {author} {\bibfnamefont {D.~C.}\ \bibnamefont {Marinescu}},
  \bibinfo {author} {\bibfnamefont {A.}~\bibnamefont {Manolescu}}, \ and\
  \bibinfo {author} {\bibfnamefont {V.}~\bibnamefont {Gudmundsson}},\ }\href
  {\doibase 10.1103/PhysRevB.83.155427} {\bibfield  {journal} {\bibinfo
  {journal} {Phys. Rev. B}\ }\textbf {\bibinfo {volume} {83}},\ \bibinfo
  {pages} {155427} (\bibinfo {year} {2011})}\BibitemShut {NoStop}%
\bibitem [{\citenamefont {Sonin}(2007{\natexlab{a}})}]{PhysRevLett.99.266602}%
  \BibitemOpen
  \bibfield  {author} {\bibinfo {author} {\bibfnamefont {E.~B.}\ \bibnamefont
  {Sonin}},\ }\href {\doibase 10.1103/PhysRevLett.99.266602} {\bibfield
  {journal} {\bibinfo  {journal} {Phys. Rev. Lett.}\ }\textbf {\bibinfo
  {volume} {99}},\ \bibinfo {pages} {266602} (\bibinfo {year}
  {2007}{\natexlab{a}})}\BibitemShut {NoStop}%
\bibitem [{\citenamefont {Sun}\ \emph {et~al.}(2008)\citenamefont {Sun},
  \citenamefont {Xie},\ and\ \citenamefont {Wang}}]{PhysRevB.77.035327}%
  \BibitemOpen
  \bibfield  {author} {\bibinfo {author} {\bibfnamefont {Q.-f.}\ \bibnamefont
  {Sun}}, \bibinfo {author} {\bibfnamefont {X.~C.}\ \bibnamefont {Xie}}, \ and\
  \bibinfo {author} {\bibfnamefont {J.}~\bibnamefont {Wang}},\ }\href {\doibase
  10.1103/PhysRevB.77.035327} {\bibfield  {journal} {\bibinfo  {journal} {Phys.
  Rev. B}\ }\textbf {\bibinfo {volume} {77}},\ \bibinfo {pages} {035327}
  (\bibinfo {year} {2008})}\BibitemShut {NoStop}%
\bibitem [{\citenamefont {Sheng}\ and\ \citenamefont
  {Chang}(2006)}]{PhysRevB.74.235315}%
  \BibitemOpen
  \bibfield  {author} {\bibinfo {author} {\bibfnamefont {J.~S.}\ \bibnamefont
  {Sheng}}\ and\ \bibinfo {author} {\bibfnamefont {K.}~\bibnamefont {Chang}},\
  }\href {\doibase 10.1103/PhysRevB.74.235315} {\bibfield  {journal} {\bibinfo
  {journal} {Phys. Rev. B}\ }\textbf {\bibinfo {volume} {74}},\ \bibinfo
  {pages} {235315} (\bibinfo {year} {2006})}\BibitemShut {NoStop}%
\bibitem [{\citenamefont {Matos-Abiague}\ and\ \citenamefont
  {Berakdar}(2005)}]{PhysRevLett.94.166801}%
  \BibitemOpen
  \bibfield  {author} {\bibinfo {author} {\bibfnamefont {A.}~\bibnamefont
  {Matos-Abiague}}\ and\ \bibinfo {author} {\bibfnamefont {J.}~\bibnamefont
  {Berakdar}},\ }\href {\doibase 10.1103/PhysRevLett.94.166801} {\bibfield
  {journal} {\bibinfo  {journal} {Phys. Rev. Lett.}\ }\textbf {\bibinfo
  {volume} {94}},\ \bibinfo {pages} {166801} (\bibinfo {year}
  {2005})}\BibitemShut {NoStop}%
\bibitem [{\citenamefont {Pershin}\ and\ \citenamefont
  {Piermarocchi}(2005)}]{PhysRevB.72.245331}%
  \BibitemOpen
  \bibfield  {author} {\bibinfo {author} {\bibfnamefont {Y.~V.}\ \bibnamefont
  {Pershin}}\ and\ \bibinfo {author} {\bibfnamefont {C.}~\bibnamefont
  {Piermarocchi}},\ }\href {\doibase 10.1103/PhysRevB.72.245331} {\bibfield
  {journal} {\bibinfo  {journal} {Phys. Rev. B}\ }\textbf {\bibinfo {volume}
  {72}},\ \bibinfo {pages} {245331} (\bibinfo {year} {2005})}\BibitemShut
  {NoStop}%
\bibitem [{\citenamefont {Kibis}(2011)}]{PhysRevLett.107.106802}%
  \BibitemOpen
  \bibfield  {author} {\bibinfo {author} {\bibfnamefont {O.~V.}\ \bibnamefont
  {Kibis}},\ }\href {\doibase 10.1103/PhysRevLett.107.106802} {\bibfield
  {journal} {\bibinfo  {journal} {Phys. Rev. Lett.}\ }\textbf {\bibinfo
  {volume} {107}},\ \bibinfo {pages} {106802} (\bibinfo {year}
  {2011})}\BibitemShut {NoStop}%
\bibitem [{\citenamefont {Kibis}\ \emph {et~al.}(2013)\citenamefont {Kibis},
  \citenamefont {Kyriienko},\ and\ \citenamefont
  {Shelykh}}]{PhysRevB.87.245437}%
  \BibitemOpen
  \bibfield  {author} {\bibinfo {author} {\bibfnamefont {O.~V.}\ \bibnamefont
  {Kibis}}, \bibinfo {author} {\bibfnamefont {O.}~\bibnamefont {Kyriienko}}, \
  and\ \bibinfo {author} {\bibfnamefont {I.~A.}\ \bibnamefont {Shelykh}},\
  }\href {\doibase 10.1103/PhysRevB.87.245437} {\bibfield  {journal} {\bibinfo
  {journal} {Phys. Rev. B}\ }\textbf {\bibinfo {volume} {87}},\ \bibinfo
  {pages} {245437} (\bibinfo {year} {2013})}\BibitemShut {NoStop}%
\bibitem [{\citenamefont {Zhu}\ and\ \citenamefont
  {Berakdar}(2008)}]{PhysRevB.77.235438}%
  \BibitemOpen
  \bibfield  {author} {\bibinfo {author} {\bibfnamefont {Z.-G.}\ \bibnamefont
  {Zhu}}\ and\ \bibinfo {author} {\bibfnamefont {J.}~\bibnamefont {Berakdar}},\
  }\href {\doibase 10.1103/PhysRevB.77.235438} {\bibfield  {journal} {\bibinfo
  {journal} {Phys. Rev. B}\ }\textbf {\bibinfo {volume} {77}},\ \bibinfo
  {pages} {235438} (\bibinfo {year} {2008})}\BibitemShut {NoStop}%
\bibitem [{\citenamefont {Jonasson}\ \emph {et~al.}(2012)\citenamefont
  {Jonasson}, \citenamefont {Tang}, \citenamefont {Goan}, \citenamefont
  {Manolescu},\ and\ \citenamefont {Gudmundsson}}]{1367-2630-14-1-013036}%
  \BibitemOpen
  \bibfield  {author} {\bibinfo {author} {\bibfnamefont {O.}~\bibnamefont
  {Jonasson}}, \bibinfo {author} {\bibfnamefont {C.-S.}\ \bibnamefont {Tang}},
  \bibinfo {author} {\bibfnamefont {H.-S.}\ \bibnamefont {Goan}}, \bibinfo
  {author} {\bibfnamefont {A.}~\bibnamefont {Manolescu}}, \ and\ \bibinfo
  {author} {\bibfnamefont {V.}~\bibnamefont {Gudmundsson}},\ }\href
  {http://stacks.iop.org/1367-2630/14/i=1/a=013036} {\bibfield  {journal}
  {\bibinfo  {journal} {New Journal of Physics}\ }\textbf {\bibinfo {volume}
  {14}},\ \bibinfo {pages} {013036} (\bibinfo {year} {2012})}\BibitemShut
  {NoStop}%
\bibitem [{\citenamefont {Jaynes}\ and\ \citenamefont
  {Cummings}(1963)}]{1443594}%
  \BibitemOpen
  \bibfield  {author} {\bibinfo {author} {\bibfnamefont {E.}~\bibnamefont
  {Jaynes}}\ and\ \bibinfo {author} {\bibfnamefont {F.~W.}\ \bibnamefont
  {Cummings}},\ }\href {\doibase 10.1109/PROC.1963.1664} {\bibfield  {journal}
  {\bibinfo  {journal} {Proceedings of the IEEE}\ }\textbf {\bibinfo {volume}
  {51}},\ \bibinfo {pages} {89} (\bibinfo {year} {1963})}\BibitemShut {NoStop}%
\bibitem [{\citenamefont {Wu}\ and\ \citenamefont
  {Yang}(2007)}]{PhysRevLett.98.013601}%
  \BibitemOpen
  \bibfield  {author} {\bibinfo {author} {\bibfnamefont {Y.}~\bibnamefont
  {Wu}}\ and\ \bibinfo {author} {\bibfnamefont {X.}~\bibnamefont {Yang}},\
  }\href {\doibase 10.1103/PhysRevLett.98.013601} {\bibfield  {journal}
  {\bibinfo  {journal} {Phys. Rev. Lett.}\ }\textbf {\bibinfo {volume} {98}},\
  \bibinfo {pages} {013601} (\bibinfo {year} {2007})}\BibitemShut {NoStop}%
\bibitem [{\citenamefont {Sornborger}\ \emph {et~al.}(2004)\citenamefont
  {Sornborger}, \citenamefont {Cleland},\ and\ \citenamefont
  {Geller}}]{Sornborger04:052315}%
  \BibitemOpen
  \bibfield  {author} {\bibinfo {author} {\bibfnamefont {A.~T.}\ \bibnamefont
  {Sornborger}}, \bibinfo {author} {\bibfnamefont {A.~N.}\ \bibnamefont
  {Cleland}}, \ and\ \bibinfo {author} {\bibfnamefont {M.~R.}\ \bibnamefont
  {Geller}},\ }\href {\doibase 10.1103/PhysRevA.70.052315} {\bibfield
  {journal} {\bibinfo  {journal} {Phys. Rev. A}\ }\textbf {\bibinfo {volume}
  {70}},\ \bibinfo {pages} {052315} (\bibinfo {year} {2004})}\BibitemShut
  {NoStop}%
\bibitem [{\citenamefont {Frustaglia}\ and\ \citenamefont
  {Richter}(2004)}]{PhysRevB.69.235310}%
  \BibitemOpen
  \bibfield  {author} {\bibinfo {author} {\bibfnamefont {D.}~\bibnamefont
  {Frustaglia}}\ and\ \bibinfo {author} {\bibfnamefont {K.}~\bibnamefont
  {Richter}},\ }\href {\doibase 10.1103/PhysRevB.69.235310} {\bibfield
  {journal} {\bibinfo  {journal} {Phys. Rev. B}\ }\textbf {\bibinfo {volume}
  {69}},\ \bibinfo {pages} {235310} (\bibinfo {year} {2004})}\BibitemShut
  {NoStop}%
\bibitem [{\citenamefont {Ying-Fang}\ \emph {et~al.}(2004)\citenamefont
  {Ying-Fang}, \citenamefont {Yong-Ping},\ and\ \citenamefont
  {Jiu-Qing}}]{0256-307X-21-11-006}%
  \BibitemOpen
  \bibfield  {author} {\bibinfo {author} {\bibfnamefont {G.}~\bibnamefont
  {Ying-Fang}}, \bibinfo {author} {\bibfnamefont {Z.}~\bibnamefont
  {Yong-Ping}}, \ and\ \bibinfo {author} {\bibfnamefont {L.}~\bibnamefont
  {Jiu-Qing}},\ }\href {http://stacks.iop.org/0256-307X/21/i=11/a=006}
  {\bibfield  {journal} {\bibinfo  {journal} {Chinese Physics Letters}\
  }\textbf {\bibinfo {volume} {21}},\ \bibinfo {pages} {2093} (\bibinfo {year}
  {2004})}\BibitemShut {NoStop}%
\bibitem [{\citenamefont {Aronov}\ and\ \citenamefont
  {Lyanda-Geller}(1993)}]{PhysRevLett.70.343}%
  \BibitemOpen
  \bibfield  {author} {\bibinfo {author} {\bibfnamefont {A.~G.}\ \bibnamefont
  {Aronov}}\ and\ \bibinfo {author} {\bibfnamefont {Y.~B.}\ \bibnamefont
  {Lyanda-Geller}},\ }\href {\doibase 10.1103/PhysRevLett.70.343} {\bibfield
  {journal} {\bibinfo  {journal} {Phys. Rev. Lett.}\ }\textbf {\bibinfo
  {volume} {70}},\ \bibinfo {pages} {343} (\bibinfo {year} {1993})}\BibitemShut
  {NoStop}%
\bibitem [{\citenamefont {Yi}\ \emph {et~al.}(1997)\citenamefont {Yi},
  \citenamefont {Qian},\ and\ \citenamefont {Su}}]{PhysRevB.55.10631}%
  \BibitemOpen
  \bibfield  {author} {\bibinfo {author} {\bibfnamefont {Y.-S.}\ \bibnamefont
  {Yi}}, \bibinfo {author} {\bibfnamefont {T.-Z.}\ \bibnamefont {Qian}}, \ and\
  \bibinfo {author} {\bibfnamefont {Z.-B.}\ \bibnamefont {Su}},\ }\href
  {\doibase 10.1103/PhysRevB.55.10631} {\bibfield  {journal} {\bibinfo
  {journal} {Phys. Rev. B}\ }\textbf {\bibinfo {volume} {55}},\ \bibinfo
  {pages} {10631} (\bibinfo {year} {1997})}\BibitemShut {NoStop}%
\bibitem [{\citenamefont {Wang}\ and\ \citenamefont
  {Vasilopoulos}(2005)}]{PhysRevB.72.165336}%
  \BibitemOpen
  \bibfield  {author} {\bibinfo {author} {\bibfnamefont {X.~F.}\ \bibnamefont
  {Wang}}\ and\ \bibinfo {author} {\bibfnamefont {P.}~\bibnamefont
  {Vasilopoulos}},\ }\href {\doibase 10.1103/PhysRevB.72.165336} {\bibfield
  {journal} {\bibinfo  {journal} {Phys. Rev. B}\ }\textbf {\bibinfo {volume}
  {72}},\ \bibinfo {pages} {165336} (\bibinfo {year} {2005})}\BibitemShut
  {NoStop}%
\bibitem [{\citenamefont {Tang}\ and\ \citenamefont {Chu}(1999)}]{Tang99:1830}%
  \BibitemOpen
  \bibfield  {author} {\bibinfo {author} {\bibfnamefont {C.~S.}\ \bibnamefont
  {Tang}}\ and\ \bibinfo {author} {\bibfnamefont {C.~S.}\ \bibnamefont {Chu}},\
  }\href {\doibase 10.1103/PhysRevB.60.1830} {\bibfield  {journal} {\bibinfo
  {journal} {Phys. Rev. B}\ }\textbf {\bibinfo {volume} {60}},\ \bibinfo
  {pages} {1830} (\bibinfo {year} {1999})}\BibitemShut {NoStop}%
\bibitem [{\citenamefont {Zhou}\ and\ \citenamefont {Li}(2005)}]{Zhou17:6663}%
  \BibitemOpen
  \bibfield  {author} {\bibinfo {author} {\bibfnamefont {G.}~\bibnamefont
  {Zhou}}\ and\ \bibinfo {author} {\bibfnamefont {Y.}~\bibnamefont {Li}},\
  }\href@noop {} {\bibfield  {journal} {\bibinfo  {journal} {J. Phys.: Condens.
  Matter}\ }\textbf {\bibinfo {volume} {17}},\ \bibinfo {pages} {6663}
  (\bibinfo {year} {2005})}\BibitemShut {NoStop}%
\bibitem [{\citenamefont {Jung}\ \emph {et~al.}(2012)\citenamefont {Jung},
  \citenamefont {Na},\ and\ \citenamefont {Reichl}}]{Jung85:023420}%
  \BibitemOpen
  \bibfield  {author} {\bibinfo {author} {\bibfnamefont {J.-W.}\ \bibnamefont
  {Jung}}, \bibinfo {author} {\bibfnamefont {K.}~\bibnamefont {Na}}, \ and\
  \bibinfo {author} {\bibfnamefont {L.~E.}\ \bibnamefont {Reichl}},\
  }\href@noop {} {\bibfield  {journal} {\bibinfo  {journal} {Phys. Rev. A}\
  }\textbf {\bibinfo {volume} {85}},\ \bibinfo {pages} {023420} (\bibinfo
  {year} {2012})}\BibitemShut {NoStop}%
\bibitem [{\citenamefont {Tang}\ and\ \citenamefont {Chu}(2000)}]{Tang00:127}%
  \BibitemOpen
  \bibfield  {author} {\bibinfo {author} {\bibfnamefont {C.~S.}\ \bibnamefont
  {Tang}}\ and\ \bibinfo {author} {\bibfnamefont {C.~S.}\ \bibnamefont {Chu}},\
  }\href@noop {} {\bibfield  {journal} {\bibinfo  {journal} {Physica B}\
  }\textbf {\bibinfo {volume} {292}},\ \bibinfo {pages} {127} (\bibinfo {year}
  {2000})}\BibitemShut {NoStop}%
\bibitem [{\citenamefont {Zhou}\ \emph {et~al.}(2003)\citenamefont {Zhou},
  \citenamefont {Yang}, \citenamefont {Xiao},\ and\ \citenamefont
  {Li}}]{Zhou68:155309}%
  \BibitemOpen
  \bibfield  {author} {\bibinfo {author} {\bibfnamefont {G.}~\bibnamefont
  {Zhou}}, \bibinfo {author} {\bibfnamefont {M.}~\bibnamefont {Yang}}, \bibinfo
  {author} {\bibfnamefont {X.}~\bibnamefont {Xiao}}, \ and\ \bibinfo {author}
  {\bibfnamefont {Y.}~\bibnamefont {Li}},\ }\href@noop {} {\bibfield  {journal}
  {\bibinfo  {journal} {Phys. Rev. B}\ }\textbf {\bibinfo {volume} {68}},\
  \bibinfo {pages} {155309} (\bibinfo {year} {2003})}\BibitemShut {NoStop}%
\bibitem [{\citenamefont {Spohn}(1980)}]{Spohn53:569}%
  \BibitemOpen
  \bibfield  {author} {\bibinfo {author} {\bibfnamefont {H.}~\bibnamefont
  {Spohn}},\ }\href@noop {} {\bibfield  {journal} {\bibinfo  {journal} {Rev.
  Mod. Phys.}\ }\textbf {\bibinfo {volume} {52}},\ \bibinfo {pages} {569}
  (\bibinfo {year} {1980})}\BibitemShut {NoStop}%
\bibitem [{\citenamefont {Gurvitz}\ and\ \citenamefont
  {Prager}(1996)}]{Gurvitz96:15932}%
  \BibitemOpen
  \bibfield  {author} {\bibinfo {author} {\bibfnamefont {S.~A.}\ \bibnamefont
  {Gurvitz}}\ and\ \bibinfo {author} {\bibfnamefont {Y.~S.}\ \bibnamefont
  {Prager}},\ }\href@noop {} {\bibfield  {journal} {\bibinfo  {journal} {Phys.
  Rev. B}\ }\textbf {\bibinfo {volume} {53}},\ \bibinfo {pages} {15932}
  (\bibinfo {year} {1996})}\BibitemShut {NoStop}%
\bibitem [{\citenamefont {van Kampen}(2001)}]{Kampen01:00}%
  \BibitemOpen
  \bibfield  {author} {\bibinfo {author} {\bibfnamefont {N.~G.}\ \bibnamefont
  {van Kampen}},\ }\href@noop {} {\emph {\bibinfo {title} {Stochastic Processes
  in Physics and Chemistry}}},\ \bibinfo {edition} {2nd}\ ed.\ (\bibinfo
  {publisher} {North-Holland, Amsterdam},\ \bibinfo {year} {2001})\BibitemShut
  {NoStop}%
\bibitem [{\citenamefont {Harbola}\ \emph {et~al.}(2006)\citenamefont
  {Harbola}, \citenamefont {Esposito},\ and\ \citenamefont
  {Mukamel}}]{Harbola06:235309}%
  \BibitemOpen
  \bibfield  {author} {\bibinfo {author} {\bibfnamefont {U.}~\bibnamefont
  {Harbola}}, \bibinfo {author} {\bibfnamefont {M.}~\bibnamefont {Esposito}}, \
  and\ \bibinfo {author} {\bibfnamefont {S.}~\bibnamefont {Mukamel}},\ }\href
  {http://link.aps.org/abstract/PRB/v74/e235309} {\bibfield  {journal}
  {\bibinfo  {journal} {Phys. Rev. B}\ }\textbf {\bibinfo {volume} {74}},\
  \bibinfo {pages} {235309} (\bibinfo {year} {2006})}\BibitemShut {NoStop}%
\bibitem [{\citenamefont {Bruder}\ and\ \citenamefont
  {Schoeller}(1994)}]{PhysRevLett.72.1076}%
  \BibitemOpen
  \bibfield  {author} {\bibinfo {author} {\bibfnamefont {C.}~\bibnamefont
  {Bruder}}\ and\ \bibinfo {author} {\bibfnamefont {H.}~\bibnamefont
  {Schoeller}},\ }\href {\doibase 10.1103/PhysRevLett.72.1076} {\bibfield
  {journal} {\bibinfo  {journal} {Phys. Rev. Lett.}\ }\textbf {\bibinfo
  {volume} {72}},\ \bibinfo {pages} {1076} (\bibinfo {year}
  {1994})}\BibitemShut {NoStop}%
\bibitem [{\citenamefont {Braggio}\ \emph {et~al.}(2006)\citenamefont
  {Braggio}, \citenamefont {K\"{o}nig},\ and\ \citenamefont
  {Fazio}}]{Braggio06:026805}%
  \BibitemOpen
  \bibfield  {author} {\bibinfo {author} {\bibfnamefont {A.}~\bibnamefont
  {Braggio}}, \bibinfo {author} {\bibfnamefont {J.}~\bibnamefont {K\"{o}nig}},
  \ and\ \bibinfo {author} {\bibfnamefont {R.}~\bibnamefont {Fazio}},\
  }\href@noop {} {\bibfield  {journal} {\bibinfo  {journal} {Phys. Rev. Lett.}\
  }\textbf {\bibinfo {volume} {96}},\ \bibinfo {pages} {026805} (\bibinfo
  {year} {2006})}\BibitemShut {NoStop}%
\bibitem [{\citenamefont {Moldoveanu}\ \emph {et~al.}(2009)\citenamefont
  {Moldoveanu}, \citenamefont {Manolescu},\ and\ \citenamefont
  {Gudmundsson}}]{Moldoveanu09:073019}%
  \BibitemOpen
  \bibfield  {author} {\bibinfo {author} {\bibfnamefont {V.}~\bibnamefont
  {Moldoveanu}}, \bibinfo {author} {\bibfnamefont {A.}~\bibnamefont
  {Manolescu}}, \ and\ \bibinfo {author} {\bibfnamefont {V.}~\bibnamefont
  {Gudmundsson}},\ }\href {http://stacks.iop.org/1367-2630/11/073019}
  {\bibfield  {journal} {\bibinfo  {journal} {New Journal of Physics}\ }\textbf
  {\bibinfo {volume} {11}},\ \bibinfo {pages} {073019} (\bibinfo {year}
  {2009})}\BibitemShut {NoStop}%
\bibitem [{\citenamefont {Breuer}\ \emph {et~al.}(1999)\citenamefont {Breuer},
  \citenamefont {Kappler},\ and\ \citenamefont
  {Petruccione}}]{PhysRevA.59.1633}%
  \BibitemOpen
  \bibfield  {author} {\bibinfo {author} {\bibfnamefont {H.-P.}\ \bibnamefont
  {Breuer}}, \bibinfo {author} {\bibfnamefont {B.}~\bibnamefont {Kappler}}, \
  and\ \bibinfo {author} {\bibfnamefont {F.}~\bibnamefont {Petruccione}},\
  }\href {\doibase 10.1103/PhysRevA.59.1633} {\bibfield  {journal} {\bibinfo
  {journal} {Phys. Rev. A}\ }\textbf {\bibinfo {volume} {59}},\ \bibinfo
  {pages} {1633} (\bibinfo {year} {1999})}\BibitemShut {NoStop}%
\bibitem [{\citenamefont {Arnold}\ \emph {et~al.}(2011)\citenamefont {Arnold},
  \citenamefont {Siegmund},\ and\ \citenamefont
  {Pankratov}}]{0953-8984-23-33-335601}%
  \BibitemOpen
  \bibfield  {author} {\bibinfo {author} {\bibfnamefont {T.}~\bibnamefont
  {Arnold}}, \bibinfo {author} {\bibfnamefont {M.}~\bibnamefont {Siegmund}}, \
  and\ \bibinfo {author} {\bibfnamefont {O.}~\bibnamefont {Pankratov}},\ }\href
  {http://stacks.iop.org/0953-8984/23/i=33/a=335601} {\bibfield  {journal}
  {\bibinfo  {journal} {Journal of Physics: Condensed Matter}\ }\textbf
  {\bibinfo {volume} {23}},\ \bibinfo {pages} {335601} (\bibinfo {year}
  {2011})}\BibitemShut {NoStop}%
\bibitem [{\citenamefont {Tan}\ and\ \citenamefont
  {Inkson}(1996)}]{0268-1242-11-11-001}%
  \BibitemOpen
  \bibfield  {author} {\bibinfo {author} {\bibfnamefont {W.-C.}\ \bibnamefont
  {Tan}}\ and\ \bibinfo {author} {\bibfnamefont {J.~C.}\ \bibnamefont
  {Inkson}},\ }\href {http://stacks.iop.org/0268-1242/11/i=11/a=001} {\bibfield
   {journal} {\bibinfo  {journal} {Semiconductor Science and Technology}\
  }\textbf {\bibinfo {volume} {11}},\ \bibinfo {pages} {1635} (\bibinfo {year}
  {1996})}\BibitemShut {NoStop}%
\bibitem [{\citenamefont {Shi}\ \emph {et~al.}(2006)\citenamefont {Shi},
  \citenamefont {Zhang}, \citenamefont {Xiao},\ and\ \citenamefont
  {Niu}}]{PhysRevLett.96.076604}%
  \BibitemOpen
  \bibfield  {author} {\bibinfo {author} {\bibfnamefont {J.}~\bibnamefont
  {Shi}}, \bibinfo {author} {\bibfnamefont {P.}~\bibnamefont {Zhang}}, \bibinfo
  {author} {\bibfnamefont {D.}~\bibnamefont {Xiao}}, \ and\ \bibinfo {author}
  {\bibfnamefont {Q.}~\bibnamefont {Niu}},\ }\href {\doibase
  10.1103/PhysRevLett.96.076604} {\bibfield  {journal} {\bibinfo  {journal}
  {Phys. Rev. Lett.}\ }\textbf {\bibinfo {volume} {96}},\ \bibinfo {pages}
  {076604} (\bibinfo {year} {2006})}\BibitemShut {NoStop}%
\bibitem [{\citenamefont {Bray-Ali}\ and\ \citenamefont
  {Nussinov}(2009)}]{PhysRevB.80.012401}%
  \BibitemOpen
  \bibfield  {author} {\bibinfo {author} {\bibfnamefont {N.}~\bibnamefont
  {Bray-Ali}}\ and\ \bibinfo {author} {\bibfnamefont {Z.}~\bibnamefont
  {Nussinov}},\ }\href {\doibase 10.1103/PhysRevB.80.012401} {\bibfield
  {journal} {\bibinfo  {journal} {Phys. Rev. B}\ }\textbf {\bibinfo {volume}
  {80}},\ \bibinfo {pages} {012401} (\bibinfo {year} {2009})}\BibitemShut
  {NoStop}%
\bibitem [{\citenamefont {Rashba}(2003)}]{PhysRevB.68.241315}%
  \BibitemOpen
  \bibfield  {author} {\bibinfo {author} {\bibfnamefont {E.~I.}\ \bibnamefont
  {Rashba}},\ }\href {\doibase 10.1103/PhysRevB.68.241315} {\bibfield
  {journal} {\bibinfo  {journal} {Phys. Rev. B}\ }\textbf {\bibinfo {volume}
  {68}},\ \bibinfo {pages} {241315} (\bibinfo {year} {2003})}\BibitemShut
  {NoStop}%
\bibitem [{\citenamefont {Drouhin}\ \emph {et~al.}(2011)\citenamefont
  {Drouhin}, \citenamefont {Fishman},\ and\ \citenamefont
  {Wegrowe}}]{PhysRevB.83.113307}%
  \BibitemOpen
  \bibfield  {author} {\bibinfo {author} {\bibfnamefont {H.-J.}\ \bibnamefont
  {Drouhin}}, \bibinfo {author} {\bibfnamefont {G.}~\bibnamefont {Fishman}}, \
  and\ \bibinfo {author} {\bibfnamefont {J.-E.}\ \bibnamefont {Wegrowe}},\
  }\href {\doibase 10.1103/PhysRevB.83.113307} {\bibfield  {journal} {\bibinfo
  {journal} {Phys. Rev. B}\ }\textbf {\bibinfo {volume} {83}},\ \bibinfo
  {pages} {113307} (\bibinfo {year} {2011})}\BibitemShut {NoStop}%
\bibitem [{\citenamefont {Bottegoni}\ \emph {et~al.}(2012)\citenamefont
  {Bottegoni}, \citenamefont {Drouhin}, \citenamefont {Fishman},\ and\
  \citenamefont {Wegrowe}}]{PhysRevB.85.235313}%
  \BibitemOpen
  \bibfield  {author} {\bibinfo {author} {\bibfnamefont {F.}~\bibnamefont
  {Bottegoni}}, \bibinfo {author} {\bibfnamefont {H.-J.}\ \bibnamefont
  {Drouhin}}, \bibinfo {author} {\bibfnamefont {G.}~\bibnamefont {Fishman}}, \
  and\ \bibinfo {author} {\bibfnamefont {J.-E.}\ \bibnamefont {Wegrowe}},\
  }\href {\doibase 10.1103/PhysRevB.85.235313} {\bibfield  {journal} {\bibinfo
  {journal} {Phys. Rev. B}\ }\textbf {\bibinfo {volume} {85}},\ \bibinfo
  {pages} {235313} (\bibinfo {year} {2012})}\BibitemShut {NoStop}%
\bibitem [{\citenamefont {Sonin}(2007{\natexlab{b}})}]{PhysRevB.76.033306}%
  \BibitemOpen
  \bibfield  {author} {\bibinfo {author} {\bibfnamefont {E.~B.}\ \bibnamefont
  {Sonin}},\ }\href {\doibase 10.1103/PhysRevB.76.033306} {\bibfield  {journal}
  {\bibinfo  {journal} {Phys. Rev. B}\ }\textbf {\bibinfo {volume} {76}},\
  \bibinfo {pages} {033306} (\bibinfo {year} {2007}{\natexlab{b}})}\BibitemShut
  {NoStop}%
\bibitem [{\citenamefont {Meijer}\ \emph {et~al.}(2002)\citenamefont {Meijer},
  \citenamefont {Morpurgo},\ and\ \citenamefont
  {Klapwijk}}]{PhysRevB.66.033107}%
  \BibitemOpen
  \bibfield  {author} {\bibinfo {author} {\bibfnamefont {F.~E.}\ \bibnamefont
  {Meijer}}, \bibinfo {author} {\bibfnamefont {A.~F.}\ \bibnamefont
  {Morpurgo}}, \ and\ \bibinfo {author} {\bibfnamefont {T.~M.}\ \bibnamefont
  {Klapwijk}},\ }\href {\doibase 10.1103/PhysRevB.66.033107} {\bibfield
  {journal} {\bibinfo  {journal} {Phys. Rev. B}\ }\textbf {\bibinfo {volume}
  {66}},\ \bibinfo {pages} {033107} (\bibinfo {year} {2002})}\BibitemShut
  {NoStop}%
\bibitem [{\citenamefont {Moln\'ar}\ \emph {et~al.}(2004)\citenamefont
  {Moln\'ar}, \citenamefont {Peeters},\ and\ \citenamefont
  {Vasilopoulos}}]{PhysRevB.69.155335}%
  \BibitemOpen
  \bibfield  {author} {\bibinfo {author} {\bibfnamefont {B.}~\bibnamefont
  {Moln\'ar}}, \bibinfo {author} {\bibfnamefont {F.~M.}\ \bibnamefont
  {Peeters}}, \ and\ \bibinfo {author} {\bibfnamefont {P.}~\bibnamefont
  {Vasilopoulos}},\ }\href {\doibase 10.1103/PhysRevB.69.155335} {\bibfield
  {journal} {\bibinfo  {journal} {Phys. Rev. B}\ }\textbf {\bibinfo {volume}
  {69}},\ \bibinfo {pages} {155335} (\bibinfo {year} {2004})}\BibitemShut
  {NoStop}%
\bibitem [{\citenamefont {Byers}\ and\ \citenamefont
  {Yang}(1961)}]{PhysRevLett.7.46}%
  \BibitemOpen
  \bibfield  {author} {\bibinfo {author} {\bibfnamefont {N.}~\bibnamefont
  {Byers}}\ and\ \bibinfo {author} {\bibfnamefont {C.~N.}\ \bibnamefont
  {Yang}},\ }\href {\doibase 10.1103/PhysRevLett.7.46} {\bibfield  {journal}
  {\bibinfo  {journal} {Phys. Rev. Lett.}\ }\textbf {\bibinfo {volume} {7}},\
  \bibinfo {pages} {46} (\bibinfo {year} {1961})}\BibitemShut {NoStop}%
\bibitem [{\citenamefont {Viefers}\ \emph {et~al.}(2004)\citenamefont
  {Viefers}, \citenamefont {Koskinen}, \citenamefont {Deo},\ and\ \citenamefont
  {Manninen}}]{Viefers20041}%
  \BibitemOpen
  \bibfield  {author} {\bibinfo {author} {\bibfnamefont {S.}~\bibnamefont
  {Viefers}}, \bibinfo {author} {\bibfnamefont {P.}~\bibnamefont {Koskinen}},
  \bibinfo {author} {\bibfnamefont {P.~S.}\ \bibnamefont {Deo}}, \ and\
  \bibinfo {author} {\bibfnamefont {M.}~\bibnamefont {Manninen}},\ }\href
  {\doibase 10.1016/j.physe.2003.08.076} {\bibfield  {journal} {\bibinfo
  {journal} {Physica E: Low-dim. Systems and Nanostr.}\ }\textbf {\bibinfo
  {volume} {21}},\ \bibinfo {pages} {1 } (\bibinfo {year} {2004})}\BibitemShut
  {NoStop}%
\bibitem [{\citenamefont {Whitney}(2008)}]{Whitney08:175304}%
  \BibitemOpen
  \bibfield  {author} {\bibinfo {author} {\bibfnamefont {R.~S.}\ \bibnamefont
  {Whitney}},\ }\href {\doibase 10.1088/1751-8113/41/17/175304} {\bibfield
  {journal} {\bibinfo  {journal} {J. Phys. A: Math. Theor.}\ }\textbf {\bibinfo
  {volume} {41}},\ \bibinfo {pages} {175304} (\bibinfo {year}
  {2008})}\BibitemShut {NoStop}%
\bibitem [{\citenamefont {Gudmundsson}\ \emph {et~al.}(2012)\citenamefont
  {Gudmundsson}, \citenamefont {Jonasson}, \citenamefont {Tang}, \citenamefont
  {Goan},\ and\ \citenamefont {Manolescu}}]{Gudmundsson85:075306}%
  \BibitemOpen
  \bibfield  {author} {\bibinfo {author} {\bibfnamefont {V.}~\bibnamefont
  {Gudmundsson}}, \bibinfo {author} {\bibfnamefont {O.}~\bibnamefont
  {Jonasson}}, \bibinfo {author} {\bibfnamefont {C.-S.}\ \bibnamefont {Tang}},
  \bibinfo {author} {\bibfnamefont {H.-S.}\ \bibnamefont {Goan}}, \ and\
  \bibinfo {author} {\bibfnamefont {A.}~\bibnamefont {Manolescu}},\ }\href@noop
  {} {\bibfield  {journal} {\bibinfo  {journal} {Phys. Rev. B}\ }\textbf
  {\bibinfo {volume} {85}},\ \bibinfo {pages} {075306} (\bibinfo {year}
  {2012})}\BibitemShut {NoStop}%
\bibitem [{\citenamefont {Gudmundsson}\ \emph {et~al.}(2009)\citenamefont
  {Gudmundsson}, \citenamefont {Gainar}, \citenamefont {Tang}, \citenamefont
  {Moldoveanu},\ and\ \citenamefont {Manolescu}}]{1367-2630-11-11-113007}%
  \BibitemOpen
  \bibfield  {author} {\bibinfo {author} {\bibfnamefont {V.}~\bibnamefont
  {Gudmundsson}}, \bibinfo {author} {\bibfnamefont {C.}~\bibnamefont {Gainar}},
  \bibinfo {author} {\bibfnamefont {C.-S.}\ \bibnamefont {Tang}}, \bibinfo
  {author} {\bibfnamefont {V.}~\bibnamefont {Moldoveanu}}, \ and\ \bibinfo
  {author} {\bibfnamefont {A.}~\bibnamefont {Manolescu}},\ }\href
  {http://stacks.iop.org/1367-2630/11/i=11/a=113007} {\bibfield  {journal}
  {\bibinfo  {journal} {New Journal of Physics}\ }\textbf {\bibinfo {volume}
  {11}},\ \bibinfo {pages} {113007} (\bibinfo {year} {2009})}\BibitemShut
  {NoStop}%
\bibitem [{sup()}]{supplmat}%
  \BibitemOpen
  \href@noop {} {\emph {\bibinfo {title} {See Supplemental Material at [URL
  will be inserted by publisher] for the time- and space-dependence of the spin
  current densities $\mathbf{j}^{\gamma}(\mathbf{r},t)$ without photon cavity
  and of the spin photocurrent densities $\mathbf{j}_{\rm
  ph}^{\gamma,p}(\mathbf{r},t)$ of photon field polarization $p=x,y$. The spin
  polarization $\gamma=x,y,z$, the Rashba coefficient $\alpha=5$~meV\,nm and
  the Dresselhaus coefficient $\beta=0$. A vector of length $a_w$ corresponds
  to $1.25\times 10^{-3} \rm{meV}/a_w$.}}}\BibitemShut {Stop}%
\end{thebibliography}
%

%
\end{document}